\documentclass[12pt]{article}

\usepackage{latexsym}
\usepackage{amsthm}
\usepackage{amsmath}
\usepackage[dvips]{graphicx}
\usepackage{amssymb}
\usepackage{amsbsy}

\newcommand{\nn}{\nonumber}

\newcommand{\del}{\partial}

\newcommand{\be}{\begin{eqnarray}}
\newcommand{\ee}{\end{eqnarray}}

\begin{document}

\baselineskip=18pt

\setcounter{footnote}{0}
\setcounter{figure}{0}
\setcounter{table}{0}

\begin{titlepage}
\vspace{1cm}

\begin{center}

{\Large \bf A Duality For The S Matrix}

\vspace{0.8cm}

{\bf N. Arkani-Hamed$^a$, F. Cachazo$^b$, C. Cheung$^a$ and J. Kaplan$^a$}

\vspace{.5cm}

{\it $^{a}$ School of Natural Sciences, Institute for Advanced Study, \\  Princeton, NJ 08540, USA}

{\it $^{b}$ Perimeter Institute for Theoretical Physics, \\ Waterloo, Ontario N2J W29, CA}

\end{center}

\vspace{1cm}

\begin{abstract}

We propose a dual formulation for the S Matrix of ${\cal N} = 4$ SYM. 
The dual provides a basis for the ``leading singularities'' of
scattering amplitudes to all orders in perturbation theory, which are sharply defined, IR safe data that
uniquely determine the full amplitudes at tree level and 1-loop, and are
conjectured to do so at all loop orders. The scattering amplitude for
$n$ particles in the sector with $k$ negative helicity gluons is
associated with a simple integral over the space of $k$ planes in $n$
dimensions, with the action of parity and cyclic symmetries manifest. 
The residues of the integrand compute a basis for the leading
singularities. A given leading singularity is associated with a
particular choice of integration contour, which we explicitly identify
at tree level and 1-loop for all NMHV amplitudes as well as the 8
particle N$^2$MHV amplitude. We also identify a number of 2-loop leading
singularities for up to 8 particles. There are a large number of
relations among residues which follow from the multi-variable
generalization of Cauchy's theorem known as the ``global residue
theorem".  These relations imply highly non-trivial identities
guaranteeing the equivalence of many different representations of the
same amplitude. They also enforce the cancellation of non-local poles as well as consistent
infrared structure at loop level. Our conjecture connects the physics of
scattering amplitudes to a particular subvariety in a Grassmannian;
space-time locality is reflected in the topological properties of this
space.

\end{abstract}

\bigskip
\bigskip

\end{titlepage}

\section{A Dual Theory for the S Matrix}

Scattering amplitudes in gauge theories and gravity possess
extraordinary structures that are completely invisible in the
textbook formulation of quantum field theory.  Much of the recent
progress in understanding scattering amplitudes was triggered by
Witten's twistor string theory for ${\cal N} = 4$ SYM amplitudes
\cite{Witten:2003nn}. The stimulus provided by this proposal led to
a number of powerful new methods for explicitly computing
amplitudes, many of which arose from a fresh examination of older
ideas in field theory and S Matrix theory. These include the BCFW
recursion relations for tree amplitudes
\cite{bcf,bcfw,Bedford:2005yy,Cachazo:2005ca,Benincasa:2007qj,ArkaniHamed:2008yf,Cheung:2008dn},
along with their supersymmetric extension \cite{simplest, Brandhuber:2008pf}. Amongst
other things, the recursion relations and their explicit solution
\cite{Drummond:2008cr} have allowed for a direct verification of
the remarkable dual conformal symmetry \cite{Drummond:2006rz,Alday:2007hr} of SYM amplitudes at tree
level \cite{Drummond:2008bq}. Furthermore, as was realized quite early on by Hodges
\cite{Hodges:2005bf, Hodges:2005aj, Hodges:2006tw}
and rigorously understood more recently in (2,2) signature \cite{ArkaniHamed:2009si,lioneldavid}, the BCFW recursion relations
are most naturally formulated in twistor space \cite{Penrose:1967wn,
Penrose:1968me,Penrose:1972ia, Penrose:1999cw}, where a beautiful
``Hodges diagram" formalism exposes structures in the amplitudes that are obscured in momentum space.
All of these amazing properties strongly suggest the existence of a ``weak-weak" dual theory which directly computes on-shell scattering amplitudes without resorting to the conventional evolution through space-time. In this paper we propose a dual formulation for the S Matrix of ${\cal N} = 4$ SYM along these lines.

\subsection{What Should the Dual Compute?}

As a preface to our proposal, it is worth discussing what properties should be expected of a dual theory for the S Matrix on general grounds. The most basic question of all is simple: what should such a theory compute? At tree level, scattering amplitudes are well-defined rational functions of the kinematical invariants, but loop level amplitudes suffer from infrared divergences. Is there any sharply defined data about loop amplitudes that might serve as the output of a putative dual theory?

A look at the general structure of scattering amplitudes at 1-loop offers a very natural answer to this question.
As is well-known, any 1-loop amplitude in four dimensions can be written as a linear combination of a basis of scalar  integrals with rational coefficients \cite{basis}.  A generic theory will have scalar box, triangle and bubble integrals, as well as purely rational terms. All of the infrared divergences are isolated in the basis of scalar integrals; the rational coefficients then carry all the information about scattering amplitudes at 1-loop, and can therefore serve as sharp ``data" to be computed by a well-defined theory.

As reviewed and discussed at length in \cite{simplest}, these rational functions have a beautiful and very physical interpretation. Loop
level amplitudes have branch cuts as a function of the kinematical
invariants, and so it is natural to examine the discontinuities
across these branch cuts. The discontinuities themselves can have
branch cuts as a function of other kinematical invariants and
the procedure can be continued, finally arriving at an object which is aptly named the discontinuity
across a leading singularity, which is the highest co-dimension singularity possible. In what follows we will often abuse nomenclature and use the phrase ``leading singularity" to mean ``discontinuity across the leading singularity". Just as the textbook unitarity cut is computed by putting two propagators on-shell, the leading singularity at 1-loop in four dimensions is computed by putting four propagators on-shell \cite{lead}.

A related point of view is to consider the sum of Feynman diagrams
in a 1-loop amplitude which share four given propagators. For an
$n$-point amplitude, the four propagators in the loop separate the
$n$-particles into four sets. Let $K_1,K_2,K_3$ and $K_4$ be the sum
of the momenta in each set. It is possible to put all these Feynman
diagrams together as a single integral of the form
\be
\label{diver}
\int \frac{d^4\ell}{\ell^2 (\ell-K_1)^2(\ell-K_1-K_2)^2(\ell+K_4)^2}R(\ell)
\ee
where we have exhibited the four special propagators and where $R(\ell)$ is a rational function of $\ell$ and the external momenta which contains the rest of propagators and tensor structure in the Feynman diagrams. The integral (\ref{diver}) has IR divergences and in a general theory it could have UV divergences as well. All divergences come from choosing the region of integration to be $\mathbb{R}^4$. If, instead, we choose to interpret the integral as a contour integral in $\mathbb{C}^4$ with contour $T^4\backsimeq (S^1)^4$ given by $|\ell^2|=\epsilon$, $|(\ell-K_1)^2|=\epsilon$, $|(\ell-K_1-K_2)^2|=\epsilon$ and $|(\ell+K_4)^2|=\epsilon$ with $\epsilon$ a small positive real number, then the integral becomes finite. In mathematical terms, the new contour integral is computing a residue! In order to compute the residue we have to evaluate $R(\ell)$ on the points, $\ell^*$, in $\mathbb{C}^4$ which satisfy the equations with $\epsilon =0$. There are two solutions and therefore two different $T^4$ contours. The two contours are related by parity. Physically, $R(\ell^*)$ factors as the product of four tree amplitudes. Explicitly, $R(
\ell^*)\to \sum M^{\rm tree}_{ul}M_{ur}^{\rm tree}M_{bl}^{\rm tree}M_{br}^{\rm tree}$, where the subindices mean $u=$upper, $l=$left, $r$=right, and $b=$bottom, to refer to the location in the figure below. The sum is over all physical states in the theory which propagate in the internal legs \cite{lead}.

\be
\includegraphics[scale=0.9]{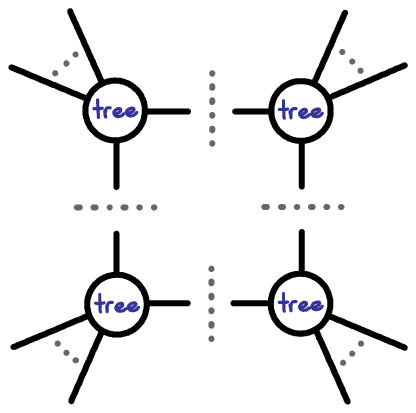} \nonumber
\ee

$R(\ell^*)$ is a leading singularity of the amplitude. At 1-loop
something special happens. In the basis of scalar integrals,
the scalar box integrals are the only objects that can be cut
in this way (with suitable normalization giving ``1"), and so the
rational coefficients of the scalar box integrals are precisely the
leading singularities at 1-loop \cite{lead}. More precisely, the coefficient of a given box integral is the sum over the two leading singularities associated with the two different solutions for $\ell^*$ or equivalently the two $T^4$ contours. 

An ansatz for the
amplitude which includes only scalar boxes with these coefficients
is an excellent initial guess for the whole amplitude, since it is
guaranteed to reproduce the leading singularity. This ansatz also
yields contributions to subleading singularities, which involve
cutting fewer lines.  In a generic theory, the set of subleading
singularities produced in this way is incomplete, and therefore the
lower-point scalar integrals and rational pieces must be present to
compensate for the difference. The remarkable feature of maximally
supersymmetric theories at 1-loop is that fixing the correct leading
singularity suffices to fix all the subleading singularities as
well, so that only the scalar box integrals appear \cite{Bern:1994zx}: \be
\includegraphics[scale=0.9]{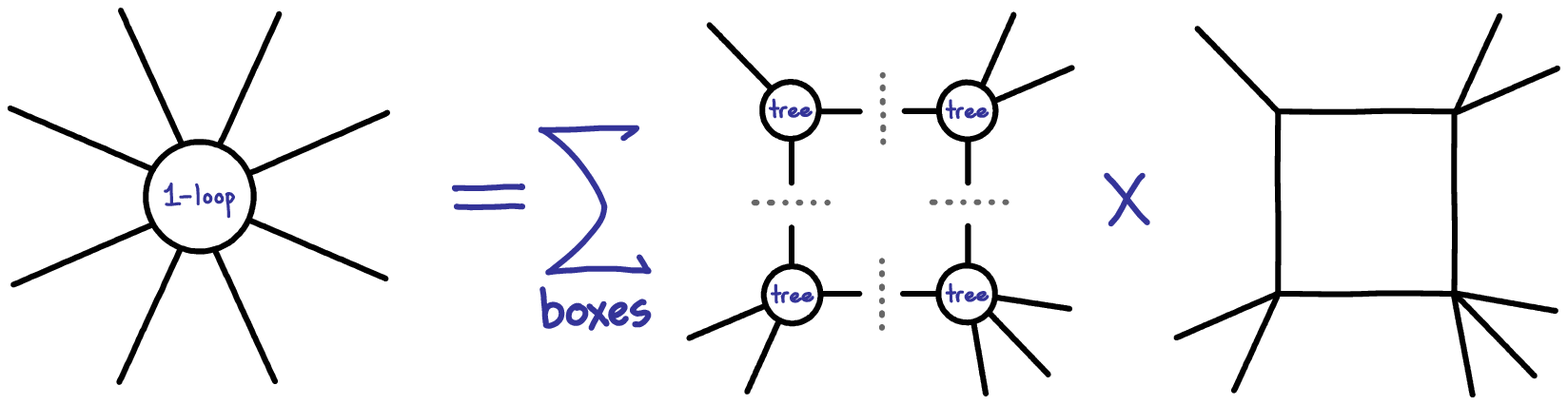} \nonumber
\ee

The procedure for moving on to higher loops is now clear \cite{Buchbinder:2005wp, Cachazo:2008dx, Bern:2007ct}. At $l$-loop order one can choose Feynman diagrams with a given set of $4l$-propagators, with each loop variable appearing in at least four of the propagators of the set. The loop integral
\be
\int \frac{d^4\ell_1\ldots d^4\ell_l}{\prod_{i=1}^{4l}P_i^2(\ell_1,\ldots, \ell_l,k_1,\ldots k_n)}R(\ell_1,\ldots ,\ell_l),
\ee
which generically suffers from IR divergences (UV divergences are absent in ${\cal N}=4$ SYM), becomes a completely well defined object if we take it to be a contour integral in $\mathbb{C}^{4l}$ with a $T^{4l}$ contour given by $|P_1^2|= \ldots = |P_{4l}^2|=\epsilon$. Once again, $R(\ell_1,\ldots, \ell_l)$ is the product of tree amplitudes summed over all possible physical on-shell states in the theory propagating in the internal lines.
As a simple illustration consider the following 2-loop 5-point leading singularity
\be
\includegraphics[scale=0.9]{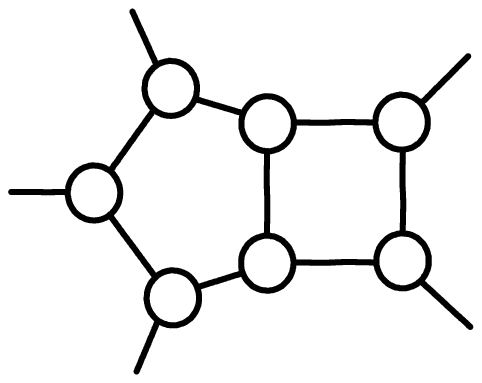} \nonumber
\ee where the open circles denote tree amplitudes and all internal
lines are on-shell.

A comment is in order here. For a given number of particles, $n$, it
is impossible to find diagrams with $4l$ propagators if $l>n-3$.
This problem is resolved by defining ``composite" leading
singularities. The theory of composite leading singularities was
developed in \cite{Buchbinder:2005wp, Cachazo:2008dx, Cachazo:2008vp,
Cachazo:2008hp}. At the end of section 3.2 we illustrate it with a
four-particle amplitude at 2-loop order.

Leading singularities, including composite ones, constitute perfectly well-defined data associated with scattering amplitudes at all orders in perturbation theory. It is natural to conjecture that, for maximally supersymmetric theories, the leading singularities continue to fully determine the amplitude at any loop order as well; for ${\cal N} = 4$ SYM, some strong evidence for this proposal has been provided at 2- and 3- loop order  \cite{Buchbinder:2005wp, Cachazo:2008vp, Cachazo:2008hp, Spradlin:2008uu}. We will optimistically assume that this conjecture is true, and that the leading singularities not only constitute well-defined data about the S Matrix at all orders in perturbation theory, but also suffice to completely determine it.

It is important to mention that the direct connection between
leading singularities and scalar integral coefficients at one-loop
does not generalize to higher loops. At higher loops, there is not a
unique basis of integrals. In the literature of higher loop ${\cal
N}=4$ amplitudes the main emphasis has been on expressing loop
amplitudes as a sum over scalar integrals and generalizations, which
include loop momentum dependent numerators, with coefficients which
are rational functions of the kinematical invariants (see e.g. \cite{Bern:2006ew, Bern:2007ct}). In
\cite{Cachazo:2008vp}, it was proposed that the coefficients appearing in the linear combination of scalar integrals can be completely determined 
by correctly matching all the leading singularities. 
This has been tested for $5$ and
$6$-point amplitudes at $2$-loop and $5$-points at 3-loops \cite{Cachazo:2008vp, Cachazo:2008hp, Spradlin:2008uu}\footnote{The five-particle and the parity even part of the six particle 2-loop amplitudes where originally computed using the unitarity-based method in \cite{zzbb}.}.
It is not the purpose of this paper to determine amplitudes in terms of
a given basis of scalar integrals; rather we aim to determine the
leading singularities directly, which are physical and basis
independent. Only at one-loop will we interchangeably talk about
leading singularities and scalar boxes, with the hope that this will
not generate confusions when discussing two-loop leading
singularities.

Summarizing, we have arrived at a natural picture for what a dual theory for the S
Matrix should compute: it should produce the leading singularities
of scattering amplitudes at any loop level.

\subsection{What Symmetries Should It Manifest?}

We will use spinor-helicity variables $\lambda_a,\tilde \lambda_a$ associated with the momentum $p_a$ as $p^{\alpha \dot \alpha}_a = \lambda_a^\alpha \tilde \lambda^{\dot \alpha}_a$ where $a=1, \cdots, n$ indexes the $n$ external particles \cite{Dixon:1996wi}. Since all helicity states are in the same supermultiplet, we can also label the external particles by Grassmann variables $\tilde \eta_a$ so that the amplitude is a smooth function of $(\lambda_a,\tilde \lambda_a,\tilde \eta_a)$. Now, the usual local formulation of ${\cal N}=4$ SYM makes two
symmetries manifest. The Bose symmetry of the non-supersymmetric theory is enhanced to a cyclic symmetry of the color-stripped superamplitude \cite{Mangano:1990by}, which simply acts as $g: a \to a+1$ (mod $n$).  The other symmetry is parity.

Nonetheless, even at tree-level both of these symmetries are obscured by the BCFW formalism.
Manifest cyclicity and parity are broken due to the choice
of two special external lines $i_*,j_*$ and the deformation of $\lambda_{i_*}$ and $\tilde
\lambda_{j_*}$ while leaving $\tilde \lambda_{i_*},\lambda_{j_*}$ untouched. Since cyclic symmetry and parity  are not manifest, they appear as highly non-trivial identities relating different BCFW
representations of the same amplitude.

The simplest non-vanishing amplitudes in SYM theory are in the sector with two negative or positive helicity gluons, the so-called MHV and 
$\overline{{\rm MHV}}$ amplitudes. In this case the superamplitude is determined by the famous Parke-Taylor amplitude \cite{Parke:1986gb}, and is manifestly cyclic and parity symmetric. The first example illustrating the tension with manifest cyclic and parity symmetries occurs in the sector with 3 gluons of a given helicity (the NMHV amplitudes) and the simplest case is the 6 particle amplitude. Consider for instance the alternating helicity
configuration. Applying the usual BCFW deformation on $\lambda_1,
\tilde \lambda_2$ yields a three-term representation for this amplitude, which
manifests the (obvious) symmetry $g^2: a \to a+2$ shifting the particle
index by two units:
\begin{equation}
\label{6BCFW}
M^{+-+-+-}_{{\rm BCFW}} = (1 + g^2 + g^4) \,\left[\frac{\langle 4 6 \rangle^4 [1 3]^4}{[12][23]\langle 45 \rangle \langle 56 \rangle (p_4 + p_5 + p_6)^2}
\times \frac{1}{\langle 6 |5 + 4|3] \langle 4|5 + 6 |1]} \right]
\end{equation}
Now with the supersymmetric form of the BCFW recursion relations, we
can also consider the parity conjugate of this deformation, where
$\lambda_2,\tilde \lambda_1$ are deformed. This yields a
parity-conjugated form of the same amplitude given by
\begin{equation}
\label{6PBCFW}
M^{+-+-+-}_{{\rm P(BCFW)}} =(1 + g^2 + g^4) \left[\frac{[3|(2+4)|6 \rangle^4}{[23][34]\langle 56 \rangle \langle 61 \rangle (p_5 + p_6 + p_1)^2}
\times \frac{1}{\langle 1 |6 + 5|4] \langle 5|6+ 1 |2]}\right]
\end{equation}
This is not manifestly the same expression as $M_{{\rm BCFW}}$.
The statement of parity invariance is instead a highly
non-trivial 6-term identity \be \label{hexagon}M_{{\rm BCFW}} =
M_{{\rm P(BCFW)}} \ee It is straightforward  to see that the same identity is enforced by demanding cyclic
symmetry of the superamplitude--indeed it is one component of the
supersymmetric expression of cyclicity. At higher points,
increasingly complicated identities must be satisfied: a 12 term
identity for the 7 point NMHV amplitude, a 40 term identity for the
8-pt $N^2$MHV amplitudes and so on.

The simplicity of the BCFW construction and its clear connection to deeper structures in twistor space suggest that this representation of the amplitudes is secretly being computed by a dual theory. If this is the case, it makes no sense for any particular BCFW representation to be privileged; we should expect that parity and cyclic symmetries are manifest in the dual theory. However physical amplitudes should not be output in a unique form, but should instead somehow be associated with equivalence classes, any given representative of which is not manifestly symmetric. Some powerful mathematical structure must therefore be at work
guaranteeing the equivalence of these different representations and
producing, for instance, the remarkable 6,12,40 and higher term identities we mentioned above.

\subsection{What Miracles Should It Perform?}

There is an even more basic reason to suspect a powerful structure at work: any dual theory for the S Matrix, making no reference to space-time, must provide an understanding of how local space-time physics  emerges. The AdS/CFT correspondence \cite{adscft} has made us accustomed to the holographic generation of spatial dimensions, but the details of this mapping are shrouded in the mysteries of the strongly coupled CFT. On the other hand, in a ``weak-weak" dual theory for the S Matrix, the mechanism allowing the emergence of space-time has nowhere to hide, and must be visible perturbatively.

It is important to first understand how the presence of a bulk space-time with local interactions is encoded in the S Matrix. An important consequence of space-time locality is that the amplitude has a highly constrained singularity structure. At tree level, the amplitudes can only have ``physical poles", of the form
\begin{equation}
\frac{1}{(p_{a_1} + \cdots + p_{a_m})^2}
\end{equation}
This is clear from Feynman diagrams. However, one can easily
imagine other types of poles, for instance of the form
\begin{equation}
\label{spurious}
\frac{1}{\langle 1 | 2 + 3 |4]}
%\, \, {\rm or} \, \, \frac{1}{[7|(5+6)|4\rangle[7|(1+2)|3\rangle \langle 6|(4+5)(2+3)|1\rangle}
\end{equation}
These cannot possibly arise from Feynman diagrams, and  do
not admit any sort of local space-time interpretation. The most basic indication of space-time locality is therefore that these sorts of non-local ``unphysical poles" must be absent from the amplitudes.

Note however that the individual terms in the BCFW form
of the amplitudes in equation (\ref{6BCFW}) {\it do} have unphysical
poles which we highlighted by separating from the physical poles as an explicit factor.
The presence of unphysical poles in each term is in perfect accord with the idea that these
expressions are being produced by a dual theory with no regard for
manifest space-time locality. But the particular combination of
objects appearing in the full amplitude must ensure that the
unphysical poles cancel, rendering them ``spurious". Simply looking
at equation (\ref{6BCFW}), the cancellation of the unphysical poles
looks somewhat miraculous. However, this cancellation follows
directly from the remarkable 6-term identity
$M_{{\rm BCFW}} = M_{{\rm P(BCFW)}}$ we just discussed in the previous subsection.
This is because the
unphysical poles appearing in $M_{\rm BCFW}$ are different than the ones
appearing in $M_{\rm P(BCFW)}$.

We therefore see that the consistency of the BCFW rules with local space-time physics is highly non-trivial,
since the individual pieces in a BCFW expression for scattering amplitudes can not arise
from local physics. However, in this example the same powerful mathematical
structure guaranteeing the equivalence of different
representations of the amplitude also ensures the absence of
unphysical poles and makes a local
spacetime interpretation possible. 

Recently, Hodges has given a remarkable
interpretation of NMHV amplitudes as volumes of polytopes in the twistor space associated with the space on which dual conformal transformations act as
conformal transformations \cite{Hodges:2009hk}. His representation
keeps the cyclic symmetries manifest and beautifully explains the cancellation
of spurious poles. We are seeking a similar understanding which does not
make heavy use of the dual space, since amongst other things, we are ultimately interested in
describing gravity, where this dual space is unlikely to play a central role. See also \cite{Korchemsky:2009hm} for another interesting discussion of spurious poles in tree amplitudes.

Beyond tree-level, there are many other relations which must be satisfied by scattering amplitudes  in order to be consistent with a local space-time description. At 1-loop, the leading double-logarithmic IR divergence of the amplitude must be proportional to the tree amplitude.  Working in dimensional regularization, this implies \cite{IReq, catani}
\be
\label{IRfirst}
\left. M^{\rm 1 \hbox{-} loop}_n\right|_{\rm IR}  = -\frac{1}{\epsilon^2}\sum_{i=1}^n (-s_{i,i+1})^{\epsilon} M^{\rm tree}_n.
\ee
This ``IR equation" guarantees that the IR divergence of the 1-loop amplitude has the
correct physical interpretation of accounting for the soft emission
of gluons in the tree-level process. As we reviewed in the previous
subsection, one-loop amplitudes can be written as linear
combinations of scalar box integrals; each integral can have IR
divergences that depend on the kinematical invariants of the
particular box under consideration, and thus there are many IR
equations associated with the double logarithms of different
kinematical invariants. Equation (\ref{IRfirst}) thus becomes a set
of relations constraining the various scalar box coefficients.

These equations naturally split into two types. One set constrains a
combination of scalar boxes with double logarithms of invariants of the form $(p_i+p_{i+1})^2$ to be proportional to the tree amplitude, while the remaining ones
constrain combinations of scalar box coefficients to vanish. A particular linear combination of these equations \cite{Roiban:2004ix,simplest} directly yields the BCFW recursion relations
in their parity-symmetric form  $\frac{1}{2}$ [{\rm BCFW} + {\rm P(BCFW)}] \cite{bcf} (indeed the BCFW recursion relations were first discovered by conjecturing that the two pieces were equal \cite{bcf}). 
This has a pleasing implication. We motivated our discussion of
leading singularities by the search for IR safe data at loop level,
but of course the tree amplitudes are also perfectly well-defined
rational functions, and it may have seemed odd to have two different
kinds of sharp data associated with the S Matrix. Fortunately this
is not the case--a subset of the leading singularities contain all
the information in the tree amplitudes as well.

The IR equations are a reflection of both locality and unitarity in the scattering amplitudes, and
given that they directly translate into a statement about relations
among leading singularities, a dual theory that computes leading
singularities should provide a direct understanding of them. 

\subsection{Our Conjecture}

In a series of papers, we have been pursuing clues to a dual theory for the S Matrix, largely by studying tree-level amplitudes. In our previous paper \cite{ArkaniHamed:2009si} (see also \cite{lioneldavid}) we showed that the BCFW construction of tree amplitudes is most naturally formulated in twistor space. While the twistor space interpretation has been extremely useful for organizing and identifying hidden patterns in the BCFW representation of tree amplitudes, twistor space alone does not provide a formulation with manifest cyclic and parity symmetry that illuminates the crucial cancellation of unphysical poles.  That said, in twistor space and momentum space, we saw that amplitudes can be written in a novel form which we called the ``link representation", offering a hope for unifying the different BCFW terms into a single object. Pursuing this line of thought, we were led to consider the following object as an ansatz for the $n$-particle tree amplitude in the sector with $k$ negative helicity gluons:
\begin{equation} \label{first}
\int \frac{d^{k \times n} C_{\alpha a}}{(12\cdots k) \,
(23\cdots (k+1)\, ) \, \cdots (n 1 \cdots (k-1)\,)} \prod_{\alpha =
1}^k \delta^{4|4}(C_{\alpha a} {\cal W}_a)
\end{equation}
where the ${\cal W}_a$ are twistor variables, obtained by
fourier-transforming with respect to the $\lambda_a$, ${\cal W} = (W
| \tilde \eta) = (\tilde \mu, \tilde \lambda| \tilde \eta)$,  and
\begin{equation}
(m_1 \cdots m_k) \equiv \epsilon^{\alpha_1 \cdots \alpha_k} C_{\alpha_1 m_1} \cdots C_{\alpha_k m_k} \end{equation}
This expression can be thought of as a unified form of the link representation, as we discuss in further detail in the appendix. The first striking aspect of this formula is its simplicity, as the integrand is comprised of a single term.
Furthermore the cyclic symmetry of the formula is manifest; as we will see with a tiny bit of work, parity is manifest as well.

We will study this object at great length for the rest of this paper. Going back to momentum space is trivial, and the resulting expression is computed as a multidimensional contour integral. As usual with contour integrals, there is really no integral to be done, and we are instead interested in the residues of the integrand at the location of each of its singularities.
Our tree-level motivations had led us to expect that the residues of this object would compute terms in the BCFW expansion of tree amplitudes, and this was correct for MHV amplitudes and the 6 particle NMHV amplitude. Continuing to 7 points, we identified all terms appearing in the tree amplitudes amongst the residues, but were puzzled to find new objects which were not needed for the tree amplitudes. To our great surprise, we found that these objects could be identified as the leading singularities of the 7 point amplitude at 1-loop! This also resolved a little mystery at 6 particles: the tree amplitude only involves a particular linear combination of residues, which left us wondering why some information was evidently ``wasted".  The 1-loop interpretation allowed us to interpret all of these residues as leading singularities. Moreover, while at MHV level our formula only produced the tree amplitude, all the 1-loop leading singularities of the 1-loop MHV amplitudes are proportional to the MHV tree amplitude and are thus consistent with a 1-loop interpretation.  Subsequently, we identified residues corresponding to  the 1-loop leading singularities for all NMHV amplitudes and the 8 particle N$^2$MHV amplitude, including the coefficient of the completely IR finite ``four-mass" boxes.

Further investigations found the pattern repeating itself: beginning with the 8 particle NMHV amplitude, some of the residues did not appear in any of the 1-loop leading singularities. We were able to identify these as certain 2-loop leading singularities! Since full 2 loop amplitudes for beyond 5 particles are not yet known, we cannot yet perform a simple check that {\it all} two-loop leading singularities are present, but given the striking pattern we have seen it is a very plausible conjecture, which if true further suggests that the residues are computing the leading singularities at all loop level. One immediate multi-loop prediction of this conjecture is that leading singularities of MHV loop amplitudes should always be proportional to the tree amplitude; this is true for all cases that have been computed so far. Another prediction is that the leading singularities occurring in the 6 and 7 particle amplitudes at two loops and beyond should be the same
as the objects occurring in the 1-loop leading singularities; this also appears to be the case for the 2-loop 6 particle amplitude \cite{vergu}.

It appears that we have found a rather remarkable object that knows about the S Matrix for ${\cal N} = 4$ SYM at all loop level. As we will discuss in detail, the expression is also intimately connected to central ideas in algebraic geometry: Grassmannians, higher-dimensional residue theorems, intersection theory, and the Schubert calculus. This part of mathematics has not yet played a particularly central role in physics; it is both startling and exciting to find it sitting at the heart of scattering amplitudes in Yang-Mills theory. 

A given leading singularity is associated with a specific combination of residues or, equivalently, a choice of integration contour. In this paper we identify the contours corresponding to the leading singularities for all NMHV amplitudes and also the 8 pt N$^2$MHV amplitude, leaving a complete exploration of the leading singularities for general amplitudes to future work. 
In the cases we consider, different representations for the amplitude correspond to different contours that can be smoothly deformed into each other: the remarkable identities guaranteeing the equality of all these representations, which at tree level also enforce the absence of unphysical poles, follow directly from the higher-dimensional analog of Cauchy's theorem known as the ``global residue theorem". The relations implied by the 1-loop ``infrared equations" also follow directly from the global residue theorem.

We believe that our conjecture represents a first direct look at the dual theory for the S Matrix that we have been seeking. 
The way in which such a simple formula manages to reproduce the incredibly complicated expressions appearing in scattering amplitudes is very striking--they are clearly being computed in a completely different way than local quantum field theory!
Further explorations into its properties should lead to a more complete and physical formulation of the duality.
With this goal in mind, we will study the properties of equation (\ref{first}) with the aim of understanding what it is trying to tell us about ${\cal N} = 4$ SYM.

\section{The Proposal}

Let us begin by defining some notation. We are interested in the single-trace $n$ particle color stripped amplitudes in ${\cal N} = 4$ SYM,
\begin{equation}
M_n({\lambda_a,\tilde \lambda_a, \tilde \eta_a}), \,\,\, a = 1, \cdots, n.
\end{equation}
Here we have chosen to label each particle by a $|\tilde \eta \rangle$ Grassmann coherent state \cite{Nair:1988bq, simplest}, where the $|\tilde \eta = 0\rangle$ state is $|+
\rangle$. The amplitude decomposes into the sum
\begin{equation}
M_n = \sum_{k = 0}^{n} M_{n;k}
\end{equation}
where the $M_k$ have charge $4k$ under the $U(1)_R$ symmetry where $\tilde \eta$ have charge 1. As is well-known $M_{k=0,1}$ and $M_{k=n,n-1}$ vanish; $M_{n;k}$ are precisely the N$^{k-2}$MHV or, equivalently the $\overline{{\rm N}^{n - 2 - k}{\rm MHV}}$ amplitudes. We will suppress the trivial dependence on the gauge coupling in everything that follows.

Let us return to our object of study, this time giving it a name:
\be {\cal L}_{n;k}({\cal W}_a) = \int \frac{d^{k \times n} C_{\alpha
a}}{(12\cdots k) \, (23\cdots (k+1) \,) \, \cdots (n 1 \cdots (k-1)
\,)} \prod_{\alpha = 1}^k \delta^{4|4}(C_{\alpha a} {\cal W}_a) \ee
where as before the ${\cal W}_a$ are twistor variables  obtained by
fourier-transforming with respect to the $\lambda_a$, so ${\cal W} =
(W | \tilde \eta) = (\tilde \mu, \tilde \lambda| \tilde \eta)$ and
\be (m_1  \cdots m_k) \equiv \epsilon^{\alpha_1 \cdots \alpha_k}
C_{\alpha_1 m_1} \cdots C_{\alpha_k m_k} \ee Thinking of $C_{\alpha
a}$ as a $k \times n$ matrix, $(m_1 \cdots m_k)$ is the determinant
of the $k \times k$ matrix made by only keeping the $k$ columns
$m_1, \cdots, m_k$. We'll use standard terminology and refer to
these as the ``minors" of $C$.

We use the notation ${\cal L}_{n;k}$ both because our claim is that this object computes the ``${\cal L}$eading singularities", as well to emphasize its relation to the ``${\cal L}$ink" representation.
Note that the $4k$ Grassmann $\delta$ functions ensure that this expression has R-charge $4k$ as needed. The expression is manifestly superconformally invariant. It is manifestly cyclically
invariant; in a moment we will also see that it is also parity invariant. However, at the moment the integral is rather formal and needs a proper definition.

\subsection{``Gauge Fixing" GL$(k)$}

To begin, notice that the product of $\delta^{4|4}$ functions impose $k$ linear relations on the $n$ ${\cal W}_a$'s. These $\delta$ functions are invariant under a GL$(k)$
transformation
\be
C_{\alpha a} \to L^{\beta}_{\alpha} C_{\beta a}
\ee
where $L^\beta_\alpha$ is any $k \times k$ matrix.

A quick note before we proceed further: ultimately, we will be interested in complex kinematical variables, and our discussion will be valid for any space-time signature.  However, for the moment we will keep things simple by defining our integral with real variables in $(2,2)$ signature. In order not to clutter our expressions with explicit references to real and complex variables, we will write GL$(k)$ instead of GL$(k,\mathbb{R})$ or GL$(k,\mathbb{C})$. Similarly, we will refer to the 4D Lorentz group as SL$(2) \times$ SL(2).

It is very natural for the integral to be invariant under a full $GL(k)$ symmetry, while the objects $(m_1 \cdots m_k)$ are invariant under the $SL(k)$ subgroup. With a total of $n$ minors in the denominator, the integrand weighted with the $d^{k \times n} C$ measure has the full $GL(k)$ symmetry. This completely fixes the form of the
integrand up to our choice that the $(m_1 \cdots m_k)$ consist of cross products of consecutive $C$ 's of the form $(j \, (j+1) \cdots (j+k -1) \,)$. This seems to be the most natural choice, it is also quite likely that this form is dictated by dual superconformal invariance, though we won't pursue this further here.

The $GL(k)$ symmetry is like a gauge symmetry that generates a divergence in the $d^{k\times n}C_{\alpha a}$ integral; to make sense of the integral we have to fix this gauge freedom. We do this by interpreting the $C_{\alpha a}$ as a collection of $n$ $k-$vectors.
Using $GL(k)$ we can set $k$ of these vectors to any fixed set of $k$ vectors we like. For instance we can set them to an orthonormal basis of the form
$(1,0,\cdots,0), (0,1,\cdots,0), \cdots, (0,\cdots,0,1)$. That is, we can choose a set $b_1,\cdots,b_k$, and fix
\be
C_{\alpha b_\beta } = \delta_{\alpha \beta }
\ee
Said another way, if $C_{\alpha a}$ is a $k \times n$ matrix, then we are fixing $k$ of the columns to an orthonormal basis of $k$ vectors.
This fixes $k^2$ of the $C$'s, leaving us $k \times (n-k)$ free variables. We index the $k$ particles belonging to the set $b_\beta$ with the letter
$I$, and the remaining $(n-k)$ particles with the letter $i$. The free $C_{\alpha a}$ can then be written as $c_{Ii}$.
To be explicit, consider the case $n=7,k=3$. Here $C$ starts as a
$3 \times 7$ matrix, and we can gauge fix any three of the columns to an orthonormal basis. For instance one possibility is
\be
\label{eq:gf1}
C = \left(\begin{array}{ccccccc} 1 & 0 & 0 & c_{14} & c_{15} & c_{16} & c_{17} \\ 0 & 1 & 0 & c_{24} & c_{25} & c_{26} & c_{27} \\ 0 & 0 & 1 & c_{34} & c_{35} & c_{36} & c_{37} \end{array} \right)
\ee
while another is
\be
\label{eq:gf2}
C = \left(\begin{array}{ccccccc} 1 & c_{12} & 0 & c_{14} & 0 & c_{16} & c_{17} \\ 0 & c_{32} & 1 & c_{34} & 0 & c_{36} & c_{37} \\ 0 & c_{52} & 0 & c_{54} &
1 & c_{56} & c_{57} \end{array} \right)
\ee
Other gauge fixings of the GL$(k)$ may also be convenient, however in this paper we will canonically gauge-fix in this way.

We can now define the ``gauge fixed" $k \times (n-k)$ dimensional integral \be
{\cal L}_{n;k}({\cal W}_a) = \int \frac{d^{k \times (n-k)} c_{Ii}} {(12\cdots k) \, (23\cdots (k+1) \,) \, \cdots (n 1 \cdots (k-1) \,)} \prod_I
\delta^{4|4}({\cal W}_I + c_{Ii} {\cal W}_i)
\ee
where the $(m_1 \cdots m_k)$ are computed from the $C_{\alpha a}$ in one of its gauge fixed forms.

Note that in this representation it is trivial to go to a basis where the particles indexed by $I$ are in the ${\cal Z}_I$ representation.  Simply by fourier transforming, we find a form of the link representation \cite{ArkaniHamed:2009si}
\be
{\cal L}_{n;k}({\cal W}_i,{\cal Z}_I) = \int \frac{d^{k \times (n-k)} c_{Ii}} {(12\cdots k) \, (23\cdots (k+1) \,) \, \cdots (n 1 \cdots (k-1)\,)} \exp (i c_{Ii} {\cal W}_i {\cal Z}_I)
\ee
Thus, going from a basis in which $k$ of the legs are in the $\cal Z$ representation to a different basis in which a different set of $k$ legs are in the $\cal Z$ representation simply amounts to gauge fixing the $C$ matrix in different ways. For instance the gauge fixing in equation (\ref{eq:gf1}) naturally corresponds to
the link representation of the amplitude with
${\cal Z}_1,{\cal Z}_2,{\cal Z}_3,{\cal W}_4,{\cal W}_5,{\cal W}_6,{\cal W}_7$, while the one in equation (\ref{eq:gf2}) corresponds to ${\cal Z}_1,{\cal W}_2,{\cal Z}_3,
{\cal W}_4, {\cal Z}_5,{\cal W}_6,{\cal W}_7$.

With this form we can straightforwardly go back to momentum space:
\be
\label{finalform}
{\cal L}_{n;k}(\lambda,\tilde \lambda, \tilde \eta) = \int \frac{d^{k \times (n-k)} c_{Ii} \, \delta^2(\lambda_i - c_{Ii} \lambda_I)
\delta^2(\tilde \lambda_I + c_{Ii} \tilde \lambda_i) \delta^4(\tilde \eta_I + c_{Ii} \tilde \eta_i)}
{(12\cdots k) \, (23\cdots (k+1)\,) \, \cdots (n 1 \cdots (k-1)\,)}
\ee

Let us count the number of integration variables after the delta functions have been used to fix as many of them as possible. We have $k \times (n-k)$ link variables, and a total of $2(n-k) + 2k - 4$ independent bosonic $\delta$ functions, where the $-4$ appears because four of these $\delta$ functions ultimately become 4-momentum conservation.
Therefore the number of remaining free integration variables is
\be
k (n-k) - (2n - 4) = (k - 2) \times (n - 2 - k)
\ee
Note that for $k=0,1,n-1,n$ the number of free integrations is negative. This reflects the fact that in these cases there are additional
$\delta$ functions in momentum space beyond the usual momentum-conserving one. For $k=0$, the amplitude is just proportional to the product over
all the $\delta^2(\lambda)'s$ and so vanishes for generic momenta. For $k=1$, the $\delta^2$'s force all the $\lambda$'s to be proportional and this also vanishes for generic momentum-conserving momenta, as is familiar for the three-particle amplitude.
Exactly the same argument holds for
$k = n, n-1$ by reversing the roles of $\lambda, \tilde \lambda$.

We can now state our conjecture precisely. As we have seen there is a $(k-2) \times (n - k - 2)$ dimensional plane in $c_{Ii}$ space, which are solutions to the
equations
\be
\label{eq:lambdaeqs}
 \lambda_i - c_{Ii} \lambda_I = 0, \,\, \tilde \lambda_I + c_{Ii} \tilde \lambda_i = 0                                                                                     \ee
The solutions of the these linear equations for the $c_{Ii}$ can be parametrized by variables $\tau_\gamma$, such that\be
c_{Ii}(\tau_\gamma) = c_{Ii}^* + d_{Ii \gamma} \tau_\gamma, \, \, {\rm with} \,\,  \gamma = 1, \cdots, (k-2) \times (n-k-2)
\ee
and where $c_{Ii}^*$ is any particular solution to equation (\ref{eq:lambdaeqs}).                                                                                                                                                                    Of course we can make any choice for the parametrization of the $\tau_\gamma$'s that                                                                                    we like, for instance we can pick the $\tau$'s to be any $(k-2) (n - k - 2)$ of the $c_{Ii}$, and solve for the rest of the $c$'s in
terms of them.

Now, we can explicitly pull out the momentum conserving delta function from the $\delta^2$ factors as
\be
\delta^{2}(\lambda_i - c_{Ii} \lambda_I) \delta^2(\tilde \lambda_I + c_{Ii} \tilde \lambda_i) = \delta^4(\sum_a p_a) \times J(\lambda,\tilde \lambda)
\times \int d^{(k-2)(n-k-2)} \tau_\gamma \delta(c_{Ii} - c_{Ii}(\tau_\gamma))                                                                                                  \ee
In this expression we cavalierly ignore the sign factors that arise in the Jacobian of real $\delta$ functions; the sign would only appear as an overall factor depending on $n,k$ and the external $\lambda,\tilde \lambda$ in any case; we drop it since keeping it would introduce non-analytic factors not present in the amplitudes. In the appendix we present a more detailed treatment in which we discuss the relationship of our conjecture to the link representation of \cite{ArkaniHamed:2009si}, and demonstrate that ignoring these sign factors gives the correct answer, justifying our loose treatment here.

Our fully gauge-fixed and well-defined proposal is now
\be
{\cal L}_{n;k} = L_{n;k} \times \delta^4(\sum_a p_a)
\ee
with
\be
 L_{n;k} = J \int
 \frac{d^{(k-2)\times(n-k-2)} \tau}{\left[(12\cdots k) \, (23\cdots
 (k+1)\, ) \, \cdots (n 1 \cdots (k-1)\,)\right](\tau)}
 \prod_I \delta^4(\tilde \eta_I + c_{Ii}(\tau) \tilde \eta_i)
\ee At this point we can freely complexify the $\lambda,\tilde
\lambda$ and $\tau$ variables. Since the integrand is holomorphic in
the $\tau$'s we must treat it as a contour integral in many complex
variables.

If we are interested in extracting the gluonic components of this
supersymmetric object, it is particularly convenient to ``gauge fix"
so that the indices $I$ which have been set  to the orthonormal
basis correspond to the negative helicity particles. Then, to obtain the
gluon component we simply set $\tilde \eta_i \to 0$ to get positive
helicity gluons and integrate $\int d^4 \tilde \eta_I$ for
negative helicity gluons, and the Grassmann integral just gives
1. As an example, for the 7 particle NMHV amplitude with 
helicities $1^+ 2^+ 3^- 4^+ 5^- 6^- 7^+$, we have $n=7,k=3$ and
\begin{equation}
L^{++-+--+}_{7;3} = J \int \frac{d^2 \tau}{[(123)(234)(345)(456)(567)(671)(712)](\tau)}
\end{equation}
with the gauge-fixed $C$ of the form
\be
C = \left(\begin{array}{ccccccc} c_{31}& c_{32} & 1 & c_{34} & 0 & 0 & c_{37} \\ c_{51} & c_{52} & 0 & c_{54} & 1 & 0 & c_{57} \\ c_{61} & c_{62} & 0 & c_{64} &
0 & 1 & c_{67} \end{array} \right)
\ee

\subsection{Geometric Picture}
\label{subsec:GeoPic}

The momentum space form of our conjecture in equation (\ref{finalform}) has a very nice geometric interpretation.  Indeed, we can motivate the conjecture by starting with an elementary observation about momentum conservation.
 
Let us work in the $n$-dimensional ``particle" space. The kinematical data is specified by giving $\lambda_{a \, \alpha}, \tilde \lambda_{a \, \dot{\alpha}}$. In any given Lorentz frame, we can think of $\lambda_{a \, \alpha}$ as simply labelling two $n$-vectors, $\vec{\lambda}_{\alpha=1}, \vec{\lambda}_{\alpha=2}$; similarly the $\tilde \lambda$'s correspond to two $n$-vectors $
\vec{\tilde \lambda}_{\dot{\alpha} =1}, \vec{\tilde \lambda}_{\dot{\alpha}=2}$. Since Lorentz transformations act as SL(2)$\times$ SL(2) we should really think about the 2-plane $\boldsymbol{\lambda}$
in the $n$-space, spanned by $\vec{\lambda}_{\alpha=1,2}$, and similarly the 2-plane $\boldsymbol{\tilde \lambda}$. Note that the $\boldsymbol{\lambda}, \boldsymbol{\tilde \lambda}$ planes intersect at what we can call the origin in $n$-space.
Now in this $n$-dimensional setting, momentum conservation, $\sum_a \lambda_{a \, \alpha} \tilde \lambda_{a \dot{\alpha}}= 0$ is simply the statement that $\vec{\lambda}_\alpha \cdot \vec{\tilde \lambda}_{\dot{\alpha}} = 0$, or put another way, that the $\boldsymbol{\lambda}$ and $\boldsymbol{\tilde \lambda}$ planes are orthogonal to each other.

This is a quadratic constraint on the $\boldsymbol{\lambda}, \boldsymbol{\tilde \lambda}$. It is possible to ``linearize" this constraint by introducing an additional auxiliary object. Consider a $k$-plane $\boldsymbol{C}$ passing through the origin in the $n$-dimensional space. Associated to $\boldsymbol{C}$ is a natural $(n-k)$ plane, $\boldsymbol{\tilde C}$, which is simply the orthogonal complement of $\boldsymbol{C}$. Given these objects, momentum conservation can be enforced by a pair of linear constraints: that $\boldsymbol{\tilde C}$ is orthogonal to $\boldsymbol\lambda$, and $\boldsymbol{C}$ is orthogonal to $\boldsymbol{\tilde \lambda}$. To spell out the obvious, if $\boldsymbol{\tilde C}$ is orthogonal to $\boldsymbol{\lambda}$, then $\boldsymbol{C}$ must contain the $\boldsymbol{\lambda}$ plane, and since $\boldsymbol{C}$ is forced to be orthogonal to $\boldsymbol{\tilde \lambda}$, these two constraints together enforce momentum conservation. Note that these constraints are clearly impossible to satisfy for $k=0,1,n-1$ or $n$, where either $\boldsymbol{C}$ or $\boldsymbol{\tilde C}$ is a point or a line. This gives a pretty geometrical explanation of why these amplitudes vanish.

\be
\includegraphics[scale=0.9]{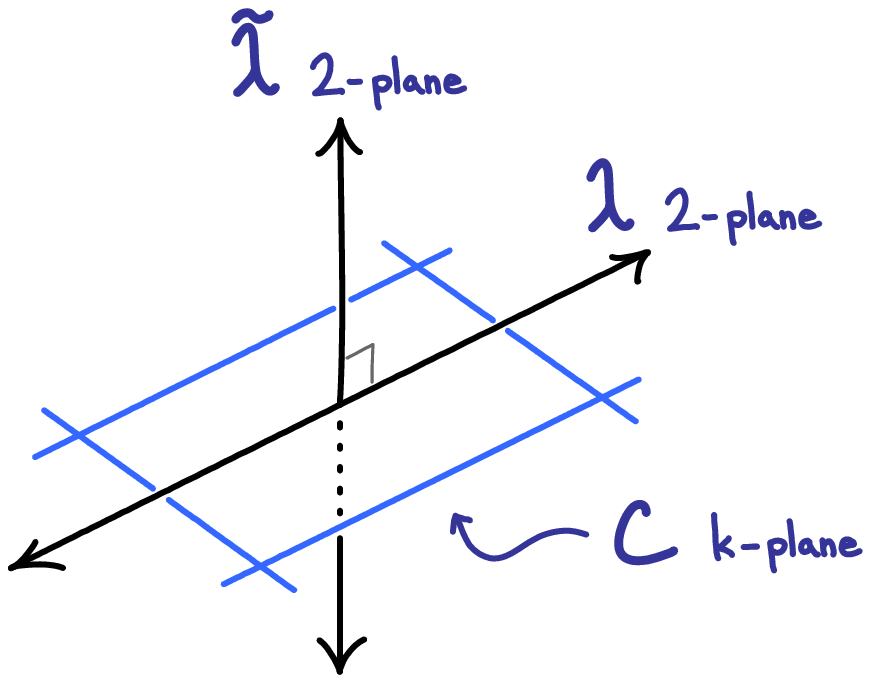} \nonumber
\ee

Let us now see how these geometric conditions are reflected in the equations. A $k$ plane in $n$ dimensions can be specified by a collection of $k$ $n$-dimensional vectors, whose span give the $k$-plane. These $k$ vectors can be grouped together into a $k \times n$ matrix $C$
\be
C = \left(\begin{array}{c} \vec{C}_1 \\ \vec{C}_2 \\ \vdots \\ \vec{C}_k \end{array} \right) =
\left(\begin{array}{ccccccccc} C_{11} & C_{12}  & \ldots & C_{1k} & C_{1,k+1} & C_{1,k+2} & \ldots & C_{1,n-1} & C_{1,n} \\ C_{21} & C_{22} & \ldots & C_{2k} & C_{2,k+1} & C_{2,k+2} & \ldots & C_{2,n-1} & C_{2,n} \\ \vdots & \vdots &  \ddots & \vdots & \vdots & \vdots & \ldots & \vdots & \vdots \\ C_{k,1} & C_{k,2} & \ldots & C_{k,k} & C_{k,k+1} & C_{k,k+2} & \ldots & C_{k,n-1} & C_{k,n} \end{array} \right)
\ee
Naturally, this is a highly redundant description of a $k$-plane since any $k \times k$ linear transformation on the $k$ vectors leaves the $k$-plane invariant. This is precisely the GL$(k)$ ``gauge symmetry" we encountered in the previous subsection. As before, we can ``gauge-fix" the $C$ matrix to put any $k$ of the columns to some fixed set values, for instance
\be
\label{choice}
C = \left(\begin{array}{ccccccccc} 1 & 0  & \ldots & 0 & c_{1,k+1} & c_{1,k+2} & \ldots & c_{1,n-1} & c_{1,n} \\ 0 & 1 & \ldots & 0 & c_{2,k+1} & c_{2,k+2} & \ldots & c_{2,n-1} & c_{2,n} \\ \vdots & \vdots &  \ddots & \vdots & \vdots & \vdots & \ldots & \vdots & \vdots \\ 0 & 0 & \ldots & 1 & c_{k,k+1} & c_{k,k+2} & \ldots & c_{k,n-1} & c_{k,n} \end{array} \right)
\ee
and we see as before that $k$-planes are specified by $n \times k - k^2 = k \times (n-k)$ parameters.

The space of $k$-planes in an $n$-dimensional space is known as the
Grassmannian $G(k,n)$, and we have just seen that $G(k,n)$ has
dimension $k\times (n-k) $. There is a clear $k \leftrightarrow
(n-k)$ symmetry here, which reflects the natural association between
the $k$-plane and its orthogonal complement $(n-k)$-plane. For the
choice of ``gauge fixing" made above, the $(n-k)$ plane $\tilde C$
is given by an $(n-k) \times n$ matrix \be \label{choicestar} \tilde
C = \left(\begin{array}{cccccccc} -c_{1,k+1} & -c_{2,k+1}  & \ldots
& -c_{k,k+1} & 1 & 0 & \ldots & 0 \\ -c_{1,k+2} & -c_{2,k+2} &
\ldots & -c_{k,k+2} & 0 & 1 & \ldots & 0 \\ \vdots & \vdots & \vdots
& \vdots & \vdots & \vdots & \ddots & \vdots \\ -c_{1,n}& -c_{2,n} &
\ldots & -c_{k,n} & 0 & 0 & \ldots & 1 \end{array} \right) \ee Note
that $-$ signs in $\tilde C$ are needed to ensure that $C \cdot
\tilde C = 0$. The linear equations (\ref{eq:lambdaeqs}) satisfied
by the $c_{Ii}$ then take the simple form \be \tilde C \cdot \lambda
= 0, \, \, C \cdot \tilde \lambda = 0 \ee

We can think of the representative $C$ given above as defining a set of co-ordinates  on the Grassmannian; in fact it is clear that this is only one chart covering an open set of all possible $k$-planes, and that different choices of ``gauge-fixing" the matrix $C$ correspond to different charts covering different open sets.
As we pointed out in the last subsection, in order to look at pure gluon amplitudes with a given helicity configuration, it is most convenient to gauge fix so that the columns corresponding to the negative helicity gluons are gauge-fixed to an orthonormal basis. This leads to a very pretty picture: the amplitude for different helicity configurations correspond to integrating the same function, but over different charts on the Grassmannian! Keeping the supersymmetric Grassmann variables manifest instead ensures that we get precisely the same answer for the superamplitude no matter which chart is chosen. We will see how this works explicitly in some examples later.

We have focused on momentum conservation in this discussion, but the superpartner of momentum conservation also follows explicitly from this geometrical picture. The amplitude is proportional to the Grassmann $\delta$ function $\delta^4(C \tilde \eta)$; since $\boldsymbol C$ must contain the $\boldsymbol{\lambda}$ plane, we can pull out an overall factor of  $\delta^8(\sum_a \lambda_a \tilde \eta_a)$, which is precisely the superpartner of the usual momentum-conserving $\delta$ function.

In the last subsection we saw that after integrating over the bosonic $\delta$ functions, we are left with $(k-2)\times (n- k - 2)$ free
variables. Not coincidentally, this is the dimension of $G(k-2,n-4)$, as is completely obvious geometrically. Recall that we are constraining $\boldsymbol{C}$ to be orthogonal to $ \boldsymbol{\tilde \lambda}$ and contain $\boldsymbol{\lambda}$. Since $\boldsymbol{C}$ must contain $\boldsymbol\lambda$, of the $k$ vectors needed to define $\boldsymbol{C}$, it is natural to choose two of them to span the plane $\boldsymbol{\lambda}$, and choose the remaining $(k-2)$ to be orthogonal to both $\boldsymbol{\lambda}$ and $\boldsymbol{\tilde \lambda}$. Thus the $k$-planes satisfying our constraints are naturally in one-to-one correspondence with a $(k-2)$ plane in the $(n-4)$ dimensional space orthogonal to $\boldsymbol{\lambda}$ and $\boldsymbol{\tilde \lambda}$. We can think of the $C(\tau_\gamma)$ matrix satisfying the constraints as a mapping from $G(k-2,n-4) \to G(k,n)$, which we are integrating with a natural measure.

This finally brings us to the interpretation of the minors $(i, \,
i+1 \, \cdots, i + k -1)$ appearing in the integration measure of
our formula. Thus far no particular basis in the $n$-dimensional
space has played a privileged role in our discussion, however, the
external states and color ordering of do make the orthonormal basis
set $e_1 = (1, 0, \cdots, 0), \, \cdots, e_n = (0, 0, \cdots, 1)$
special. The $k$ $n$-vectors defining the $k$-plane can clearly be
projected into the $k-$dimensional subspace $e_{m_1}, \cdots,
e_{m_k}$ and will generally fill some volume in this space; the
minor $(m_1 \cdots m_k)$ is  that volume. By itself this volume is
not a particularly natural geometrical object, since it is not
invariant under the full GL$(k)$ symmetry but only under SL$(k)$.
However, as we commented in the last subsection, due to the
transformation of the measure, the whole integral is nicely GL$(k)$
invariant. Indeed, this motivates the interpretation of the set of
all $n \choose k$ minors $(m_1 \cdots m_k)$ as coordinates in a
$\mathbb{P}^{{n \choose k} -1}$ projective space. This embedding of
$G(k,n)$ into $\mathbb{P}^{{n \choose k} - 1}$ is known as the
Pl\"{u}cker embedding, and the minors are referred to as the
Pl\"{u}cker coordinates.

\subsection{Manifest Cyclic and Parity Symmetries}

Before moving on to some simple examples, let us quickly show that the cyclic symmetry and parity are manifest in
our proposal. By ``are manifest", we really mean that they leave the integrand invariant; of course the choice of contour can break these symmetries.

First for the cyclic symmetry: this is trivially present in the
original form of the proposal. While the gauge fixing breaks the
manifest cyclic invariance, the underlying gauge symmetry ensures
that all the different gauge fixings yield the same cyclically
invariant result. We will see this explicitly in a number of
examples.

While the cyclic symmetry is obvious before gauge-fixing, parity is
obvious after gauge fixing. Our integral is \be {\cal L}_{n;k} =
\int \frac{d^{(n-k) \times k} C}{(12 \cdots k) (23 \cdots k+1)
\cdots (n 1 \cdots k-1)} \delta^2(\tilde C \cdot \lambda) \delta^2
(C \cdot \tilde \lambda) \delta^4(\tilde C \cdot \tilde \eta) \ee
Fourier transforming with respect to $\tilde \eta$ and swapping $k
\leftrightarrow (n-k)$ amounts to also swapping $C \leftrightarrow
\tilde C$. It is also easy to see that in any gauge fixing, the minor
$(12\cdots k)$ of $C$ is the same as the minor $(k+1 \cdots n)$ of
$\tilde C$, and thus the measure and integrand are parity invariant.  To
see this in a concrete example, consider the 6 particle MHV
amplitude. Gauge fixing the first two columns of $C$, we find that $C$
and $\tilde C$ are \be
C = \left(\begin{array}{cccccc} 1 & 0  &  c_{13} & c_{14} &  c_{15} & c_{16} \\ 0 & 1 &  c_{23} & c_{24} &  c_{25} & c_{26}  \end{array} \right) \\
\tilde C = \left(\begin{array}{cccccc} -c_{13} & -c_{23}  & 1 & 0 & 0 & 0 \\ -c_{14} & -c_{24}  & 0 & 1 & 0 & 0 \\
-c_{15} & -c_{25}  & 0 & 0 & 1 & 0 \\
-c_{16} & -c_{26}  & 0 & 0 & 0 & 1  \end{array} \right)
\ee
and  we see immediately that $(45)=(6123) = c_{14}c_{25}-c_{15}c_{24}$, and so on. The general statement is easily proven by induction.

\subsection{The Polynomial Degree of the Minors}

Let us make one final general observation, and determine the order of the polynomials in the $\tau_\gamma$ that will appear in each determinant $(m_1, \cdots, m_k)$. In general the solutions $c_{Ii}(\tau) = c_{Ii}^* + \tau_{Ii}$ where $\tau_{Ii} = d_{Ii\gamma} \tau_\gamma$ are linear in the $\tau_\gamma$ and satisfy
\be
\tau_{Ii} \lambda_I = 0, \, \tau_{Ii} \tilde \lambda_i = 0
\ee
The first equation tells us that, thinking of $\tau_{Ii}$ as a set of $\vec{\tau}_i$ $k$-vectors, all these vectors are orthogonal to the $2$-plane $\boldsymbol{\lambda}$. Therefore all the $\vec{\tau}$ lie on a $(k-2)$ dimensional subspace, and so the cross product of any
$(k-1)$ of them must vanish. Hence, in the $k$ cross-products appearing in $(m_1 \cdots m_k)$, the $\tau_{Ii}$ variables can appear at most $(k-2)$
times, and each factor in the denominator is therefore a polynomial of degree at most $(k-2)$ in the free integration variables.
Note that we could have made exactly the same argument using the parity conjugate form
of the amplitude which would tell us that the polynomial is of degree at most $(n - k) -2$.
Thus in general each factor in the denominator is a polynomial of degree min[$(k-2),(n-2-k)]$ in $(k-2) \times (n - 2 - k)$ variables.
The NMHV amplitudes are particularly simple: there are $(n-5)$ integration variables and
each term in the denominator is linear in them.

\section{First Examples}

\subsection{MHV Amplitudes: $k=2$}

In this case we set $k=2$. From our general formula we can see that there is no integration to be done and therefore $L_{n;2}$ is straightforward to evaluate. Before evaluating $L_{n;2}$ in detail, let us use this example to illustrate some of the geometrical properties we discussed in generality before. The geometrical ideas provide an intuition for motivating our formula.

\subsubsection{Direct Geometrical Evaluation}

In section 2.2, momentum conservation was expressed as a fully geometrical condition. Two fixed $2$-planes, the $\boldsymbol{\lambda}$ and $\boldsymbol{\tilde\lambda}$ plane were introduced. Momentum conservation is simply the statement that these two $2$-planes are orthogonal in $\mathbb{C}^n$.

For $k=2$, we are interested in the space of all $2$-planes that contain the $\boldsymbol{\lambda}$-plane and are orthogonal to the $\boldsymbol{\tilde\lambda}$-plane. Clearly there is only one such plane and it is must coincide with the $\boldsymbol{\lambda}$-plane itself. This means that the $2$ $n$-vectors giving the rows of the $2\times n$ matrix representation of $C$ must be linear combinations of $(\lambda^{1}_1,\lambda^2_1,\ldots, \lambda^n_1)$ and $(\lambda^1_2,\lambda^2_2,\ldots, \lambda^n_2)$. Since we are interested in the determinants of the minors, we are free to choose them to be exactly equal to the two $\lambda$ n-vectors,
\be
C = \left(\begin{array}{cccccc} C_{11} & C_{12} & C_{13} & \ldots & C_{1,n-1} & C_{1,n} \\ C_{21} & C_{22} & C_{23} & \ldots & C_{2,n-1} & C_{2,n}
\end{array}\right) = \left(\begin{array}{cccccc} \lambda^1_1 & \lambda^2_1 & \lambda^{3}_1 & \ldots & \lambda^{n-1}_1 & \lambda^n_1 \\ \lambda^1_2 & \lambda^2_2 & \lambda^3_2 & \ldots & \lambda^{n-1}_2 & \lambda^n_2
\end{array}\right).
\ee
This fixes the GL(2) ``gauge symmetry".
With this identification it is clear that the determinant of a $2\times 2$ minor made from columns $i$ and $j$ gives $(ij) = \langle i j \rangle$. Therefore, the product $(12)(23)\ldots (n-1 \, n)(n,1)$ gives rise to the usual Parke-Taylor formula for the denominator of an MHV amplitude.

The numerator is fixed by imposing the supersymmetric version of the geometric condition: $C$ must be a 2-plane which is orthogonal to a $0|4$ plane in $\mathbb{C}^{n|n}$ spanned by the 4 $n$-vectors $(\tilde\eta_1^I, \tilde\eta_2^I,\ldots, \tilde\eta_n^I)$, which is imposed by the Grassmann $\delta$ functions
\be
\delta^4 (\sum_{a=1}^n C_{1a}\tilde\eta_a)\delta^4(\sum_{a=1}^nC_{2a}\tilde\eta_a) = \delta^8(\sum_{a=1}^n \lambda_a\tilde\eta_a)
\ee
Putting everything together we find
\be
L_{n;2} = \frac{\delta^8(\sum_{a=1}^n \lambda_a\tilde\eta_a)}{\langle 1 2 \rangle \langle 2 3 \rangle\ldots \langle n 1\rangle}.
\ee
In this derivation we have made an identification which only works for $k=2$; we have embedded the ${\rm SL}(2)$ action of the Lorentz group inside the ${\rm GL}(2)$ acting on the space of $2\times n$ matrices.

\subsubsection{Evaluation Using Canonical Gauge Fixing}

Let us now show how the same formula can be obtained by the canonical gauge fixings of the ${\rm GL}(2)$.

Put $k=2$ in equation (\ref{finalform}), and gauge fix so that the index $I$ runs over the particles
$x,y$. Therefore, on the support of
$\delta^2(\lambda_i - c_{Ii} \lambda_I)$, we can solve
\be
\label{asi}
\lambda_I = c_{Ii} \lambda_I \implies c_{Ii} = \frac{\epsilon_{IJ} \langle i J \rangle}{\langle x y \rangle}
\ee
Note that we can more generally write
\be
\lambda_a = C_{\alpha a} \lambda_\alpha \implies \lambda_a \lambda_b = \langle x y \rangle \epsilon^{\alpha \beta} C_{\alpha a} C_{\beta b}
= \langle x y \rangle (a b)
\ee
where the $\alpha, \beta$ indices also range over $x,y$. It is also trivial to see that
\be
\delta^2(\lambda- c_{Ii} \lambda_I) \delta^2(\tilde \lambda_I + c_{Ii} \tilde \lambda_i) = J \, \delta^4(\sum_a p_a) \, {\rm with} \, \,
J = \langle x y \rangle^{4 - n}
\ee
Therefore
\be
\label{bose}
\frac{1}{(12)(23) \cdots (n1)} = \frac{\langle x y \rangle^4}{\langle 1 2 \rangle \langle 2 3 \rangle \cdots \langle n 1 \rangle}
\ee
The Grassmann delta functions can also be easily simplified starting with
\be
\delta^4(\tilde\eta_x+\sum_{i\neq x,i\neq y}c_{xi}\tilde\eta_i)\delta^4(\tilde\eta_y+\sum_{i\neq x,i\neq y}c_{yi}\tilde\eta_i)
\ee
and using (\ref{asi}) we obtain
\be
\label{fermi}
\frac{1}{\langle x~y\rangle^4}\delta^8(\sum_{a=1}^n \lambda_a\tilde\eta_a)
\ee
Combining (\ref{bose}) with (\ref{fermi}) we get the desired result.

Note that already with this simple example we see explicitly something that we claimed on general grounds: the final form of the amplitude is independent of how we ``gauge-fixed" the GL(2) symmetry. 

\subsection{6 Particle NMHV Amplitude}

Next, let us consider NMHV amplitudes with $k=3$. The number of integration variables is $(3 - 2) \times (n - 3 - 2) = (n-5)$. Obviously the simplest case is with $n=6$, which involves only one integration variable which we call $\tau$.
Since this is such a simple case, we will work through the computation of the residues of interest rather explicitly here. 

The relevant integral for the 6 particle NMHV amplitude is
\be
 L_{6;3} = J \int
 \frac{d\tau}{\left[(123)(234)(345)(456)(561)(612)\right](\tau)}
 \prod_I \delta^4(\tilde \eta_I + c_{Ii}(\tau) \tilde \eta_i)
\ee
To study the alternating helicity amplitude, we choose a convenient gauge fixing
\be \label{alter} C =
\left(\begin{array}{cccccc} c_{21} & 1 & c_{23} & 0 & c_{25} & 0 \\
c_{41} & 0 & c_{43} & 1 & c_{45} & 0 \\ c_{61} & 0 & c_{63} & 0 &
c_{65} & 1
\end{array}\right).
\ee

In terms of the $c_{Ii}$ appearing in this matrix, the minors are given by
\begin{eqnarray}
(123) = c_{41} c_{63} - c_{61} c_{43} \equiv \tilde c_{25}, \, (345) = \tilde c_{41}, \, (561) = \tilde c_{63}, \nonumber \\
(234) = c_{36}, \, (456) = c_{25}, (612) = c_{41}.
\end{eqnarray}
so that \be \frac{1}{(123)(234)(345)(456)(561)(612)} =
\frac{1}{c_{25} c_{41} c_{63} \tilde c_{25} \tilde c_{41} \tilde
c_{63}} \ee and \be L_{6,3} = \int \frac{d \tau}{[c_{25}
c_{63} c_{41} \tilde c_{25} \tilde c_{63} \tilde c_{41}](\tau)} \ee
Here we have introduced the notation $\tilde c_{Ii}$ to denote a
pole which maps to $c_{Ii}$, and vice-versa, under a parity
transformation.

To identify $c_{Ii}(\tau)$ explicitly, we are looking for the 1-dimensional space of solutions to the equations
\be
\lambda_i - c_{Ii} \lambda_I= 0, \, \tilde \lambda_I + c_{Ii} \tilde \lambda_i = 0
\ee
Since any three two-dimensional vectors are linearly dependent, they satisfy the ``Schouten identity" which we write in the form
\be
\epsilon_{ijk} \tilde \lambda_i [jk] = 0, \, \, \epsilon_{IJK} \lambda_I \langle J K \rangle = 0
\ee
Hence, given any particular solution $c_{Ii}^*$, we can find another solution
\be
c_{Ii}(\tau) = c_{Ii}^* + \epsilon_{ijk} \epsilon_{IJK} [jk]\langle J K \rangle \tau
\ee
With this choice it is easy to verify that the Jacobian $J$ is 1:
\begin{equation}
\prod_i \delta^2(\lambda_i - c_{Ii}\lambda_I) \prod_I \delta^2(\tilde \lambda_I + c_{Ii} \tilde \lambda_i) = 1 \times \delta^4(\sum_a \lambda_a \tilde \lambda_a) \int d \tau \delta^9 \left(c_{Ii} - c_{Ii}(\tau) \right)
\end{equation}
 We can also always choose the origin for $\tau$ so that, for example, $c_{25}(0) = 0$; this allows us to solve for the $c_{Ii}^*$. For instance we can use the $\lambda_5$ and $\tilde \lambda_2$ equations to solve for $c_{65},c_{45},c_{21},c_{23}$,
\begin{equation}
c^*_{65} = \frac{\langle 5 4 \rangle}{\langle 6 4 \rangle}, \, c^*_{45} = \frac{\langle 5 6 \rangle}{\langle 4 6 \rangle}, \, c^*_{21} = \frac{[23]}{[13]}, c^*_{23} = \frac{[21]}{[31]}
\end{equation}
and then use, say, the $\tilde \lambda_4, \tilde \lambda_6$ equations to solve for the rest of the $c$'s
\begin{equation}
c^*_{41} = \frac{\langle 6 | (5 + 4) |3]}{\langle 4 6 \rangle [1 3]}, c^*_{43} =  \frac{\langle 6 | (5 + 4) |1]}{\langle 4 6 \rangle [3 1]}; c^*_{61} = \frac{\langle 4 | (5 + 6) |3]}{\langle 6 4 \rangle [3 1]}, c^*_{63} = \frac{\langle 4 | (5 + 6) |1]}{\langle 6 4 \rangle [1 3]}
\end{equation}

In the above computation we have used a tiny bit of foresight to parametrize the one-dimensional space of solutions, but we would have arrived at precisely the same expression by brute force, for instance by picking one of the $c$'s, say $c_{25}$, to be special and solving for all the rest of the $c$'s in terms of this one. That would be a different parametrization of the 1-dimensional space of solutions with a different $J$, but of course precisely the same final answer.

As we saw on general grounds, each of the minors is linear in $\tau$, and so the integrand has 6 poles, associated with the points where each minor vanishes. Let us introduce some compact notation to denote the relevant residues. A given minor $(i, \, i+1, i+2)$ is specified by its starting point $i$. We will therefore refer to the residue at the pole corresponding to the vanishing of the minor $(i, \, i+1, \, i+2)$ as $\{ i \}$.

A short computation then yields
\be
\{ 1 \} = - \frac{[3|(2+4)|6 \rangle^4}{[23][34]\langle 56 \rangle \langle 61 \rangle (p_5 + p_6 + p_1)^2 \langle 1 |6 + 5|4] \langle 5|6+ 1 |2]}, \, \, \{3 \} = g^2 \{ 1 \}, \, \{ 5 \} = g^4 \{ 1 \}
\ee
and
\be
\{4 \} = \frac{\langle 4 6 \rangle^4 [1 3]^4}{[12][23]\langle 45 \rangle \langle 56 \rangle (p_4 + p_5 + p_6)^2
\langle 6 |5 + 4|3] \langle 4|5 + 6 |1]}, \, \, \{ 6 \} = g^2 \{ 4 \}, \{ 2 \} = g^4 \{ 4 \}
\ee
The reader will recognize the residues $\{2 \},\{4 \},\{ 6 \}$ as the three terms in the BCFW form of the 6 particle tree amplitude, and the $\{1 \},\{ 3 \}, \{ 5 \}$ residues as the negative of the three terms in the P(BCFW) form of the same amplitude:
\be
M^{+-+-+-}_{{\rm BCFW}} = \{ 2 \} + \{ 4 \} + \{ 6 \}, \, \, M^{+-+-+-}_{{\rm P(BCFW)}} = - \{1 \} - \{ 3 \} - \{ 5 \}
\ee

We can now identify the tree amplitude with a particular choice of contour in our integral; depending on what contour is chosen we can get different 
forms of the amplitude. Of course since we have the full $\tilde \eta$ dependence, we can obtain any helicity amplitude we please, but for simplicity let us start with the alternating helicity amplitude which is naturally associated with our form of the  gauge-fixing.

Let $\Gamma$ be a contour that encircles the poles where $(234),(456),(612)$ vanish, and $\tilde \Gamma$ be a contour that encircles the poles $(123),(345),(561)$. Note that under a cyclic shift, the poles contained in $\Gamma$ map to the ones contained in $\tilde \Gamma$ and vice-versa. Most naively, then, to extract a cyclically invariant object from the residues we should take the contour $\Gamma + \tilde \Gamma$; in fact there is a minus sign in the parity mapping from $\Gamma$ to $\tilde \Gamma$ and the correct, manifestly cyclically invariant contour is $\Gamma - \tilde \Gamma$. This gives the manifestly parity symmetric form [BCFW + P(BCFW)] of the amplitude. However, since the integrand vanishes as $\frac{1}{\tau^6}$ as $\tau \to \infty$, up to a factor of 2 we can also write the amplitude using only the $\Gamma$ or the $-\tilde \Gamma$ contour. These give us the BCFW and P(BCFW) forms of the amplitude individually. Applying Cauchy's theorem to the contour $\Gamma + \tilde \Gamma$ and enclosing all the poles with the same orientation gives us the remarkable 6 term identity which guarantees the equality of the BCFW and P(BCFW) forms of the tree amplitude.

\be
\includegraphics[scale=1.0]{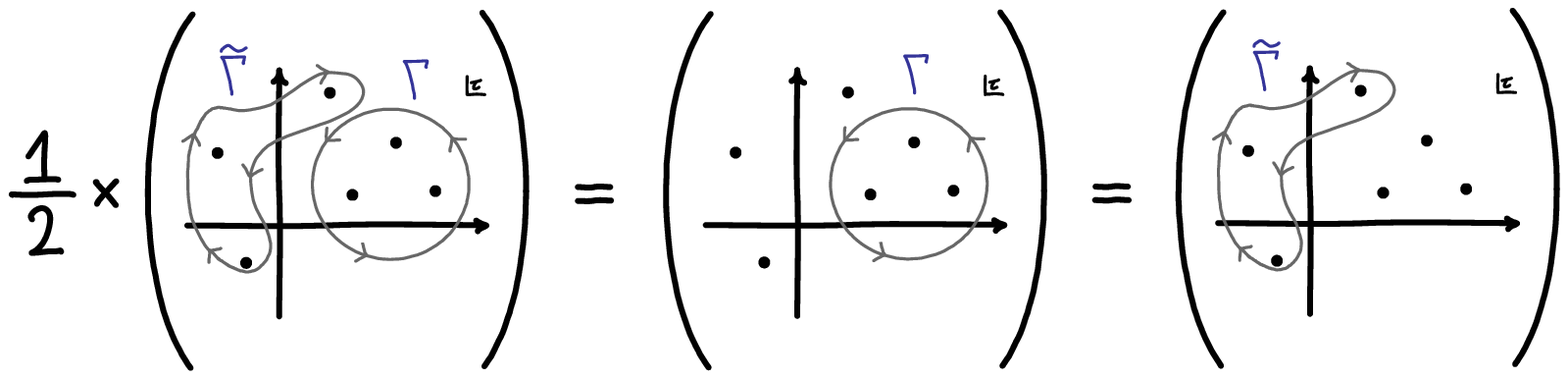} \nonumber
\ee

As we remarked in the introduction, the 6 term identity not only guarantees the cyclicity and parity of the tree amplitude, but also enforces the absence of unphysical poles. This is also easy to see from our contour integral. As we change the external kinematics, the position of the 6 poles in the $\tau$ plane move; we should only expect singularities when poles collide.  The ``unphysical" poles correspond to a collision between poles contained in e.g. $\Gamma$, or ones in $\tilde \Gamma$. For instance, looking at the residue where $(456) = c_{25} = 0$, when $\langle 6|5 + 4|3] \to 0$, the minor $(612) = c_{41} \to 0$. However clearly there is no actual singularity here, since we can always deform the contour to $-\tilde \Gamma$, none of whose encircled poles are colliding. The physical singularities, on the other hand, involve the collision of a pair of $c$ and $\tilde c$ poles; since the tree amplitude contour separates the $c$ poles from the $\tilde c$ poles, the contour is necessarily pinched between them and a singularity arises.

\be
\includegraphics[scale=0.8]{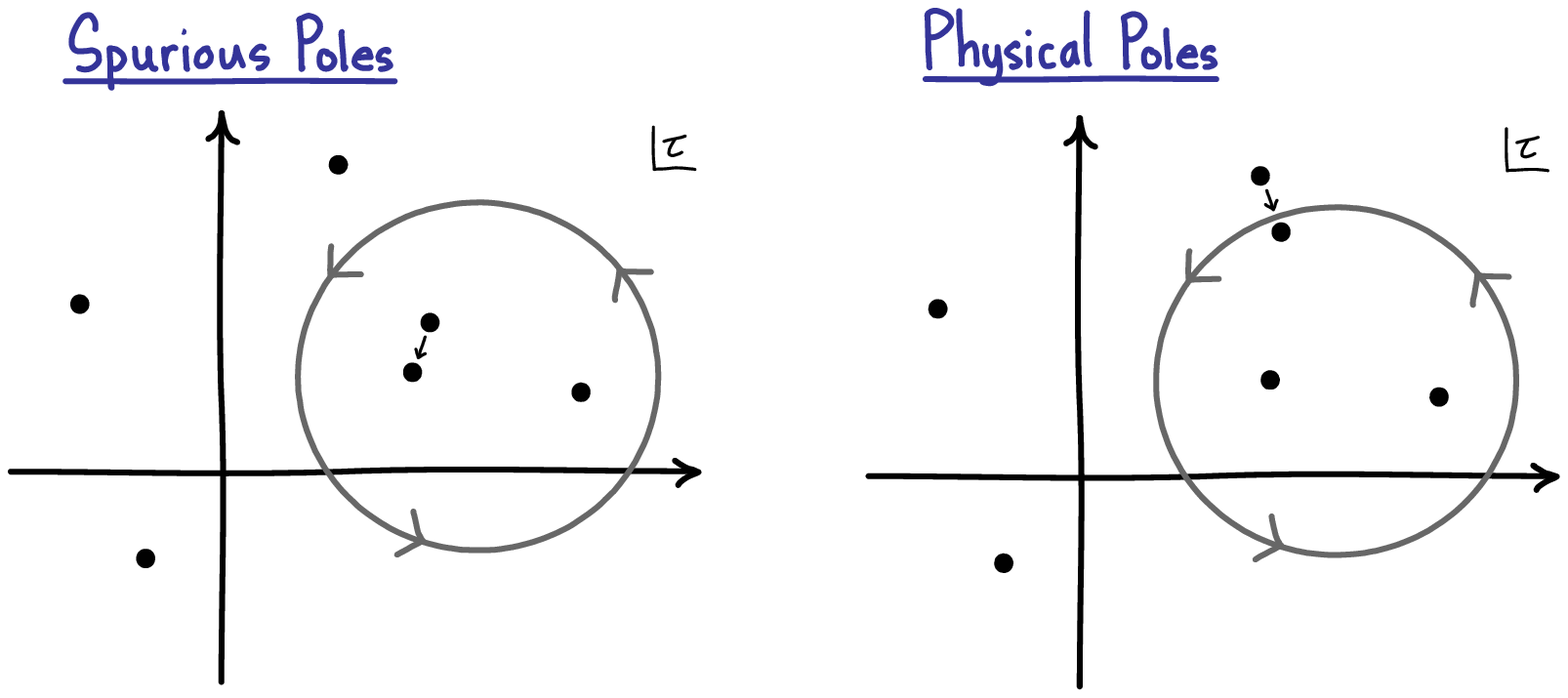} \nonumber
\ee

Let us look at the physical singularities in more detail. Given that there are 3 $c$ and 3 $\tilde c$ poles, there are 9 ways of colliding them in pairs. On physical grounds we expect poles of the form $s_{12}$,
$s_{23}$, $s_{34}$, $s_{45}$, $s_{56}$, $s_{61}$, $t_{123}$,
$t_{234}$, $t_{345}$. Note that collinear poles can be factored
$s_{i \,i+1} = \langle i\,i+1\rangle [i \,i+1]$ while $t_{i \,i+1 \,i+2}$
cannot. This means that $t$-poles must be parity invariant. Indeed,
note that a $t_{123}$ pole is reached when $(123)=\tilde c_{25}$ and
$(456) =c_{25}$ collide. The same happens for $t_{234}$ and
$t_{345}$. Working with complexified momenta means that we can consider the
poles $\langle ij\rangle$ and $[ij]$ as independent. This gives 12
poles but there are only 6 pairs of non-parity related poles that can
collide pinching the contour. This means that complex collinear
limits must also pair up. It is easy to see that this is the case.
For example, poles $\tilde c_{25}$ and $c_{41}$ collide when
$[12]\to 0$ or when $\langle 45\rangle \to 0$.

Let us quickly see how to obtain the other helicity configurations. For instance, consider the ``split-helicity" configuration $M^{---+++}$. Performing the appropriate $\tilde \eta$ integrations simply multiplies the alternating helicity integrand by a factor $c_{25}(\tau)^4$. This has the effect of removing the pole at $c_{25} = 0$, though the integrand still vanishes at infinity like $\frac{1}{\tau^2}$. Thus the BCFW form of this amplitude has only two terms, while the P(BCFW) still has three.

It is also instructive to see how all of these results can be recovered from another gauge fixing. For instance, let us consider the gauge fixing where the first three columns are set to the orthonormal basis
\be
\label{split}
C = \left(\begin{array}{cccccc} 1 & 0 & 0 & c_{14} & c_{15} & c_{16} \\ 0 & 1 & 0 & c_{24} & c_{25} & c_{26}  \\ 0 & 0 & 1 & c_{34} & c_{35} & c_{36} \end{array} \right)
\ee
This is the gauge-fixing convenient for the split-helicity configuration; we see that $(123)=1$, $(234) = c_{41}$, $(345) = c_{14}c_{25}-c_{24}c_{15}$, and so on.
From this form it is immediately apparent that, in computing the split helicity amplitude,  we only have $5$ poles in the complex $\tau$ plane,  since by the gauge-fixing the minor $(123)$ is identically set to $1$. But how are we to see all six terms needed for the alternating helicity amplitude? The answer is obvious. If we want to use this gauge-fixing to compute the alternating helicity amplitude, the appropriate $\tilde \eta$ integrals once again multiply the integrand by $c_{25}(\tau)^4$. This grows as $\tau^4$ for large $\tau$, while the denominator falls as $\tau^{-5}$, so there is a pole at infinity; this precisely gives the ``missing" residue, which we have checked correctly completes the amplitude.

\subsubsection{All Loop Leading Singularities?}

There is something a little peculiar about our discussion of the 6 particle NMHV amplitude. The residues of our object $L_{6;3}$ appear to contain more information than what is needed for the tree amplitude, which are a particular combination of the residues. It is natural to wonder whether the individual residues have any meaning. As we reviewed in our discussion of one-loop leading singularities, BCFW and P(BCFW) terms are the 1-loop leading singularities associated with scalar boxes with two adjacent massless legs. But all the scalar boxes with 6 external legs are of this form. Therefore, {\it all} the leading singularities of the 6 particle NMHV amplitude are determined by the residues. The box coefficients are given by residues as shown below
\be
\includegraphics[scale=0.7]{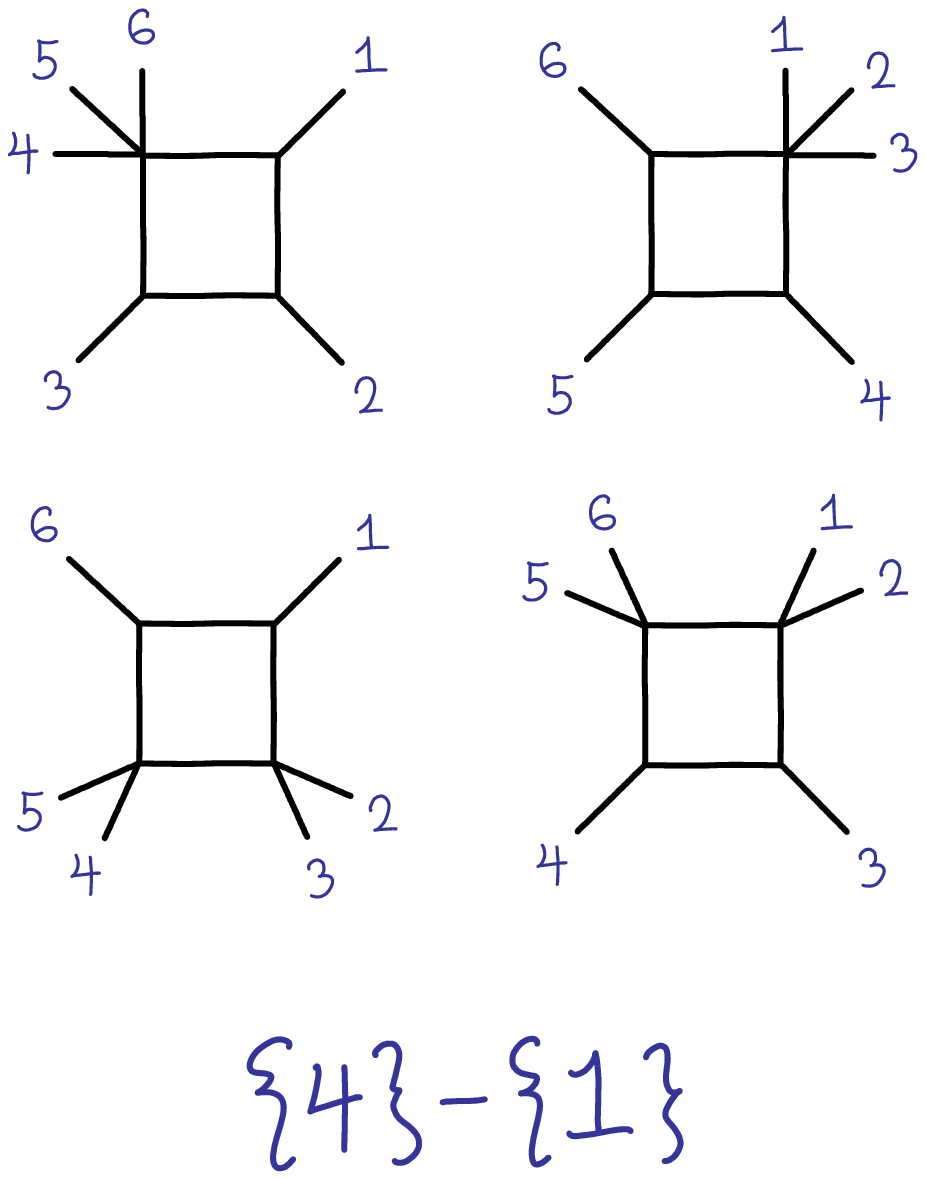} \nonumber
\ee
where all other boxes are cyclically related to these\footnote{The fact that the coefficient of these four scalar boxes is the same guarantees the IR equation from the double-logarithm of $t_{123}$.}. There are two terms since each box receives a contribution from a leading singularity and its parity conjugate.

With this in mind, we can revisit the case of the MHV amplitude, where $L_{n;2}$ involves no integrations and directly produces the tree amplitude. The one-loop MHV amplitudes in ${\cal N}=4$ were computed in \cite{Bern:1993qk} and are given by
\be
M^{1\hbox{-}{\rm loop}}_{n;{\rm MHV}} = M^{\rm tree}_{n;{\rm MHV}}\sum_{\{P,Q,R,S\}\in \{{\rm 1m,2m\hbox{-}e}\}} I_{P,Q,R,S}
\ee
where the sum is over all one-loop $1$-mass and $2$-mass-easy box integrals normalized to have leading singularity equal to one. Thus, the only non-vanishing leading singularities are proportional to the MHV tree amplitude, and one can therefore also give a 1-loop interpretation to $L_{n;k}$ for MHV amplitudes!

So far the evidence that $L_{n;k}$ is actually computing 1-loop leading singularities is rather circumstantial; we will see more dramatic and direct evidence starting with the 7 particle NMHV amplitude. But first, let us take a peek at our ultimate claim, which is that $L_{n;k }$ computes leading singularities at {\it all} loop orders. In particular, all leading singularities of two-loop amplitudes have been computed for MHV amplitudes up to 6 particles and have been found to either vanish or be proportional to the tree MHV amplitude. The reader might wonder how a two-loop computation can possibly give rise to such a simple object as a tree amplitude. In order to illustrate this we consider the four-particle 2-loop amplitude and show that it provides the prototype of what we called ``composite" leading singularities in section 1.1.

Following the discussion and the notation in section 1.1., at two-loop level and four particles, the closest we can get to $4\times 2 = 8$ propagators is to collect all Feynman diagrams with seven propagators with a ``double-box" topology. After choosing the contour where all seven propagators are on-shell, the rational function $R(\ell_1,\ell_2)$ in
\be
\int \frac{d^4\ell_1d^4\ell_2}{\prod_{i=1}^7P_i^2}R(\ell_1,\ell_2)
\ee
factors as the product of seven tree amplitudes shown in the figure. The piece on the right hand side is a one-loop leading singularity which is equal to a 4-point tree-amplitude. This four-point tree amplitude has two factorization channels. The remaining loop integral can be taken as a contour integral on a $T^4$ defined so that the $t$-channel of the four-particle tree amplitude is manifest. We are then left with a 1-loop leading singularity which evaluates to a four-particle tree amplitude once again! This is precisely the procedure used to obtain the 5 and 6-point results.

\be
\includegraphics[scale=0.9]{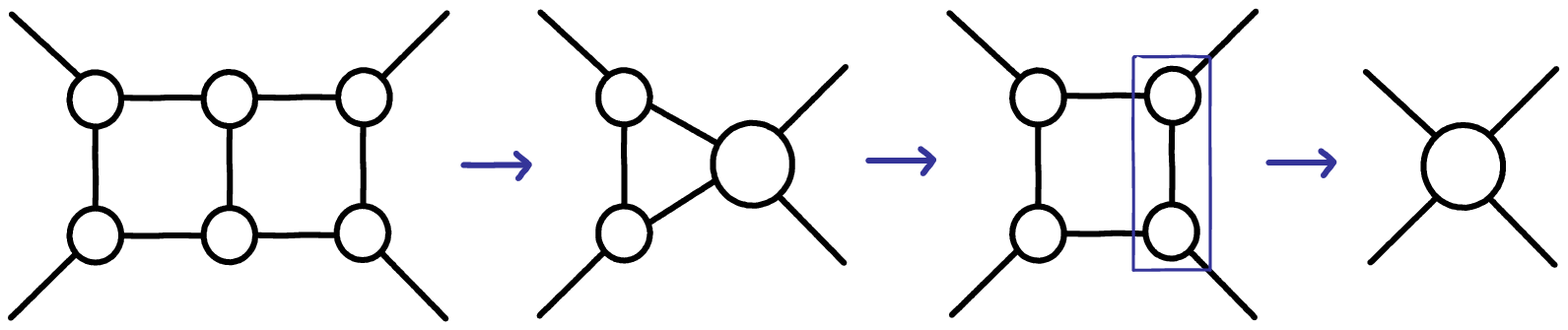} \nonumber
\ee

There is also a computation in progress for the 6 particle NMHV amplitude \cite{vergu}.
%For the MHV amplitudes which have been studied, it continues to be the case that at 2-loops, the leading singularities either vanish or are proportional to the tree MHV amplitude.
Quite remarkably, for the 6 particle NMHV amplitude, the very same rational functions we have already encountered at 1-loop determine the 2-loop leading singularities as well! Thus, the available data at higher-loop orders is non-trivially consistent with the idea that $L_{n;k}$ is indeed computing leading singularities at {\it all} loop orders. We will provide much more compelling evidence that two-loop leading singularities are present in our discussion of the 8 particle NMHV and N$^2$MHV amplitudes. However, let us first consider the striking evidence for 1-loop physics in the 7 particle NMHV amplitude.

\subsection{7 Particle NMHV Amplitude and Direct One Loop Evidence}

Consider $k=3$ and $n=7$. The number of integration variables is $2$. From the six particle example we learned how different gauge fixings lead to equivalent forms for the amplitudes. We find it most useful to consider ``gauge" choices in which there are no contributions from poles at infinity.

In this discussion we concentrate our attention on the phenomena that do not appear for $6$ particles. The most important is the definition of residues in more than one complex variables and the generalization of the residue theorem which we will find leads to the equations obtained from the IR behavior of one-loop amplitudes.

To be concrete let us look at the $(1^- 2^+ 3^- 4^+ 5^- 6^+ 7^+)$ amplitude. Then we have
\be
\label{sev}
L_{7;3} = J
\int \frac{d^2 \tau}{\left[(123)(234)(345)(456)(567)(671)(712)\right](\tau)}
\ee
where the matrix $C$ is of the form
\be
C = \left(\begin{array}{ccccccc} 1 & c_{12} & 0 & c_{14} & 0 & c_{16} & c_{17} \\ 0 &
c_{32} & 1 & c_{34} & 0 & c_{36} & c_{37} \\ 0 & c_{52} & 0 & c_{54} &
1 & c_{56} & c_{57} \end{array} \right)
\ee
As we have mentioned in this case each of the factors $(i, i+1 , i+2)$ in the denominator is linear in the two variables $(\tau_1,\tau_2)$.

\subsubsection{Multivariable Residues}

We want to interpret the integral as a contour integral.
In order to learn how to compute residues in this situation, let us consider a simple example. Consider the following function of two complex variables $z_1$ and $z_2$
\be
f(z_1,z_2) = \frac{h(z_1,z_2)}{(a z_1+b z_2 + c)(e z_1+f z_2 + g)}
\ee
where $h(z_1,z_2)$ is any function which is regular where $(a z_1+b z_2 + c)=0$ and $(e z_1+f z_2 + g)=0$.
Then we want to define a contour integral of the form
\be
{\cal I} = \int \frac{h(z_1,z_2) dz_1dz_2}{(a z_1+b z_2 + c)(e z_1+f z_2 + g)}
\ee
which can be called a residue of $f$. The most natural way to evaluate an integral of this form is to perform a change of variables $u_1 = a z_1+b z_2 + c$ and $u_2=e z_1+f z_2 + g$. Therefore, we have
\be
{\cal I} = \int \frac{du_1}{u_1}\frac{du_2}{u_2}\frac{h(z_1(u),z_2(u))}{{\rm det}\left(\frac{\del(u_1,u_2)}{\del (z_1,z_2)}\right)}
\ee
Now it is very natural to define the residue at $(u_1,u_2)=(0,0)$ as the integral over $|u_1| = \epsilon$ and $|u_2|=\epsilon$ for some small real number $\epsilon$ in complete analogy with the one-dimensional case. Therefore the residue is given by
\be
{\rm Res}[f](z_1^*,z_2^*) =  \frac{h(z_1^*,z_2^*)}{\left. {\rm det}\left(\frac{\del(u_1,u_2)}{\del (z_1,z_2)}\right)\right|_{z_1^*,z_2^*}} .
\ee
Here $z_1^*$ and $z_2^*$ are the solutions to $a z_1+b z_2 + c = 0$ and $e z_1+f z_2 + g = 0$.

Note that due to the antisymmetry of the determinant in $(u_1,u_2)$, the residue defined in this way depends not only on $(z^*_1,z^*_2)$ but also on the order in which the two vanishing factors $u_1,u_2$ are written.
This is unlike what we are used to with a single complex variable; the reason is that in the familiar case, the contour defining the residue actually encloses the pole. In the case at hand, the contour of integration used in the definition of the residue is the product of two circles, $S^1\times S^1\backsimeq T^2$ in $\mathbb{C}^2$, which does not enclose the point $(z_1^*,z_2^*)$. In the math literature this $T^2$ is called the distinguished boundary to emphasize this fact.

Using this result, let us consider a function of the form
\be
f(z_1,z_2) = \frac{g(z_1,z_2)}{p_1(z_1,z_2) p_2(z_1,z_2) \cdots p_M(z_1,z_2)}
\ee
For $M \ge 2$ and the $p_i(z_1,z_2)$ are linear. Then we have $M \choose 2$ residues, determined by putting any choice of two of the factors $p_{i_1},p_{i_2}$ to 0. The residue is given by
\be
{\rm Res}[f](z_1^*,z_2^*) =  \frac{g(z_1^*,z_2^*)}{\prod_{i \neq (i_1,i_2)} p_i(z^*_1, z^*_2) \left. {\rm det}\left(\frac{\del(p_{i_1},p_{i_2})}{\del (z_1,z_2)}\right)\right|_{z_1^*,z_2^*}} .
\ee
We can trivially generalize this discussion to functions of $N$ complex variables of the form
\be
f(z_1, \cdots, z_N) = \frac{g(z_1,\cdots,z_N)}{p_1(z_1, \cdots, z_N) \cdots p_M(z_1,\cdots,z_N)}
\ee
where the $p_i(z)$ are linear, and $M \ge N$. We can choose $N$ of denominator factors, $p_{i_1},\cdots,p_{i_N}$, and solve the linear equations $p_{i_1}(z^*_1, \cdots, z^*_N) = \cdots = p_{i_N}(z^*_1,\cdots,z^*_N)$ = 0 to determine the point $z^*_1, \cdots, z^*_N$. The residue is then defined to be
\be
{\rm Res}[f](z^*_1,\cdots,z^*_N) = \frac{g(z_1^*,\cdots,z_N^*)}{\prod_{i \neq(i_1,\cdots,i_N)} p_i(z_1^*,\cdots,z^*_N) \left. {\rm det}\left(\frac{\del(p_{i_1}, \cdots,p_{i_N})}{\del (z_1, \cdots, z_N)}\right)\right|_{z_1^*,\cdots, z_N^*}}
\ee
Note again that, since the determinant is antisymmetric in $(i_1,\cdots,i_N)$, the residue also has this antisymmetry property. So, the residue is not only associated with the point $z^*_1,\cdots,z^*_N$, but the {\it sign} of the residue is determined by the order in which the factors $p_{i_1},\cdots,p_{i_N}$ are taken.

\subsubsection{One Loop Leading Singularities}

In equation (\ref{sev}), one has seven linear factors in the denominator. This means that we have ${7 \choose 2} = 21$ different residues. We choose to denote each residue by the pair of linear factors which vanish; we denote the residue obtained by putting $(i, \, i+1, \, i+2),(j, \,j+1, \, j+2) \to 0$ as $\{i,j\}$. Note that as mentioned above this residue is naturally antisymmetric, so that $\{i, j \} = - \{j, i \}$.

The computation of the residues is completely straightforward, and only involves solving linear equations. Just to illustrate this first non-trivial example with more than one complex variable we give some details in the appendix, but for the rest of this paper we leave the computation of the residues in the capable hands of Mathematica.
Even though the residues only involve solving simple linear equations, the expressions quickly become complicated. To give two examples, we have

\be
\{1,4\} = \frac{\langle 1 3 \rangle^4 [7|(4 +6)|5\rangle^4}{s_{123} s_{456}
\langle 1 2 \rangle \langle 2 3 \rangle \langle 4 5 \rangle \langle 5 6 \rangle
[7|(5+6)|4\rangle[7|(1+2)|3\rangle \langle 6|(4+5)(2+3)|1\rangle} \nonumber
\ee
and
\be
\{2,4\} = \frac{([7|(2 + 4)|3\rangle \langle 5 4 \rangle + [76]\langle 6 5\rangle\langle 3 4
\rangle)^4}{\langle 2 3 \rangle \langle 3 4 \rangle \langle 4 5 \rangle \langle 5 6
\rangle
[7 1] [1|(2+3)|4\rangle [7|(5+6)|4\rangle
\langle 4 (5+6)(7+1)|2\rangle \langle 4|(2+3)(7+1)|6\rangle} \nonumber
\ee
We will spare the reader the sight of all 21 residues.

The connection of these 21 residues to the corresponding one-loop amplitude is the following. In \cite{Bern:2004ky}, Bern et.al.~computed the full seven-particle amplitude. In this case one has $1$-mass, $2$-mass-easy, $2$-mass-hard and $3$-mass boxes. In \cite{Bern:2004ky} it was found that all the coefficients could be written as linear combinations of basic objects. The amplitude ($-+-+-++$) has a flip symmetry which is obtained by applying $i\to {\rm mod}(6-i)+1$ to the particle labels. In \cite{Bern:2004ky}, a list of 12 basic objects was given. Out of these, 3 are flip invariant. This means that the total number including the flip images is given by $3+2\times 9 = 21$. Quite remarkably, our 21 residues map one to one to the 21 basic objects in \cite{Bern:2004ky}.

\be
\includegraphics[scale=0.7]{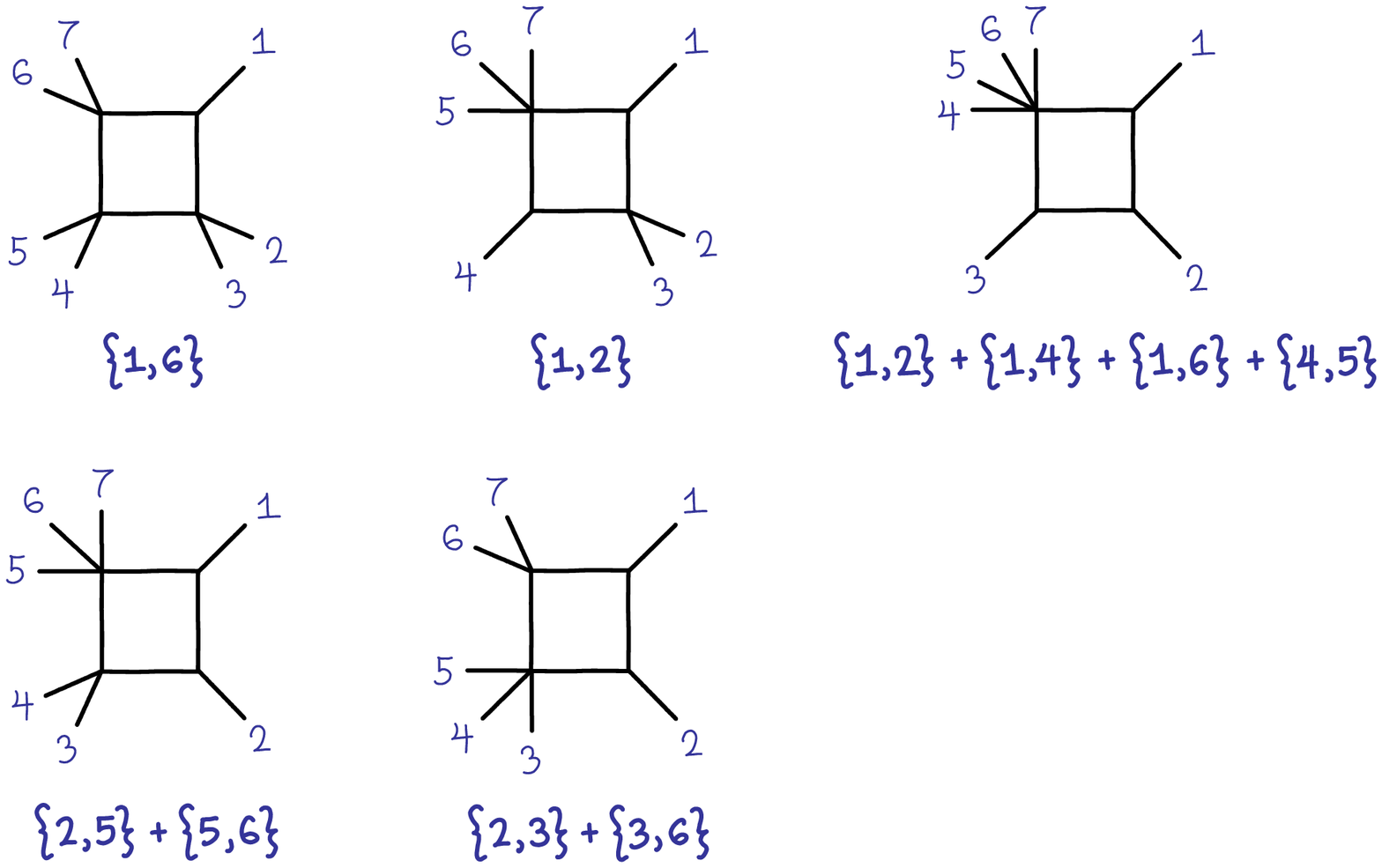} \nonumber
\ee

We can also specifically identify the residues that give the tree amplitude. Once again there is a BCFW and P(BCFW) form of the amplitude, and we can identify them amongst the residues as

\begin{eqnarray}
\label{sevbcf}
M^{-+-+-++}_{{\rm BCFW}} & = & \{1,3\}+\{1,5\}+\{1,7\}+\{3,4\}+\{3,6\}+\{5,6\}, \\
M^{-+-+-++}_{{\rm P(BCFW)}} & = & -\{2,3\}-\{2,5\}-\{2,7\}-\{4,5\}-\{4,7\}-\{6,7\}.
\end{eqnarray}

\subsubsection{Relations Between Coefficients}

The precise agreement between our 21 residues and the 21 rational functions first defined by Bern et.al.~in the 1-loop amplitude is a quite remarkable confirmation that our object is computing 1-loop amplitudes. This agreement is
even more surprising given the fact that, as noted in \cite{Bern:2004ky}, these basic objects are not linearly independent. In \cite{Bern:2004ky} four relations among the 21 basic objects were given. One of them is flip symmetric. This means that counting the flip images one gets $1+2\times 3=7$ relations. These relations come from the IR equations for the 1-loop amplitude, which as reviewed in the appendix, relate the double-logarithmic divergences of the 1-loop amplitude to the tree amplitude. These equations constrain the coefficients $B$ of the scalar boxes; partitioning the external states into 4 sets $(s_1,s_2,s_3,s_4)$, the coefficient of the box with the sums of the momenta in $s_i$ flowing into its corners is denoted by $B_{(s_1)(s_2)(s_3)(s_4)}$.

Let us look at the equation associated with the double-logarithm of $t_{123}$,
\begin{eqnarray}
\label{ires}
0 = & -B_{(1)(2)(3)(4567)} + B_{(7)(1)(23)(456)} + B_{(3)(4)(567)(12)} -\frac{1}{2}B_{(4)(5)(67)(123)} -\frac{1}{2}B_{(6)(7)(123)(45)} \\ & -B_{(4)(56)(7)(123)} + B_{(1)(23)(4)(567)} + B_{(3)(456)(7)(12)} +\frac{1}{2} B_{(3)(45)(67)(12)} + \frac{1}{2} B_{(1)(23)(45)(67)}. \nonumber
\end{eqnarray}
%\
Here e.g. $B_{(1)(2)(3)(4567)}$ is the coefficient of a box with momenta flowing through the corners given by the sum of momenta appearing in the parentheses. Using the explicit map we have just found between leading singularities and residues we find
\be\nonumber
\begin{array}{ccc}
B_{(7)(1)(23)(456)} = \{1,4\}+\{4,5\}, & \!\!\! B_{(3)(4)(567)(12)} = \! \{5,1\}+\{4,5\}, & \!\!\! B_{(4)(5)(67)(123)} =\! \{5,1\}+\{1,2\}\\
B_{(6)(7)(123)(45)}=\{1,4\}+\{7,1\}, & B_{(4)(56)(7)(123)} = \{4,5\}, & B_{(1)(23)(4)(567)}= \{ 1,2\},\\
B_{(3)(456)(7)(12)} = \{7,1\}, & B_{(3)(45)(67)(12)} = \{3,1\}, & B_{(1)(23)(45)(67)} = \{1,6\} \\
\end{array}
\ee
and
\be
B_{(1)(2)(3)(4567)} = \{1,4\}+\{4,5\}+\{1,6\}+\{1,2\}.
\ee
Using this in equation (\ref{ires}) one finds a surprising simple result
\be
\label{fino}
0 = \{1,2\}+\{1,3\}+\{1,4\}+\{1,5\}+\{1,6\}+\{1,7\}.
\ee
Clearly, by cyclic symmetry we have seven relations of the form
\be
\label{ire}
0 = \sum_{i=1}^7 \{j,i\}, \;\; j\in \{1,\ldots, 7\}
\ee
Note that $\{i,i\}=0$.

In our particular problem, we have a flip symmetry. The relation coming from $j=2$ is flip symmetric while the other six split into pairs related by a flip transformation. Very nicely, in \cite{Bern:2004ky}, three of the four relations given are six-term relations of the form (\ref{ire}) while the last one is a 12 term identity that can be shown to follow from the other six-term identities and one more of our six-term identities.

Expressing the residues as explicit rational functions of the kinematical invariants, these identities appear to be miraculous statements. Another amazing identity, completely analogous to the six-term identity we encountered with 6 particles, is the 12 term identity $M_{{\rm BCFW}} = M_{{\rm P(BCFW)}}$.

\subsubsection{Relations Arise From Generalized Residue Theorems}

The form of the IR equations (\ref{ire}) begs for a residue theorem derivation. Note that we have seven relations. In the theory of residues of a single complex variables one expects a single relation. This is the first difference between single and multidimensional residues. We will touch on these points in more detail and generality in the next section, but for the case at hand we can make do with repeated application of the usual Cauchy theorem. Our integral is of the form
\be
\int \frac{d\tau_1d\tau_2}{\prod_{i}(a_i \tau_1+b_i\tau_2 +c_i)}
\ee
in our example the product is taken for $i = 1, \cdots, 7$ but we'll imagine it for any $i=1, \cdots, M$ for $M > 2$.
Let us choose the $j^{\rm th}$ factor in the denominator and treat $\tau_1$ as a fixed variable. Therefore we can use the integral in $\tau_2$ to compute the residue at $\tau_2 = (-a_j\tau_1-c_j)/b_j$. The remaining $\tau_1$ integral is then of the form
 \be
 \int \frac{d\tau_1}{\prod_{i\neq j}(\tilde a_i \tau_1 + \tilde c_i)}
 \ee
Now the usual one dimensional residue theorem tells us that the sum over the $M-1$ residues is zero. Recalling that each residue came from first setting $(j,j+1,j+2)=0$ we find
\be
\label{firstresidue}
0 = \sum_{i}\{j,i\}.
\ee
As we saw in the previous subsection, these equations are nothing but the IR equations for the 1-loop amplitude! Thus, for the 7 particle amplitude, the IR equations which imprint both locality and unitarity in the 1-loop amplitude, are a direct consequence of the residue theorem for two complex variables.

There is a simple corollary of equation (\ref{firstresidue}) that is sometimes useful. Let $H$ be a subset of $\{1,\ldots, M\}$ and $\bar{H}$ its complement. Then the following set of equalities hold
\be
0 = \sum_{i\in H}\sum_{j}\{i,j\} = \sum_{j} \sum_{i\in H}\{i,j\} = \sum_{j\in \bar{H}}\sum_{i\in H}\{i,j\}
\ee
The last equality holds due to the anti-symmetry property. Consider the identity obtained by choosing $H =\{ 1,2,4,6\}$ and ${\bar H}=\{3,5,7\}$ for our seven-particle amplitude. The identity is
\begin{eqnarray}\nonumber
0 = && \{ 1,3\} +\{ 1,5\} +\{ 1,7\} +\{ 2,3\} +\{ 2,5\} +\{ 2,7\} + \{ 4,3\} +\{ 4,5\} +\{ 4,7\} \\ &&  +\{ 6,3\} +\{ 6,5\} +\{ 6,7\}.
\end{eqnarray}
It is easy to see that this 12-term identity is precisely the one obtained from equating the expressions for $M_{{\rm BCFW}}$ and $M_{{\rm P(BCFW)}}$ in equation (\ref{sevbcf}). In this example we appeared to need some foresight to make the correct choice for $H$ to produce the needed identity for tree amplitudes. In fact there is a more natural and systematic way of understanding these identities, which we defer to a general discussion of all NMHV amplitudes in section \ref{sec:GenNMHV}.

\subsubsection{Prediction for Higher Loops}

We have seen that every one of the residues associated with the 7 particle NMHV amplitude can be identified with 1-loop leading singularities. In complete analogy with what we saw for the MHV and 6 particle NMHV amplitudes, we then predict that the leading singularities for the 7 particle amplitude at two loops and beyond should all be determined by the 21 objects we have already identified. In the next subsection, we find more direct evidence for the presence of new objects associated with 2-loop leading singularities starting with the 8 particle NMHV amplitude.

\subsection{8 Particle NMHV and Direct Two Loop Evidence}

The expressions for the box coefficients of general 1-loop NMHV amplitudes have been determined in \cite{Drummond:2008bq}, and we will discuss the identification of these with residues in the next section. But in this subsection we look at the 8 particle NMHV amplitude specifically, because for the first time in this case not all residues are accounted for amongst the 1-loop leading singularities, and we can identify new 2-loop leading singularities.

The mapping between box coefficients and residues is given below. We use a notation for residues which is the obvious
generalization of what we introduced for the $7$ particle case: the residue $\{i,j,k\}$ is associated with putting the minors $(i, \, i+1, \, i+2), (j, \, j+1, \, j+2), (k, \, k+1, \, k+2)$ to zero, and is antisymmetric in exchanging any of the indices.
 For the 3-mass boxes and the 2-mass-hard boxes with two adjacent massless legs, we have
\be
\includegraphics[scale=0.7]{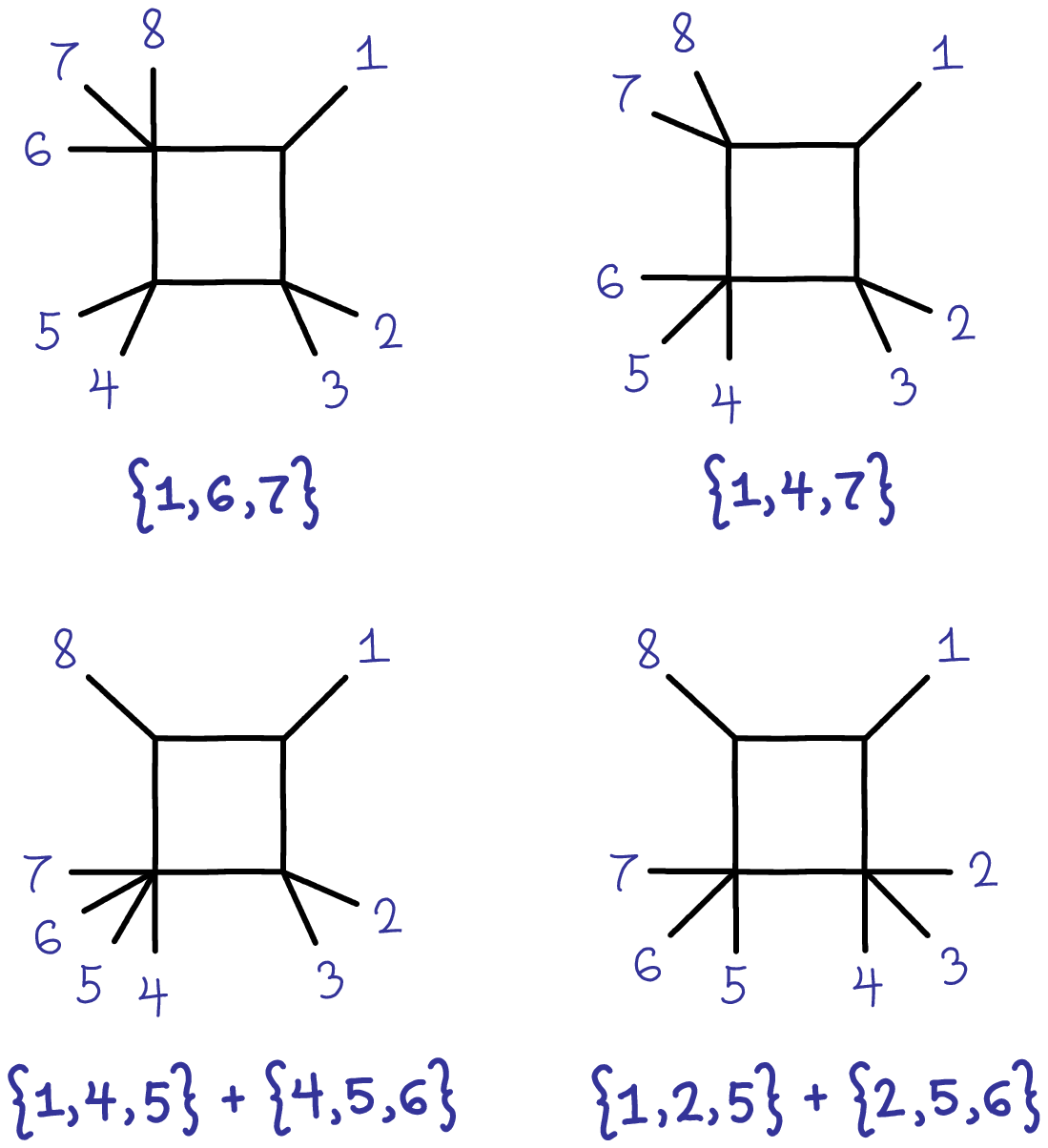} \nonumber
\ee
For the 2-mass-easy boxes with massless legs at opposite corners, and the 1-mass boxes, we have
\be
\includegraphics[scale=0.7]{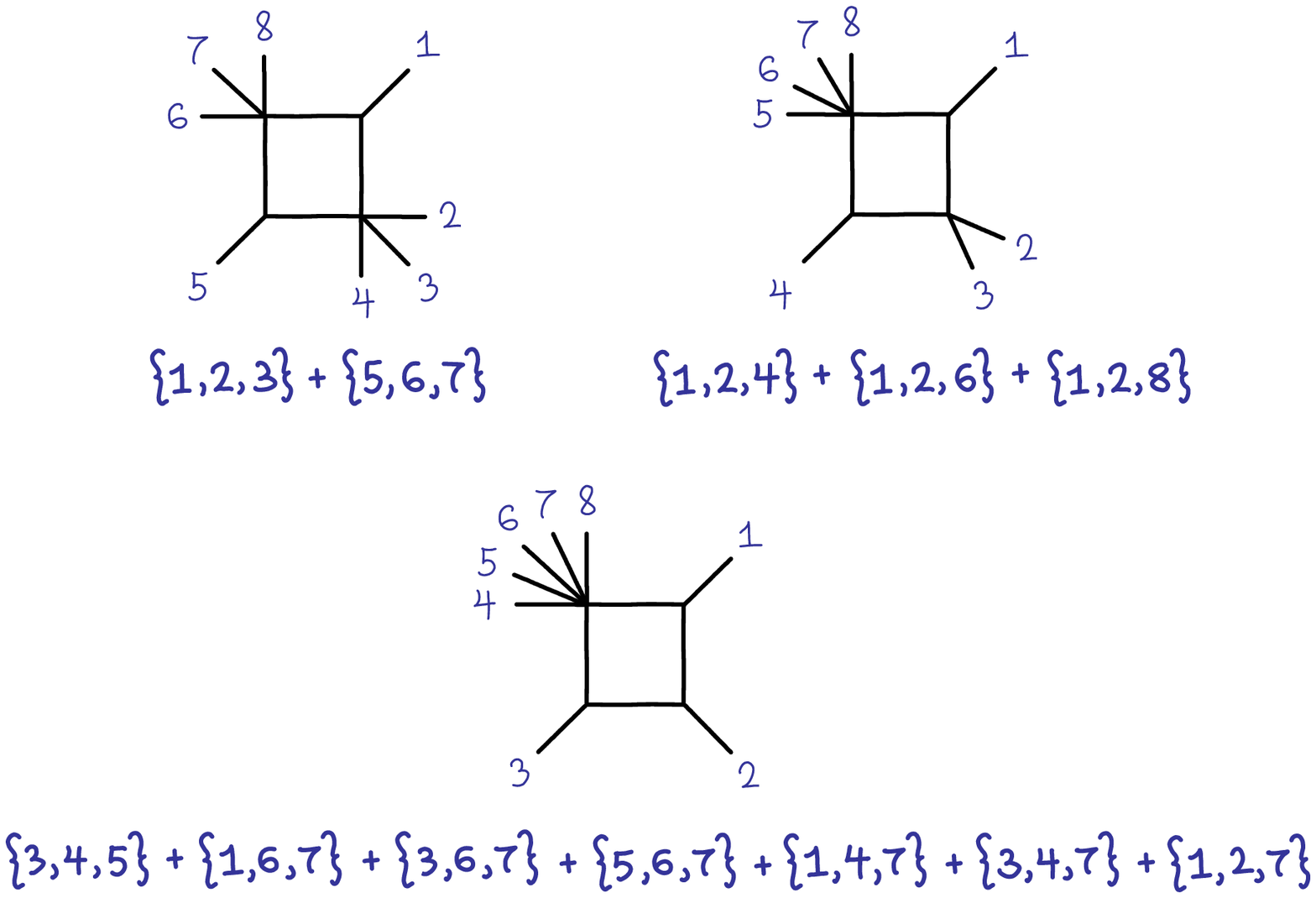} \nonumber
\ee
All other boxes are related to these by the cyclic symmetry.

We defer a discussion of the tree amplitudes to the next section, where we discuss these for all NMHV amplitudes in generality.

\subsubsection{IR Equations and Residue Theorems}

Let us take a very quick look at some of the IR equations and their origin in residue theorems. Consider the IR equation associated with a double-logarithmic dependence on $t_{1234}$. This is a particularly simple case to look at, since the 1-mass box, which has the longest expression in terms of residues, does not produce this singularity and therefore does not participate. This IR equations involves 14 box coefficients:
\begin{eqnarray}
-\frac{1}{2}B_{(7)(8)(1234)(56)} - \frac{1}{2} B_{(1)(2)(34)(5678)} -\frac{1}{2}B_{(3)(4)(5678)(12)}
- B_{(8)(1234)(5)(67)} \nonumber \\ - B_{(1)(234)(5)(678)} - B_{(1)(23)(4)(5678)} +
B_{(8)(123)(4)(567)} + \frac{1}{2} B_{(1)(234)(56)(78)} \nonumber \\ + \frac{1}{2} B_{(5)(678)(12)(34)} + \frac{1}{2}
B_{(8)(12)(34)(567)} + \frac{1}{2} B_{(4)(56)(78)(123)} + B_{(4)(5)(678)(123)} \nonumber \\ + B_{(8)(1)(234)(567)} -\frac{1}{2} B_{(5)(6)(78)(1234)} = 0
\end{eqnarray}
We should expect these to follow simply from the generalized residue theorem just as we found for 7 particles. Following the same steps as we did for the two-complex variable case, the generalized residue theorem in the present case is of the form
\be
\sum_i \{j_1,j_2,i\} = 0
\ee
Using the explicit form of the box coefficients as residues, the IR equation becomes
\begin{eqnarray}
\frac{1}{2} \{1, 2, 3\} + \{1, 2, 4\} + \frac{1}{2} \{1, 2, 5\} + \{1, 2, 6\}+ \frac{1}{2} \{1, 2, 7\} \nonumber \\ + \{1, 2, 8\}
-\frac{1}{2}\{2,
5, 6\} - \frac{1}{2}\{4, 1, 2\} - \frac{1}{2}\{4, 5, 6\} + \frac{1}{2} \{5, 6, 1\} \nonumber \\ + \{5, 6, 2\}+ \frac{1}{2}\{5, 6, 3\}+ \{5,
6, 4\}+ \frac{1}{2}\{5, 6, 7\}  + \{5, 6, 8\} \nonumber \\ - \frac{1}{2}\{6, 1, 2\}-\frac{1}{2}\{8, 1, 2\}- \frac{1}{2}\{8, 5, 6\} = 0
\end{eqnarray}
Combining terms keeping in mind the antisymmetry of the residues, this follows from combining the residue theorems as
\be
\sum_i \{1,2,i\} + \sum_i \{5,6,i\} = 0
\ee

\subsubsection{Two-Loop Evidence}

Not all of the $8 \choose 3$= 56 residues we have are determined by the 1-loop leading singularities. A quick look at the map between residues and boxes reveals that all residues which appear at 1-loop are either of the form $\{i, \, i+1, \, j\}$ or $\{i, \, i+3, \, i+6\}$, that is, the indices appearing in the residues have at least one pair separated by an odd integer. Objects of the form $\{i, \, i+2l, \, i+2m\}$ with all even differences do not appear. Now, this is not quite an invariant statement; after all, we can use the residue theorem identities to trade some residues for others. However it is easy to see that, beginning with the form of the 1-loop leading singularities we have identified, any re-writing would involve sums of {\it pairs} of the $\{i, \, i+2l, \, i+2m\}$ residues.
%For instance looking at the residue theorem involving $\sum_k \{3, 5, k\} = 0$, we have
%\be
%0 = \left(\{1,3,5\} + \{3,5,7\}\right) + \left(-\{2,3,5\} + \{3,5,4\} + \{5,6,3\}\right)
%\ee
%which allows us to trade the linear combination $\{1,3,5\} + \{3,5,7\}$ for sums of terms appearing in the 1-loop leading singularities, but we relate any {\it single} term of the form $\{i, \, i+2, \, i+4\}$ to the others. This is a general result.
Consider an arbitrary residue theorem $\sum_k \{j_1,j_2,k\}= 0$; without loss of generality we can put $j_1 =1$. Unless $j_2$ is also odd, this residue theorem will not contain any of our ``missing" residues and is irrelevant. If $j_2$ is odd, then the terms with $k$ odd correspond to our residues, and there are even number of these (in fact only two of them since the terms with $k=j_1,j_2$ vanish). Thus this simple parity argument shows us that we can never express a single residue of the form $\{i,i+2l, i+ 2m\}$ in terms of the others; these objects are not determined by the 1-loop leading singularities.

It is therefore natural to conjecture that these residues are
associated with a genuine {\it two-loop} leading singularity. Before
making this connection, we can give a natural interpretation to the
rational functions that appear in these missing residues. As we will
discuss in more detail in \cite{notes}, there is a canonical way of
beginning with an object and adding a particle to it by applying
what we call an ``inverse soft factor" \cite{Nima}. Consider an $n$
particle object $O_n^{ab\cdots}$, where we have marked two
consecutive colors $a,b$ as being special. We can define an $(n+1)$
particle object $O_{n+1}^{a c b}$ that can be thought of as
inserting the particle $c$ between $a$ and $b$. Clearly this
operation will have to involve deforming the momenta of $a,b$ in
order for the new object to conserve momentum with the addition of
$c$. In \cite{notes} we will describe this operation in a fully
supersymmetric way, but for now we will only talk about the addition
of gluons. We add a positive helicity gluon $c^+$ by defining the
object \be O^{a c^+ b}_{n+1}(\lambda_a,\tilde \lambda_a; \lambda_c,
\tilde \lambda_c; \lambda_b, \tilde \lambda_b, \cdots) =
\frac{\langle a b \rangle}{\langle a c \rangle \langle c b \rangle}
O^{ab}_n(\lambda_a, \tilde \lambda_a^\prime; \lambda_b, \tilde
\lambda_b^\prime, \cdots) \ee where \be \tilde \lambda^\prime_a =
\frac{(p_a + p_c)|b\rangle}{\langle a b \rangle}, \,
\lambda^\prime_b = \frac{(p_b + p_c)|b\rangle}{\langle b a \rangle}
\ee It is easy to see that the deformation preserves momentum
conservation with the addition of $c$, and has the correct little
group properties as well. In the soft limit $p_c \to 0$,
$\lambda^{\prime}_{ab} \to \lambda_{ab}$, and $O^{a c^+ b}_{n+1}$
reduces to $O_n^{ab}$ when stripped of the standard soft factor
$\frac{\langle a b \rangle}{\langle a c \rangle \langle c b
\rangle}$ associated with positive helicity gluons. This is why we
refer to this way of adding a particle as ``applying an inverse soft
factor". A negative helicity gluon is added in the obvious analogous
way reversing the roles of $\lambda$ and $\tilde \lambda$.

If we apply the inverse soft factor to the leading singularity
associated with a 7 particle 3 mass box, the resulting object is
naturally interpreted as a 2-loop leading singularity, allowing us
to interpret the missing residues as certain 2-loop leading
singularities! The map below is associated with adding the particle
8 between 1 and 7 in a 7 particle 3-mass box, which is associated
with the $\{3,5,7\}$ residue. \be
\includegraphics[scale=0.7]{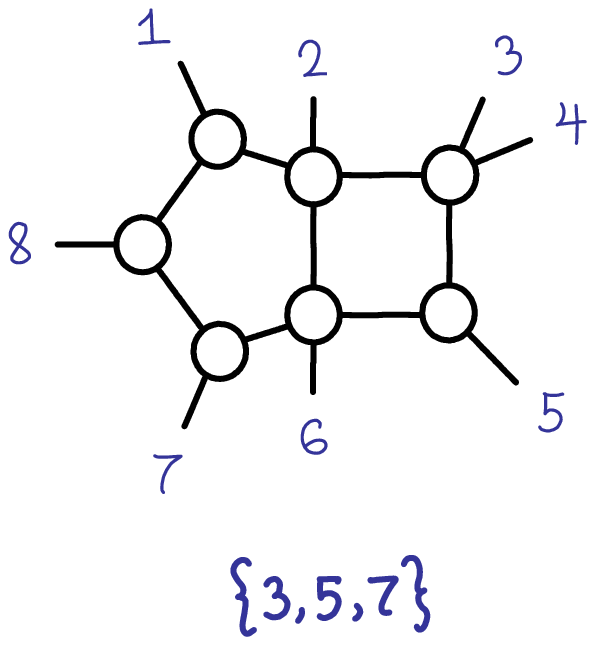} \nonumber
\ee
It is obviously beyond the scope of this paper to find all the 2-loop singularities of the 8 particle amplitude and identify all of them with our residues; as for the lower-point cases we have examined, our conjecture makes a clear prediction that all the leading singularities at 2 loops and beyond are one of the 56 residues we have identified.

\section{General NMHV Amplitudes}
\label{sec:GenNMHV}
Using the explicit calculation of the 1-loop box coefficients for all 1-loop NMHV amplitudes given in \cite{Drummond:2008bq}, we have associated all the 1-loop leading singularities with residues. A given residue for the $n$ particle NMHV amplitude is determined by giving a list $\{i_1, \cdots, i_{n-5}\}$
specifying the minors that vanish. To save space at large $n$, we can equally well specify the minors that are left out and denote the residues by
$\overline{\{j_1, j_2,j_3,j_4,j_5\}}$.

The simplest association of residues with box coefficients is for the 3 mass boxes, which are given by a single residue:

\be
\includegraphics[scale=0.7]{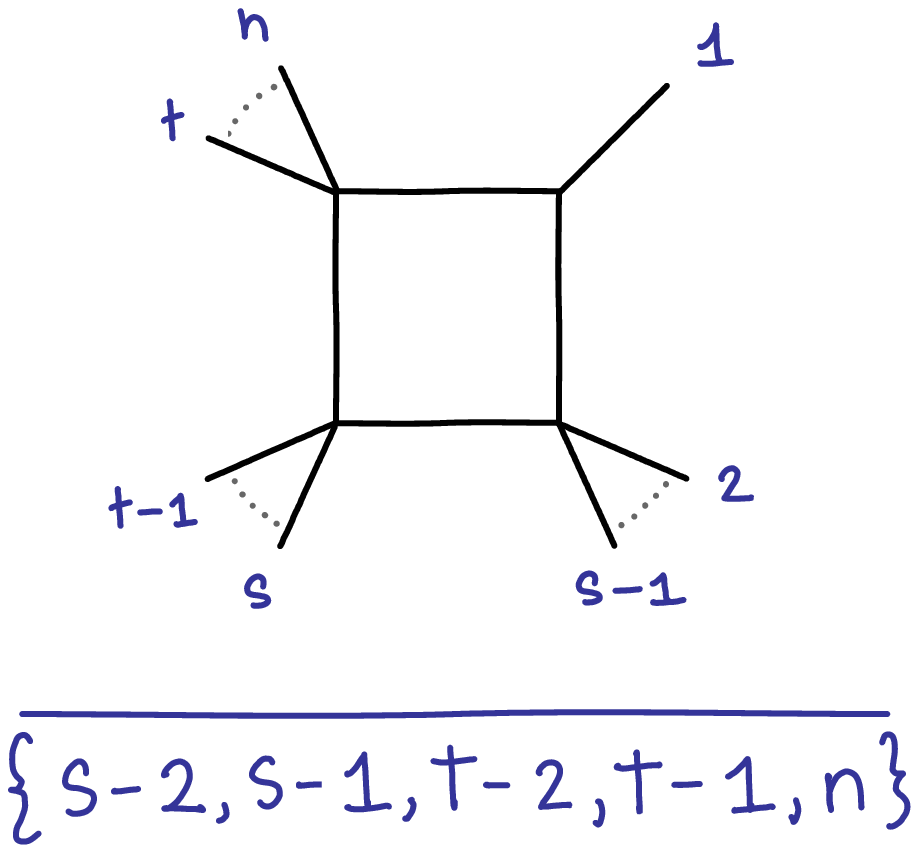} \nonumber
\ee

The ``2 mass hard" boxes, with two massive legs and two adjacent massless legs, are also very simply identified as the sum of two terms:

\be
\includegraphics[scale=0.7]{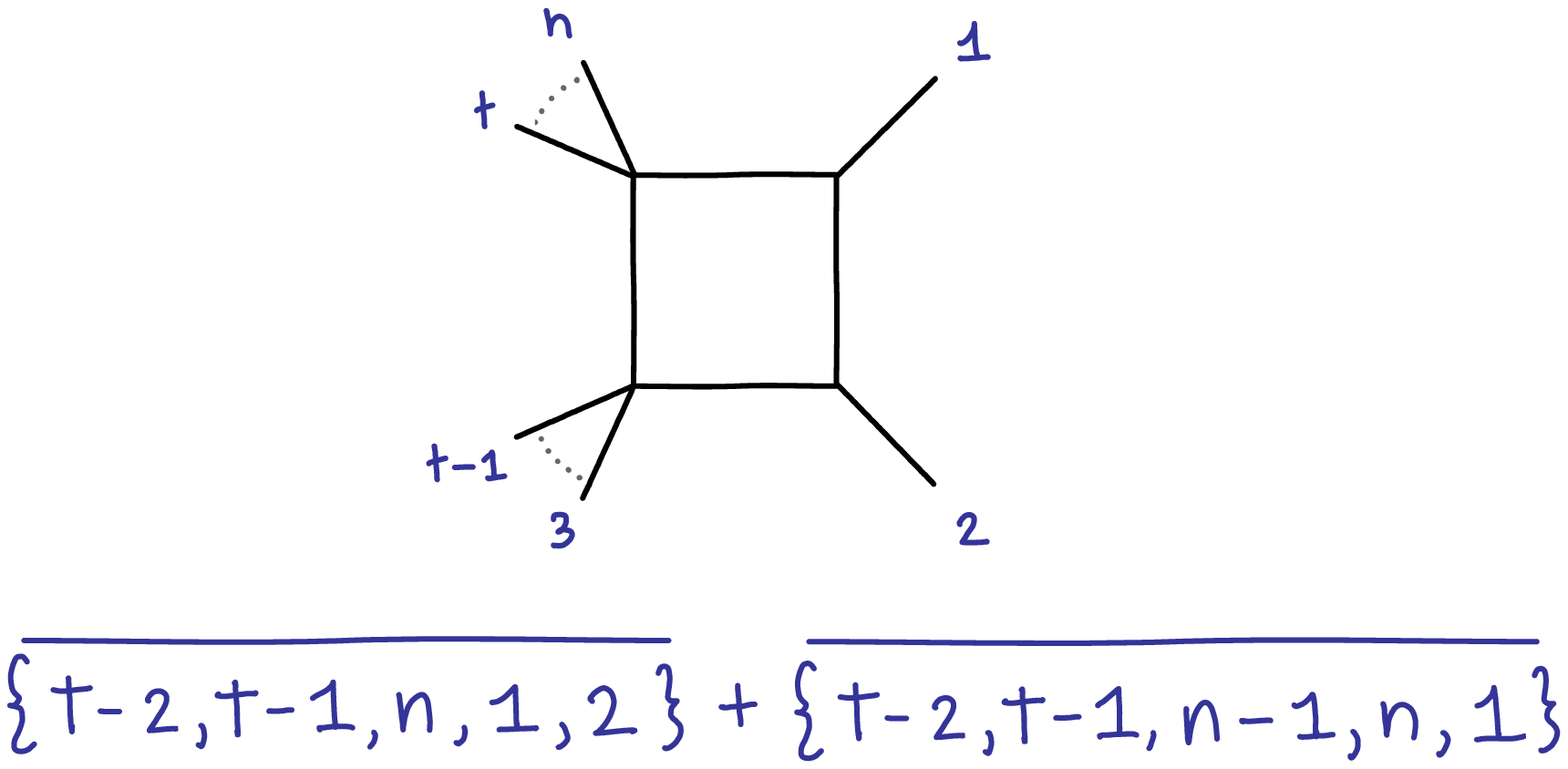} \nonumber
\ee

We will refrain from giving the full mapping from residues to the ``2 mass easy" and ``1 mass" boxes here, since merely listing the residues is not particularly illuminating. In fact we have found a beautiful structure in this map, which is easiest to study using ``Hodges diagrams" and other concepts we will more properly introduce in \cite{notes}. Elsewhere, we will also study the structure of the IR equations more systematically, and show that the same pattern we saw for $n=7,8$ persists to all $n$: the 1-loop IR equations follow from residue theorems.

\subsection{Tree Amplitudes}

Here we give a compact discussion of all tree NMHV amplitudes. Since we will be dealing with sums over many residues, it is convenient to introduce some algebraic notation for doing this. We can denote a given residue $\{i_1,i_2,\cdots,i_{n-5}\}$ by an antisymmetric product $\{i_1\} \{i_2\} \cdots \{i_{n-5}\}$. In this way we can talk about an expression like $(\{1\} + \{3\})(\{4\} + \{7\}) = \{1,4\} + \{1,7\} + \{3,4\} + \{3,7\}$. In order to write the NMHV amplitudes in a compact way, we introduce a little more notation. Let us define
\be
\{i_1\} \star \{i_2\} = \left\{\begin{array}{c} \{i_1\} \{i_2\} \, \, {\rm if} \, \, i_1 < i_2 \\  \\ 0 \, \, \, \, \, \, {\rm otherwise} \end{array}\right\}
\ee
Let us also define the formal sums ``${\cal E}$" (for ``even") and ``${\cal O}$" (for ``odd")
\be
{\cal E} = \sum_{k \, {\rm even}} \{k\}, \, {\cal O} = \sum_{k \, {\rm odd}} \{k\}
\ee
The BCFW form of the NMHV amplitude is then given by the following sum of residues:
\be
\begin{array}{cc} M^{{\rm NMHV}}_{n; {\rm BCFW}} = &
\underbrace{{\cal E} \star {\cal O} \star {\cal E} \star \cdots}\\ & (n-5) \, {\rm factors} \end{array}
\ee
Just to be explicit, we can write this out for $n=8$ as the 10 terms
\begin{eqnarray}
\{2\}\{3\}\{4\} + \{2\}\{3\}\{6\} + \{2\}\{3\}\{8\} + \{2\}\{5\}\{6\} + \{2\}\{5\}\{8\} + \{2\}\{7\}\{8\} \nonumber \\ + \{4\}\{5\}\{6\} + \{4\}\{5\}\{8\} + \{4\}\{7\}\{8\} + \{6\}\{7\}\{8\}
\end{eqnarray}
The P(BCFW) form of the amplitude is of exactly the same form, but exchanging ${\cal E} \leftrightarrow - {\cal O}$. Since the cyclic shift also exchanges ${\cal E}$ and ${\cal O}$, the expected identity $M_{{\rm BCFW}} = M_{{\rm P(BCFW)}}$ will also guarantee the cyclic invariance of the amplitude. We thus turn to proving this identity.

\subsection{Cyclic Invariance and Homology Classes}

Let's start by writing the general form of the residue theorems, which in a completely straightforward generalization of what we have seen already is of the form
\be
\sum_k \{i_1\} \{i_2\} \cdots \{i_{n-6}\} \, \{k\}= 0
\ee
We will show that these identities imply the desired statement
\be
{\cal E} \star {\cal O} \star \cdots = -(1)^{n-5} {\cal O} \star {\cal E} \star \cdots
\ee
Note that the residues involved in the left and right hand sides are completely distinct, and the unphysical poles on the left and right hand sides are different. Thus aside from cyclic invariance, this identity also enforces the absence of unphysical poles as well.

Before giving the proof in the general case, let us go back to the case of $n=7$ for
simplicity, and ask the general question: under what conditions is the object
\be
F = \sum_{j,k} f_{j,k} \{j\} \{k\} \, \, \, f_{j,k} = -f_{k,j}
\ee
cyclically invariant? Most naively, cyclic invariance requires $f_{j+1,k+1} = f_{j,k}$; however, this is too strong a requirement. The reason is that because of the residue theorems, we can add to $F$ the quantity zero in the form
\be
0 = \sum_{j,k} 2 \alpha_j \{j\} \{k\} = \sum_{j,k} (\alpha_j - \alpha_k) \{j\} \{k\}
\ee
Thus $f$ is ambiguous; because of the identities two sets of $f$ give the same object $F$:
\be
\label{ginv}
f_{j,k} \to f_{j,k} + \alpha_j - \alpha_k
\ee
This highlights an important point. As we have seen repeatedly, the amplitude is not associated with a unique contour of integration picking out a unique set of residues, but rather by an equivalence class of residues defined up to combinations that vanish due to residue theorems. We can say this more intuitively and geometrically by saying there is an equivalence class of contours that can smoothly be deformed into each other; in this sense the amplitude is associated with some homology class. Now the homology class for the tree amplitude should be cyclic, even though any individual representative may not be manifestly cyclic. It would be clearly be nice to be able to understand the cyclic symmetry properties of the homology class directly.

In order to do this, let's interpret the redundancy of equation (\ref{ginv}) as a sort of ``gauge invariance"; for the object $F$ to be cyclically invariant it is not necessary for $f$ to be cyclically invariant, only that it is cyclically invariant up to a gauge transformation. As usual, it is preferable to deal with ``gauge-invariant" quantities. In this case, if we define the difference operators $\Delta_{1,2}$ via
\be
\Delta_1 f_{j,k} = f_{j,k} - f_{j-1,k}, \, \, \Delta_2 f_{j,k} = f_{j,k} - f_{j,k-1}
\ee
then it is obvious that the ``gauge invariant" object made out of $f$ is
\be
(\Delta_1 \Delta_2) f
\ee
This ``curvature" is an invariant characterization of the homology class defining $F$. And in particular,
the quantity $F$ is cyclically invariant only if the curvature $\Delta_1 \Delta_2 f$ is cyclically invariant; the converse is
also easy to prove so this is an if and only if statement.

Now it is trivial to see that
\be
(\Delta_1 \Delta_2)  \, \{a\} \star \{b\} = [\{a\} - \{a-1\}] \star [\{b\} - \{b-1\}]
\ee
So we can look at the curvature associated with the BCFW form of the amplitude:
\begin{eqnarray}
(\Delta_1 \Delta_2) \left[\left(\{2\}+ \{4\} + \{6\}\right) \star \left(\{3\} + \{5\} + \{7\}\right)\right] \!\!\!\! & = &\nonumber \\
\left[\{2\} - \{1\} + \{4\} - \{3\} + \{6\}- \{5\}\right]\!\!\!\! & \star & \!\!\!\! \left[\{3\} - \{2\} + \{5\} - \{4\} + \{7\}- \{6\}\right] \nonumber \\
= -\left[\{1\} - \{2\} + \{3\} - \{4\} + \{5\} - \{6\} + \{7\}\right]\!\!\!\! & \star & \!\!\!\! \left[\{1\} - \{2\} + \{3\} - \{4\} + \{5\} - \{6\}+ \{7\}\right] \nonumber \end{eqnarray}
where in the last line we added (7) to the first bracket, since (7)$\star (a) = 0$, and also $(1)$ to the second bracket, since $(a) \star$(1) = 0. The last line is then manifestly cyclic invariant, since we have a single factor ``squared", and this factor goes into minus itself under a cyclic shift.

This pattern continues trivially for all $n$; the ``gauge invariant curvature" is
\be
(\Delta_1 \cdots \Delta_{n-5}) f_{j_1,\cdots,j_{n-5}}
\ee
The ``curvatures" associated with the BCFW and P(BCFW) forms of the amplitude can then easily be seen to be identical
\be\nonumber
(\Delta_1 \cdots \Delta_{n-5}) \left[{\cal E} \star {\cal O} \star \cdots \right] =
(-1)^{n-5}(\Delta_1 \cdots \Delta_{n-5}) \left[{\cal O} \star {\cal E} \star \cdots \right] =
- \left[({\cal E} - {\cal O}) \star ({\cal E} - {\cal O}) \star \cdots \right]
\ee
which establishes that $M_{{\rm BCFW}} = M_{{\rm P(BCFW)}}$.

\section{Residues in Multidimensional Complex Analysis and the Global Residue Theorem}

In moving beyond NMHV amplitudes, with $k>3$, we encounter a number of new features. Each minor becomes a higher than linear degree polynomial in several $\tau$ variables. We are interested in making sense of residues and residue theorems in this general situation. This is a basic subject in multi-variable complex analysis and algebraic geometry. The identification of residues is entirely analogous to the simple treatment we gave in discussing NMHV amplitudes. The procedure for deriving residue theorems by repeated application of the ordinary Cauchy theorem one variable at a time does not extend to the general case; however there is still a residue theorem of exactly the same type we have seen, known as the global residue theorem.
The subject of residues in multidimensional complex analysis has not made very many appearances in the physics literature (the only example we are aware of is \cite{Beasley:2003fx}). For this reason in this section we provide a short, self-contained review of the subject in the generality needed for the purposes of this paper. Most of the discussion will follow \cite{GH} and \cite{Tsikh}.

\subsection{Local Residues}

Consider a holomorphic mapping $f=(f_1,\ldots, f_n): \mathbb{C}^n\to \mathbb{C}^n$. For us each $f_i$ is in fact a polynomial of a fixed degree. Let us assume that $f$ has an isolated zero at $a\in \mathbb{C}^n$. In other words, if we choose a small enough neighborhood $U$ of $a$, then $f^{-1}(0)\cap U = \{a\}$. The local residue of a meromorphic form $\omega = h(z) dz/f_1\ldots f_n$ with $dz=dz_1\wedge\ldots \wedge dz_n$ at the point $a$ is defined by the integral
\be
\label{mapp}
{\rm res}(\omega)_a = \frac{1}{(2\pi i)^n} \int_{\Gamma_a} \frac{h dz}{f_1\ldots f_n}
\ee
over the contour
\be
\Gamma_a  = \{ z\in \mathbb{C}^n: |f_i(z)|=\epsilon_j, \; j = 1,\ldots ,n \},
\ee
where $\epsilon_j\in \mathbb{R}^+$ are small positive real numbers. Note that this is the straightforward generalization of the natural construction we used for $7$ particles\footnote{Note that $n$ here denotes the dimension of the space of free integration variables and not the number of particles!}.

The value of this integral is given by
\be
{\rm res}(\omega)_a = \frac{h(a)}{{\cal J}_f(a)}\quad {\rm with} \quad {\cal J}_f(z)=\left|\frac{\del (f_1,\ldots,f_n)}{\del (z_1,\ldots ,z_n)} \right|.
\ee

The most important piece of information we did not specify carefully in our 7-particle example was the definition of the orientation of the contour. Now that we have introduced the form notation we can define the orientation unambiguously to be such that $d({\rm arg} f_1)\wedge \ldots \wedge d({\rm arg}f_n) \geq 0$. More precisely, the cycle under consideration can be parameterized as $f_i(z) =\epsilon_i e^{i\theta_i}$, with $\theta_i = {\arg}f_i$. Therefore, the orientation is determined by the order of the $\theta_i$'s in the measure. In a sense one can think of (\ref{mapp}) as the local residue of $h(z)$ with respect to the mapping $f$. It is in this sense that the local residue is skew-symmetric with respect to permutations $p$ of the components of $f$:
\be
{\rm res}_{(f_1,\ldots f_n)} = ({\rm sgn}\;p)\; {\rm res}_{(f_{p(1)},\ldots, f_{p(n)})}.
\ee

As mentioned in the seven-particle case, the $(S^1)^n\backsimeq T^n$ contour does not enclose the point $a$ simply because it has real dimension $n$ and one needs a $2n-1$ dimensional subspace to do that (only for $n=1$ the two dimensions agree). It is interesting to note that there is an equivalent definition of the residue in terms of an integration over a $S^{2n-1}$ sphere that encloses the point $a$. The integral is of a $2n-1$ real form defined in terms of $f$ and $\bar f$. We will not make use of this and we refer the reader to \cite{GH} or \cite{Tsikh} for details. The reason we mentioned this fact is that it enters in the proof of the following results which we simply state. The proofs follow intuitively by using the $S^{2n-1}$ spheres and Stokes' theorem.

\subsubsection{Total Sum of Residues}

As we discovered in the seven-particle discussion there is a natural generalization of the residue theorem to the multidimensional case. Let us state the theorem in the generality we need.

Let $\omega = hdz/f_1\ldots f_n$ be defined by polynomials $h$ and $f_i$. Let $F_i = \{ z\in \mathbb{C}^n : f_i(z)=0\}$ be the hypersurface (i.e. $n-1$ dimensional subspace) associated with $f_i$ and $Z=F_1\cap F_2\cap\ldots \cap F_n$ be the set of zeroes of $f$. Here we assume that $Z$ is a discrete set of points. Then one defines the Global residue of $h$ with respect to the map $f$ as
\be
{\rm Res}_f(h) = \sum_{a\in Z} {\rm res}(\omega)_a.
\ee
Now, the Global Residue Theorem (GRT) states that if ${\rm deg}(h) < {\rm deg}(f_1)+\ldots + {\rm deg}(f_n) - n$ then
\be
{\rm Res}_f(h) = 0.
\ee

Just as we found in section 3.3.4, this theorem allows us to find relations among the local residues of a form $\omega$ with $f:\mathbb{C}^n\to \mathbb{C}^m$ and $m>n$. Once again we assume that $Z_\alpha = F_{\alpha_1}\cap \ldots \cap F_{\alpha_n}$ is discrete for any subset $\alpha=\{ \alpha_1,\ldots \alpha_n \} \subset \{1,\ldots, m\}$. We also assume that $Z_\alpha \cap Z_\beta = \emptyset$ for $\alpha\neq \beta$. For each $Z_\alpha$ we can define
\be
{\rm Res}_\alpha \;\omega  = \sum_{a\in Z_\alpha}{\rm res}(\omega )_a.
\ee
Note that the sum of the residues ${\rm Res}_\alpha (\omega )$ is skew-symmetric with respect to $\alpha_1,\ldots, \alpha_n$.

Now, for each set $\gamma = \{\gamma_1,\ldots ,\gamma_{n-1}\}\subset \{1,\ldots , m\}$ one has
\be
\label{genGRT}
\sum_{j\in \{1,\ldots, [\gamma],\ldots ,m\}} {\rm Res}_{(\gamma, j)}\omega
 = 0
\ee
where $[\gamma]$ in the range of the sum means excluding the set $\gamma$. This follows easily from the GRT.

A consequence of this is that for any partition of the set $\{ 1,\ldots, m\}$ into $n$ disjoint subsets $J_1,\ldots, J_n$ we also have the relation
\be
\sum_{j_1\in J_1,\ldots, j_n\in J_n}{\rm Res}_{(j_1,\ldots ,j_n)}\omega = 0
\ee

We can summarize the implications of the global residue theorem for
our particular application as follows. Let $D = (k-2)(n-k-2)$ be the
number of free variables. To define a residue, we solve the
equations putting $D$ minors $(j_1 \cdots j_1 + k -1) = \cdots =(j_D
\cdots j_D + k - 1)=0$. Since these are in general polynomial
equations of degree higher than one, there will be many solutions,
which we index by a variable $A$. The corresponding residue is
denoted as $\{j_1, \cdots, j_D\}_A$. The global residue theorem is
then the statement that for any partition of the set $\{ 1,
\cdots,n\}$ into $D$ disjoint subsets $J_1, \cdots,J_D$, \be \sum_A
\sum_{j_1\in J_1,\ldots, j_n\in J_D} {\{j_1,\ldots ,j_D\}}_A = 0 \ee
as long as there are no poles at infinity. Due to the antisymmetry
property of the residues, these all follow from taking linear
combinations of the set of equations \be \sum_A \sum_i \{j_1,
\cdots, j_{D-1}, i\} = 0 \ee

Note that in this discussion we have tacitly assumed that $n > D$ in
order to even be able to define the residues. It is however clear
that for large enough $n \sim 2k$ that $D \sim k^2 \gg n$. We will
discuss how to think about residues in this case in section 7.

\section{N$^2$MHV Amplitudes}

Up until now we have only studied NMHV amplitudes. There are two
features of NMHV amplitudes which make them special. The first is
that for $k=3$ each factor in the denominator is linear. The second
is that at one-loop it is easy to see from the quadruple cut formula
for box coefficients that all four-mass boxes vanish trivially. In
this section we will study N$^2$MHV amplitudes, which are the
simplest case where the minors are not linear but quadratic
polynomials in $2(n-6)$ variables. Most of our discussion will focus
on the 8 particle amplitude, which is also the simplest case in
which four-mass box coefficients are non-vanishing. At the end of
the section we will also take a peek at the 9 particle amplitude.

This section is divided into five parts. In the first we concentrate
on showing how a four-mass box coefficient appears as a residue in
our formula for the 8 particle N$^2$MHV amplitude. Four-mass boxes
are special because they are completely IR finite; furthermore
unlike all other boxes we have seen so far, where each of the two
leading singularities have been rational functions of the
kinematical variables, for the 4 mass boxes each leading singularity
has a square-root dependence on the kinematical variables. Even more
than our explicit identification of all NMHV box coefficients, this
gives smoking-gun evidence that our conjecture is correctly
computing 1-loop leading singularities. Since our main goal here is
simply to verify the presence of 4 mass boxes, in this part we
study the simplest ``split-helicity" configuration.

In the second part, we concentrate on the alternating helicity configuration and map out all residues coming from setting four of the eight denominators to zero. In other words, we solve ${8\choose 4 } = 70$ equations; we find that each set of equations has two solutions, giving a total of  $140$ residues. Many of the objects familiar from the tree and 1-loop alternating helicity amplitudes can be identified with these residues; in particular we identify the objects $T,U,V$ introduced in \cite{bcf} to express the BCFW form tree-level 8-point amplitude, as well as P$(T)$,P($U$) and P$(V)$ in the P(BCFW) form \cite{Hodges:2006tw}. Using the same ``curvature" notions we used in our analysis of general NMHV tree amplitudes, we can easily derive the $M_{{\rm BCFW}} = M_{{\rm P(BCFW)}}$ equality, which is a remarkable 40 term identity.

Moving on, however, we find that some of the one-loop leading singularities are no-where to be found in this list of 140 residue. These include the two other objects $W,X$ that naturally occur \cite{bcf} in the BCFW form of the tree-amplitude. Where could these objects be hiding? In the third part of this section we will see the answer is that they are associated with a new kind of residue. Mathematicians refer to these as  ``composite residues", and indeed their appearance will be ubiquitous for all $k>4$ and $n>10$; in this section we will give a very simple intuitive explanation of what these residues are and how to find them, and defer a general discussion of their properties to the next section where the residues of maps $f:\mathbb{C}^n\to \mathbb{C}^p$ with $p<n$ are considered. Armed with the more general notion of composite residues, we proceed to identify {\it all} the 1-loop box coefficients for the 8 particle N$^2$MHV amplitude.

However, just as we found for the 8 particle NMHV amplitude, in the fourth part of this section we show that even some of the ``usual" residues are not associated with any 1-loop leading singularities. It is again natural to think that these are naturally computing 2-loop leading singularities. There is one more piece of evidence in favor of this interpretation. Up to now we have seen that the residue theorems are associated with 1-loop IR equations. Now, there are clearly global residue theorems involving the residues associated with 4 mass boxes. However since the 4 mass scalar boxes are completely IR finite, they should not make any appearance in 1-loop IR equations. Interestingly, however, we find that precisely the residue theorems involving the 4 mass boxes {\it also} involve the ``mystery" residues with no 1-loop interpretation. The natural interpretation is that these are in fact {\it two loop} IR equations, which relate the IR divergent part of the two-loop amplitude to the 1-loop amplitude! An amusing application of this residue theorem is an expression for the 4-mass box coefficient which is explicitly a rational function of the kinematical invariants.

We end by taking a brief look at the 9 particle amplitude, and
by showing that the set of residues include information that is
unambiguously absent at 1-loop, further reinforcing our claim that
all leading singularities at all loop order are being correctly
computed.

\subsection{Four-Mass Box Coefficient}
\label{subsec:4m}

Let us start with a brief comment about the computation of the coefficient using the quadruple cut approach. Consider the four-mass box with external legs given by $K_1 = p_8+p_1$, $K_2 = p_2+p_3$, $K_3=p_4+p_5$ and $K_4=p_6+p_7$. Cutting all four propagators of the box give rise to the following equations for the loop momentum $\ell$,
\be
\ell^2=0, \quad (\ell-K_1)^2=0, \quad (\ell-K_1-K_2)^2=0, \quad (\ell+K_4)^2 = 0.
\ee
One way to solve these equation is to parameterize $\ell_{a\dot a} = (\lambda_3+\alpha\lambda_4)_a(\beta\tilde\lambda_3+\gamma\tilde\lambda_4)_{\dot a}$. Plugging this into the equations we find two solutions obtained from solving a quadratic equation. The discriminant of the quadratic equation is given by
\be
\Delta = 1-2(\rho_1+\rho_2)+(\rho_1-\rho_2)^2\; {\rm with}\; \rho_1 = \frac{K_1^2K_3^2}{K_{12}^2K_{23}^2}\;\; {\rm and} \;\; \rho_2 = \frac{K_2^2K_4^2}{K_{12}^2K_{23}^2}.
\ee
This means that $\alpha$, $\beta$ and $\gamma$ are of the form $a\pm b\sqrt{\Delta}$ for some $a$ and $b$ rational functions of the kinematical invariants.

For simplicity consider the one-loop amplitude
$M(1^-,2^-,3^-,4^-,5^+,6^+,7^+,8^+)$. In this case the box we are
studying is the only non-zero one. A non-trivial test of our
proposal is that out of all residues of the integral with
denominator $(2345)(3456)\ldots (8123)$ only one combination of four
factors should give rise to residues located at points with a square
root dependence on the kinematical invariants. Indeed we have
checked that $(2345)=(4567)=(6781)=(8123)=0$ is the only combination
that gives a quadratic equation and therefore two residues located
at points that depend on a square root. The argument of the square
root turns out to be precisely equal to $\Delta$!

The coefficient we are interested in is computed in the quadruple cut approach as the product of four four-particle tree-amplitudes (normalized with the Jacobian of the 4-mass box). As a sample let us write two of the four amplitudes
\be
\frac{\langle 23\rangle^3}{\langle 3~\ell-p_{23}\rangle\langle \ell-p_{23}~\ell\rangle\langle \ell~2\rangle},\qquad \frac{\langle \ell-p_{23}~4\rangle^3}{\langle 45\rangle\langle 5~\ell-p_{2345}\rangle\langle \ell-p_{2345}~\ell-p_{23}\rangle}.
\ee
The product of all four amplitudes can be simplified to give $B(\ell)$ as
\be\nonumber
\frac{\langle 23\rangle^3[67]^3[4|\ell-p_{23}|4\rangle^3\langle 1~\ell\rangle^3}{\langle 45\rangle\langle 81\rangle\langle \ell~2\rangle\langle 3|\ell-p_{23}|4][4|p_{23}|\ell\rangle\langle\ell|p_{18}|7]\langle 5|\ell-p_{2345}|6]\langle 8|(\ell-p_{18})(\ell-p_{2345})(\ell-p_{23})|4]}
\ee
The coefficient of the box is then $B(\ell^{(+)}) + B(\ell^{(-)})$ where $\ell^{\pm}$ are the two solutions to the cut equations.

We have numerically verified that the residues at the two points in $\mathbb{C}^4$ which are solutions of $(2345)=(4567)=(6781)=(8123)=0$ match $B(\ell^{(+)})$ and $B(\ell^{(-)})$ respectively! Extending our now familiar notation for residues, we denote these two solutions by $\{2468\}_1$ and $\{2468\}_2$, and so
\be
B(\ell^{(+)}) = \{2468\}_1, \, \, B(\ell^{(-)}) = \{2468\}_2
\ee

A final comment is in order here. The global residue theorem implies
that $B(\ell^{(+)}) + B(\ell^{(-)})$ is equal to the sum of other
residues. As an illustration let us choose $f_1=(2345)$,
$f_2=(4567)$, $f_3=(6781)$ and $f_4=(8123)(3456)(5678)(7812)$.
Recall that $(1234)=1$ in the natural gauge choice for this helicity
configuration. The residue theorem then gives
\be
B(\ell^{(+)}) + B(\ell^{(-)}) = - \sum_{A} \left( \{2463\}_A +
\{2465\}_A + \{2467\}_A \right)
\ee
As we have mentioned, all the residues
other than the ones determining $B(\ell^{\pm})$ are rational
functions of the kinematical invariants; they are determined by
solving linear equations and do not contain square roots. We
therefore obtain a remarkable form of the 4 mass box coefficients
directly as a rational function of the kinematical invariants.

\subsection{Tree Amplitude}

We now turn our attention to discussing the alternating helicity amplitude with assignment $1^+ 2^- 3^+ 4^- 5^+ 6^- 7^+ 8^-$, beginning with a discussion of the tree amplitude. In \cite{bcf}, the 20 terms of BCFW form of this amplitude were determined in terms of five basic objects $T,U,V,W,X$, together with their cyclic images. Explicitly, these 5 objects are given by
\be
T\! &=&\! \frac{\text{[13]}^4 \text{$\langle $46$\rangle $}^4
   \text{$\langle $68$\rangle $}^4}{\text{[12]} \text{[23]}
   \text{$\langle $45$\rangle $} \text{$\langle $56$\rangle
   $}\text{$\langle
   $67$\rangle $} \text{$\langle $78$\rangle $} \text{$\langle $6$|$4+5$|$3]} \text{$\langle
   $6$|$7+8$|$1]} \text{$\langle
   $6$|$(7+8)$$(1+2+3)$|$4$\rangle $}
   \text{$\langle $8$|$(1+2+3)$$(4+5)$|$6$\rangle $}} \nn\\
U\! &=&\! \frac{\text{[13]}^4 \text{[57]}^4 \text{$\langle
   $48$\rangle $}^4}{\text{[12]} \text{[23]} \text{[56]}
   \text{[67]} \text{$\langle $4$|$2+3$|$1]} \text{$\langle
   $4$|$5+6$|$7]} \text{$\langle $8$|$1+2$|$3]}
   \text{$\langle $8$|$6+7$|$5]} t_1^{[3]}t_5^{[3]} } \nn\\
V\! &=& \! \frac{\text{[13]}^4 \text{$\langle $46$\rangle $}^4
   \text{$\langle $8$|$1+2+3$|$7]}^4}{\text{[12]}\text{[23]}\text{$\langle $45$\rangle $}
   \text{$\langle $56$\rangle $}
   \text{[1$|$(2+3)$$(4+5+6)$|$7]}  \text{$\langle
   $4$|$5+6$|$7]}  \text{$\langle
   $8$|$1+2$|$3]} \text{$\langle
   $8$|$(1+2+3)$$(4+5)$|$6$\rangle $} t_1^{[3]}t_4^{[3]}t_8^{[4]}
   } \nn \\
W\! &=& \! \frac{\text{[35]}^4 \text{$\langle $6$|$8+1+2$|$3]}^4
   \text{$\langle $82$\rangle $}^4}{\text{[34]} \text{[45]}\text{$\langle $12$\rangle
   $}\text{$\langle $67$\rangle
   $} \text{$\langle
   $81$\rangle $} \text{[5$|K_6^{[2]}K_8^{[3]}|$3]}
  \text{$\langle $2$|K_8^{[3]}$(6+7)$$(4+5)$|$3]}
   \text{$\langle $6$|$4+5$|$3]}  \text{$\langle $7$|$$K_8^{[3]}$$|$3]}  \text{$\langle $8$|$1+2$|$3]}
   t_8^{[3]} }
   \nn \\
X\! &=&\! \frac{\text{[35]}^4 \text{$\langle $82$\rangle
   $}^4}{\text{[34]} \text{[45]} \text{[56]} \text{$\langle
   $78$\rangle $} \text{$\langle $81$\rangle $}\text{$\langle
   $12$\rangle $} \text{$\langle $2$|$3+4+5$|$6]}
   \text{$\langle $7$|$8+1+2$|$3]}
   t_7^{[4]}} \nn\\
\ee
Here we use the notation of \cite{bcf} where $K_i^{[m]}=p_i + \cdots + p_{i + m -1}$ and $t_i^{[m]} = (p_i + \cdots + p_{i + m -1})^2$.
In \cite{bcf} it was also noted that, using a peculiar-looking 6 term identity, the tree amplitude could be written purely in terms of $T,U,V$, together with their cyclic images. As we will explain in \cite{notes}, this identity is in fact nothing other than our 6 term identity for the 6 particle amplitude, dressed into a 6 term identity for the 8 particles by the application of two ``inverse soft factors". At any rate, the nicest form of the alternating helicity amplitude is then given by
\be
M^{+-+-+-+-}_{{\rm BCFW}} = (1 + g + \cdots + g^7)\left[T + V\right] + (1 + g + \cdots + g^3) U
\ee
Here $g$ is the operation that sends $i \to i+1$ and conjugates $\langle \rangle \leftrightarrow []$.
We have only 4 images of $U$ since it has a flip symmetry.
The P(BCFW) form of the amplitude is of exactly the same form, with $T,U,V \to$P($T$),P($U$),P($V$), given by
\be
P(g^4T)\!\!\! &=&\!\!\! \frac{(\text{[75]} \text{$\langle
   $2$|$7+8$|$6]}+\text{[76]} \text{$\langle
   $2$|$1+3$|$5]})^4}{\text{[45]} \text{[56]} \text{[67]}
   \text{[78]} \text{$\langle $12$\rangle $} \text{$\langle
   $23$\rangle $} \text{$\langle $3$|$4+5$|$6]}
   \text{$\langle $1$|$8+7$|$6]} \text{$[
   $4$|$(3+2+1)$$(8+7)$|$6$] $} \text{$[
   $6$|$(5+4)$$(3+2+1)$|$8$] $}} \nn \\
P(g^4U)\!\!\! &=&\!\!\! \frac{\text{$\langle $2$|$1+3$|$5+7$|$6$\rangle
   $}^4}{\text{$\langle $12$\rangle $} \text{$\langle
   $23$\rangle $} \text{$\langle $56$\rangle
   $} \text{$\langle $67$\rangle $}\text{$\langle $1$|$2+3$|$4]}
   \text{$\langle $7$|$6+5$|$4]}  \text{$\langle
   $3$|$2+1$|$8]} \text{$\langle $5$|$6+7$|$8]}
   t_1^{[3]} t_5^{[3]} } \nn \\
P(g^4V)\!\!\! &=&\!\!\! \frac{\text{$\langle
   $2$|$(1+3)$$(8+1+2+3)$$(4+6)$|$5]}^4}{\text{[45]} \text{[56]} \text{$\langle
   $12$\rangle $} \text{$\langle $23$\rangle $}\langle\text{7$|$(6+5+4)$$(3+2)$|$1}\rangle
   \text{$\langle $7$|$6+5$|$4]} \text{$\langle
   $3$|$2+1$|$8]} \text{$[
   $6$|$(5+4)$$(3+2+1)$|$8$] $} t_1^{[3]}t_4^{[3]}t_8^{[4]}
} \nn
\ee
Clearly, the expected identity
\be
M^{+-+-+-+-}_{{\rm BCFW}} = M^{+-+-+-+-}_{{\rm P(BCFW)}}
\ee
is an amazingly non-trivial 40 term identity!

Let us try and identify these with residues. As we mentioned before, the residues are associated with putting 4 of the minors to 0; this amounts to solving four coupled quadratic equations. Somewhat surprisingly, it turns out there are always {\it two} solutions to these equations. We will understand this fact more deeply when we discuss the relationship of our conjecture to algebraic geometry and the Schubert calculus in section \ref{sec:SchubCalc}. This means that we should label the residues as $\{i_1,i_2,i_3,i_4\}_{1,2}$, where the subscript denotes the two solutions associated with putting the four minors beginning with $i_1,i_2,i_3,i_4$ to zero.

Now, we find a very pleasant surprise in identifying $T,U,V$ and P($T$),P($U$),P($V$) with residues. A term and its Parity conjugate are the two different solutions associated with a single residue! Specifically, we find
\begin{eqnarray}
T = \{3,4,5,6\}_1 &,& {\rm P}(T) = - \{3,4,5,6\}_2 \nonumber \\
U = \{4,5,8,1\}_1 &,& {\rm P}(U) = - \{4,5,8,1\}_2 \nonumber \\
V = \{4,5,8,3\}_1 &,& {\rm P}(V) = - \{4,5,8,3\}_2
\end{eqnarray}
Under the conjugation operation $g$, there is a minus sign in the
identification, so that e.g. $gT = -\{2,3,4,5\}$. The 40 term
identity then becomes
\be\nonumber \sum_{A=1,2} \left[(1 - g +
\cdots - g^7) \left({\{3,4,5,6\}}_A + {\{4,5,8,3\}}_A\right) + (1 -
g + \cdots - g^3) {\{4,5,8,1\}}_A\right] = 0
\ee
This looks exactly
like a residue theorem! In fact, the global residue theorems for our
case are linear combinations of the statements \be \sum_{i; A=1,2}
{\{j_1,j_2,j_3,i\}}_A = 0 \ee Thus in order to check whether a sum
of the form \be F = \sum_A \sum_{i,j,k,l} f_{i,j,k,l}
{\{i,j,k,l\}}_A \ee vanishes, it suffices to see whether the
``curvature"  vanishes \be (\Delta_1 \Delta_2 \Delta_3 \Delta_4)
f_{i,j,k,l} = 0 \ee A completely straightforward computation shows
that the ``curvature" of the summand in the 40 term identity indeed
vanishes, establishing the validity of $M_{{\rm BCFW}} =
M_{{\rm P(BCFW)}}$ in this case.

\subsection{A New Class of Residues}
We have identified $T,U,V$ and their parity conjugates, which are
everything needed for the tree amplitude. However, $W,X$ do appear
directly in the BCFW formula, and are some of the 1-loop leading
singularities. Indeed $W$, its flipped cousin $rW$ (where $r: i \to
9 - i$) and their parity conjugates are given by \be W =
\{7,8,4,6\}_1, P(W) = \{7,8,4,6\}_2 \, ; \, rW = \{3,4,6,2\}_1, \,
P(rW) = \{3,4,6,2\}_2 \ee while the parity conjugates are given by
the second roots. However, examining all 140 residues, we could not
find any objects that matched $X$! Where could $X$ possibly be
hiding?

The answer is quite remarkable, and illustrates the rich structure of our conjecture. The definition we have given for residues is valid for completely generic rational functions of many variables. However we have a very special set of functions, and an interesting phenomenon can occur that allows us to define residues in a more interesting way. We can illustrate the main idea by
considering a  function of 3 complex variables $x,y,z$ of the form
\be
\frac{1}{x(x+ yz)}
\ee
There are only two polynomial factors in the denominator while the definition of residue we have been using would require three polynomials to determine the three variables.  However, something special happens on the locus where the first polynomial $x$ vanishes: the second polynomial $(x + y z) \to yz$ factorizes, so it seems sensible to define the residue of this function at $x=y=z=0$ to be 1.

We find exactly the same structure in our problem! We find that in some cases, in putting only two of the minors $(i, \, i+1, \, i+2, i+3) = 0$ and $(j, \, j+1, \, j+2, \, j+3)=0$, a third minor $(k, \, k+1, \, k+2, \, k+3)$ factorizes into a product of two linear factors in the four $\tau$'s.  Therefore we can define a residue using only these three minors. These residues are called ``composite residues" by mathematicians and will be discussed more systematically in the next section. We find that there are still two composite residues. We use a notation $\{i, j, k^2\}_{1,2}$ to denote these residues.

In order to see the composite residues explicitly in our example, it is easiest to use the gauge fixing where
\be
C = \left(\begin{array}{cccccccc} 1 & 0  & 0 & 0 & c_{15} & c_{16} & c_{17} & c_{18} \\ 0 & 1 & 0 & 0 & c_{25} & c_{26} & c_{27} & c_{28} \\ 0 & 0 &  1 & 0 & c_{35} & c_{36}  & c_{37} & c_{38} \\ 0 & 0 & 0 & 1 & c_{45} & c_{46} & c_{47} & c_{48} \end{array} \right)
\ee
Consider the first four determinants of the minors that enter in our formula,
\be
\label{nino}
\begin{array}{l}
(1234) = 1, \\
(2345) = c_{15}, \\
(3456) = \left| \begin{array}{cc} c_{15} & c_{16} \\ c_{25} & c_{26} \end{array} \right|, \\ \\
(4567) = \left| \begin{array}{ccc} c_{15} & c_{16} & c_{17} \\ c_{25} & c_{26} & c_{27} \\ c_{35} & c_{36} & c_{37} \end{array} \right|.
\end{array}
\ee
We will now show that, it is possible to define a residue using only three minors $(2345)$, $(3456)$, $(4567)$. Recall that the $c_{Ii}$ are linear in the $\tau_A$. Now, note that on the locus of $(2345) = 0$, $(3456)$ factorizes into the product of two terms:
\be
(2345) = 0 \implies c_{15} = 0 \implies (3456) = -c_{16} c_{25}
\ee
So just using these two minors allows us to solve for 3 $\tau$'s, by setting $c_{15} = c_{16} = c_{25} = 0$. Then $(4567)$ is a quadratic
polynomial in the remaining $\tau$ variable, which we can set to zero finding two solutions. The residue is defined in the usual way, with the relevant Jacobian being the one determined by the four objects we are setting to zero $\partial(c_{15},c_{16},c_{25},(4567))/\partial(\tau_1,\cdots,\tau_4)$.
The two composite residues in this example are denoted by $\{23^24\}_{1,2}$.

Having identified all the sorts of residues we should be looking for, we can now identify all the 1-loop leading singularities! For instance the
object $X$ is
\be
X = \{6, 7^2, 8\}_1
\ee
We can now give the map from residues to all 1-loop leading singularities. We have implicitly given all the coefficients for the 2 mass hard and 1 mass boxes in giving the BCFW and P(BCFW) form of the tree amplitude. The four mass boxes are given as
\be
\includegraphics[scale=0.7]{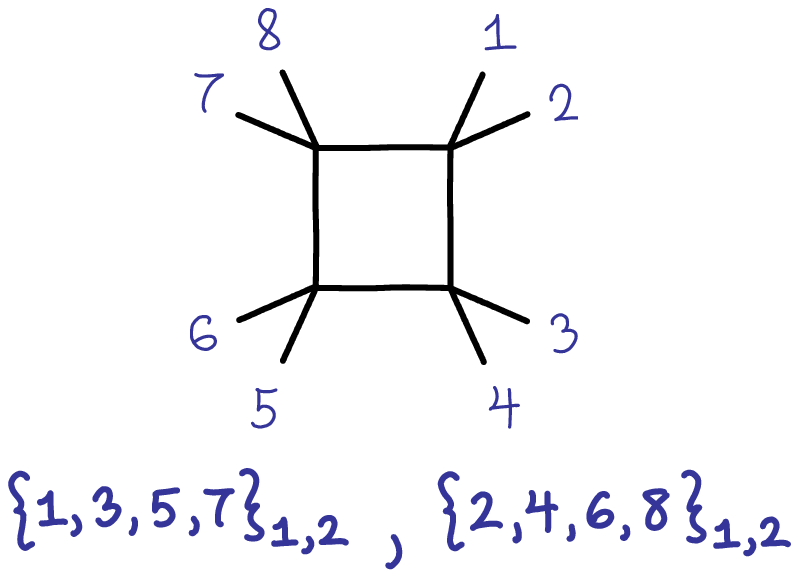} \nonumber
\ee
The 3 mass and 2 mass easy boxes are given as
\be
\includegraphics[scale=0.7]{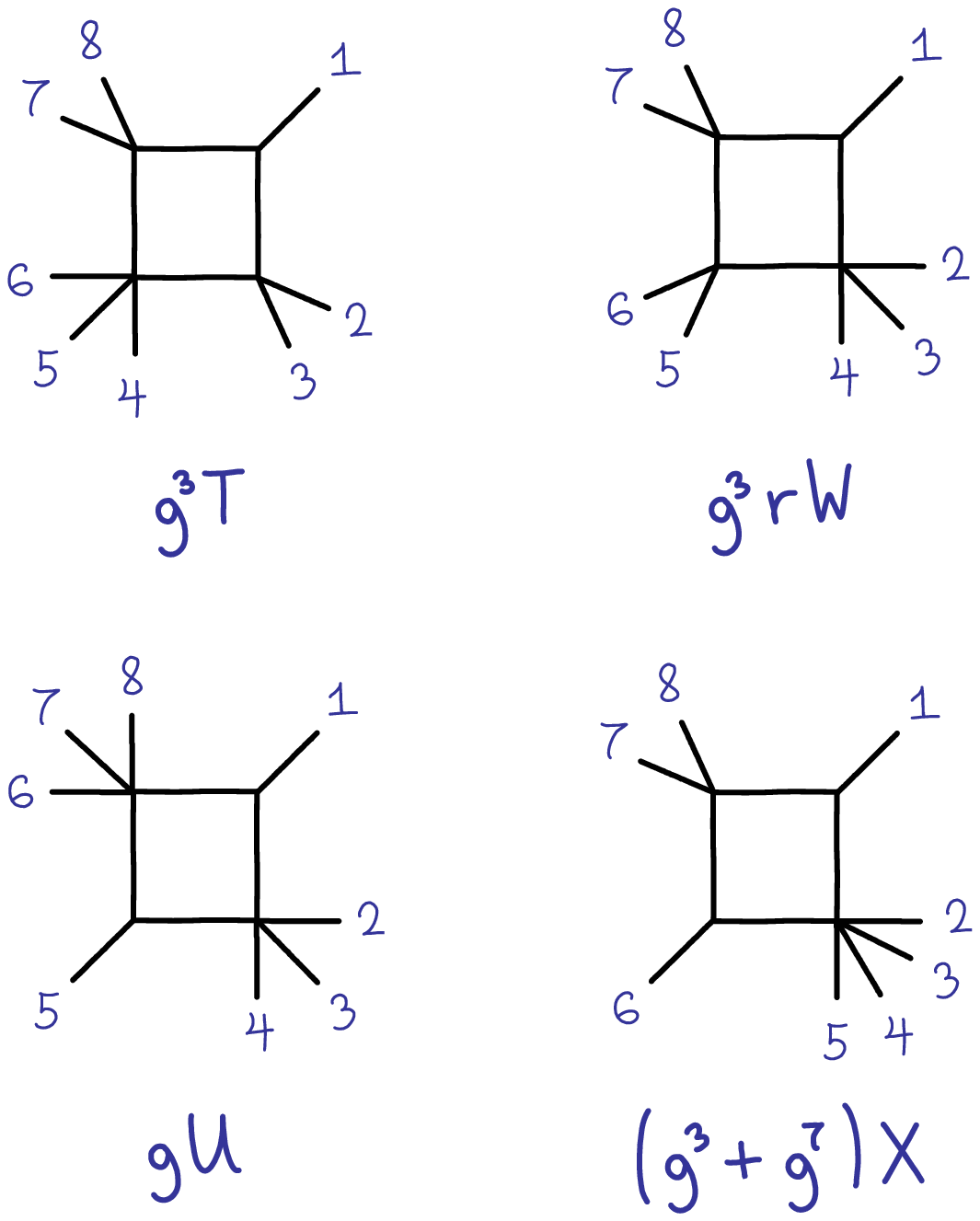} \nonumber
\ee
where here the operation $r$ is a reflection, sending $i \to 9 - i$.

\subsection{Two Loop Leading Singularities}
Just as we saw in the 8 particle NMHV amplitude, even amongst the
140 ``usual" residues, there are objects we can not identify with
any 1-loop leading singularities, of the form $\{1,2,4,6\}_A$ and
$\{1,2,5,7\}_A$ together with their cyclic images. It is again
natural to associate these with two-loop leading singularities.
Following the same strategy as for the NMHV case, we look for these
residues by applying ``inverse soft factors" on the 7 particle
1-loop leading singularities, and then associate these with a
certain 2-loop leading singularity. Doing this we find the mapping

\be
\includegraphics[scale=0.7]{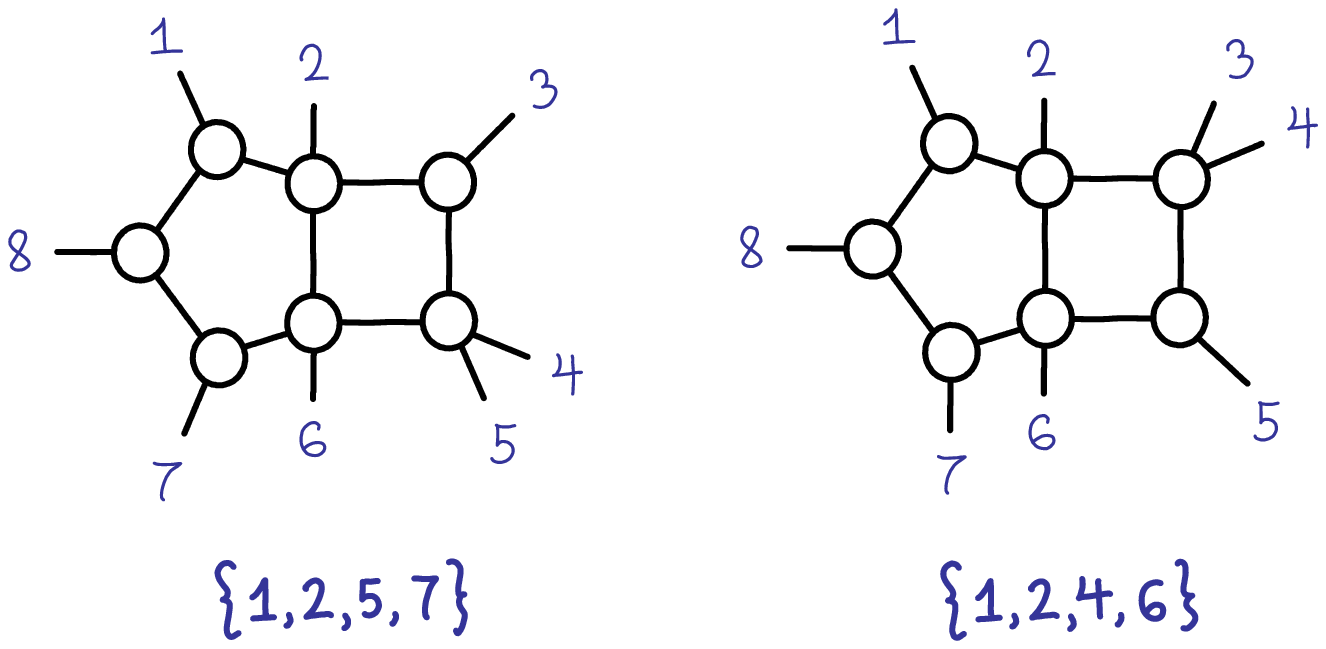} \nonumber
\ee

There is one more piece of evidence that these residues are
associated with two-loop physics. Let us return to the global
residue theorem involving the 4 mass box coefficient
$\{1,3,5,7\}_A$, which we can write as $\sum_{i;A} \{1,k,5,7\} = 0$.
Let us group the terms as \be \sum_A \left(\{1,3,5,7\}_A +
\{1,6,5,7\}_A + \{1,8,5,7\}_A \right)= \sum_A \left( - \{1,2,5,7\}_A
+ \{4,5,7,1\}_A \right) \ee The second two terms on the LHS are just
cyclic images of $W$, so the left hand side groups together objects
that appear at 1-loop, while the RHS has the residues with no known
1-loop interpretation. What could the physical interpretation of
this residue theorem be? So far, all residue theorems have been
related to IR equations, but as we have stressed the 4 mass box is
completely IR finite and would not participate in any 1-loop IR
equation. However, recognizing the terms on the RHS as giving
two-loop singularities allows a very natural interpretation: the
residue theorem should be thought of as a 2-loop IR equation, which
relates the 2-loop IR divergences to the 1-loop amplitude. IR finite
4 mass boxes can then occur as a part of this 1 loop amplitude, just
as the 1-loop IR equations relate the IR divergent part of the
1-loop amplitude to the IR finite tree amplitude. It would be nice
to check this picture in detail. The 2-loop IR equations have been
understood by Catani \cite{catani}; proceeding further would require
us to identify all the 2-loop leading singularities and reconstruct
the amplitude from them, which is beyond the scope of the present
work.

Finally, as we already mentioned, this global residue theorem allows
us to write the 4-mass box coefficient in a new way, manifestly
as a sum of rational terms. We will not give the explicit expression
here, and simply mention that all the terms appearing in this
equation are explicitly known: the 1-loop residues are related to
$W$ and P$(W)$, and the 2-loop residues are obtained by applying an
inverse soft factor, defined in section 3.4.2, to the 7-particle 3-mass box coefficients.

\subsection{9 Particle N$^2$MHV amplitude}
We end with very brief comments about the 9 particle amplitude. We
will not present a detailed mapping between residues and leading
singularities; instead our goal is to give another sharp proof that
the residues contain information beyond one loop.

The strategy for doing this is simple. The best way of describing it
is to let all the initial data encoded in the spinors $\lambda$ and
$\tilde\lambda$ of each particle to be rational numbers, i.e, to
think about the $n$-vectors defining the $\boldsymbol{\lambda}$ and
$\boldsymbol{\tilde\lambda}$-planes as vectors in $\mathbb{Q}^n$.
Suppose that all the residues are secretly associated only with
one-loop leading singularities. Now, we know that the one-loop
leading singularity is a product of tree amplitudes, which are
rational functions of the kinematical invariants. However the values
of the cut momenta $\ell^{\pm}$ can be irrational; indeed as we
remarked in our discussion of the 4 mass box coefficients for the 8
particle N$^2$MHV amplitude, $\ell^{\pm} = A \pm B \sqrt{\Delta}$
where the discriminant $\Delta$ is reproduced below:

\be \label{asum} \Delta = 1-2(\rho_1+\rho_2)+(\rho_1-\rho_2)^2\;
{\rm with}\; \rho_1 = \frac{K_1^2K_3^2}{K_{12}^2K_{23}^2}\;\; {\rm
and} \;\; \rho_2 = \frac{K_2^2K_4^2}{K_{12}^2K_{23}^2}. \ee
Note that $\rho_1$ and $\rho_2$ belong to $\mathbb{Q}$ by
assumption.

This says that for the residues associated with 1-loop leading
singularities, the number field $E$ of solutions for the residues
can only be a quadratic extension of $\mathbb{Q}$, moreover, the
only square roots that appear can only be of the form
$\sqrt{\Delta}$ for some box. Furthermore, with 9 particles, there
are only 9 four-mass boxes and these are cyclically related, so the
list of possible square roots that can appear is a small one. We
have checked all the usual (non-composite) residues, and found that
only $\{1,3,5,7,8,9\}_A$ and its cyclic partners contain a square
root, and that the argument of the square root is precisely the
$\Delta$ associated with the box. However moving to the composite
residues, we find a wide array of ``new" square roots, which can
definitely not be expressed in the number field only containing
rationals and the square roots of the one loop $\Delta$'s. This
proves conclusively that the residues contains a great deal of
information beyond one loop.

\section{Composite Residues and Large $k$ Amplitudes}

As we already mentioned at the end of section 5, the number of free
variables of integration in our formula, $(k-2)(n-k-2)$, is in
general larger than the number of determinant factors in the
denominator which is $n$. This means that for large $n$ and $k$, the
``ordinary" definition of residues we have been using, which
requires at least as many polynomial factors as complex variables,
can not be used. Instead, the new sort of ``composite" residues we
already encountered in the previous section will be needed. That is,
we will find that on the zero locus of a given minor, the subsequent
minor factors into pieces allowing us to solve for more of the
$\tau$ variables. In the first part of this section we show that
doing this allows us to solve for all the $\tau$ variables and
determine residues. We leave a more complete analysis, as well as a
physical interpretation of these residues, to future work. In the
second part of this section, we briefly discuss the way in which
these residues are discussed in the mathematical literature.

\subsection{The Large $n=2k$ Limit}

Let us concentrate on the most extreme case, i.e., consider $n=2k$. In this case we have $(k-2)^2$ integration variables. Consider once again the gauge fixing where
\be
C = \left(\begin{array}{ccccccccc} 1 & 0  & \ldots & 0 & c_{1,k+1} & c_{1,k+2} & \ldots & c_{1,2k-1} & c_{1,2k} \\ 0 & 1 & \ldots & 0 & c_{2,k+1} & c_{2,k+2} & \ldots & c_{2,2k-1} & c_{2,2k} \\ \vdots & \vdots &  \ddots & \vdots & \vdots & \vdots & \ldots & c_{k-1,2k-1} & c_{k-1,2k} \\ 0 & 0 & \ldots & 1 & c_{k,k+1} & c_{k,k+2} & \ldots & c_{k,2k-1} & c_{k,2k} \end{array} \right)
\ee

Following the same analysis as in the eight particle case, let us list the first four of the determinant factors,
\be
\begin{array}{l}
(12\ldots \, k) = 1, \\
(23\ldots \, k+1) = c_{1,k+1}, \\
(34\ldots \,  k+2) = \left| \begin{array}{cc} c_{1,k+1} & c_{1,k+2} \\ c_{2,k+1} & c_{2,k+2} \end{array} \right|, \\ \\
(45\ldots \, k+3) = \left| \begin{array}{ccc} c_{1,k+1} & c_{1,k+2} & c_{1,k+3} \\ c_{2,k+1} & c_{2,k+2} & c_{2,k+3} \\ c_{3,k+1} & c_{3,k+2} & c_{3,k+3} \end{array} \right|.
\end{array}
\ee
Let us denote the free variables by $\tau=(\tau_1,\ldots, \tau_{(k-2)^2})\in \mathbb{C}^{(k-2)^2}$. Each factor is at most a polynomial of degree $(k-2)^2-2$.
Now, let's see what happens as we begin to set minor factors to 0. As we saw already,
\be
(234 \ldots\, (k+1)) = 0 \implies (345\ldots \, (k+2)) =  - c_{1,k+2} c_{2,k+1}
\ee
Note that $c_{1,k+2} c_{2,k+1}$ is the product of the elements on the ``secondary" or ``skew" diagonal of the minor $(345\ldots \, (k+2))$. Now, putting all three of these factors to zero, we find
\be
(456 \ldots \, (k+3)) = c_{1,k+3} c_{2,k+2} c_{3,k+1}
\ee
Now it is easy to see the pattern. Once we use all factors in a given step, the determinant of the next minor reduces to the product of the entries in the secondary diagonal of the minor.

Note that the matrices increase in size until one of them becomes a $k\times k$ matrix and then start decreasing. At this point one has to be careful to remember that even though we seem to get $k$ linear factors coming from the $k$ $c_{Ii}$'s in the secondary diagonal, we know from the general analysis in section 4.3 that the order cannot be larger than $k-2$. This means that $2$ of the factors become trivial, i.e., $\tau$-independent.

Now we can count the number of factors we can set to zero to define residues. Using the first $k-2$ determinants (excluding $(123\ldots\, k)$) we find $1+2+\ldots +(k-2) = (k-1)(k-2)/2$ factors. Using the last $k-2$ determinants we also find the same number. Therefore we have $(k-1)(k-2)$ factors which is larger that then number of free variables. This gives many different residues.
Once again, the precise counting of these residues as well as their precise physical interpretation falls outside the scope of this paper and it is left for future investigations.

\subsection{Composite Residues and Residual Forms}

Mathematicians naturally encounter composite residues in studying mappings $f:\mathbb{C}^n\to \mathbb{C}^p$ consisting of $p$ irreducible polynomials in $n$ variables, with $n>p$.  It is clear the previous discussion of residues does apply straightforwardly in this case. In \cite{Tsikh} the concept of residual currents and in \cite{Yuzhakov} residue forms are introduced as a way to study these cases. We find the residue forms to be naturally adapted for our problem. Let us give the definition of a residue form. Let $S=\{z\in\mathbb{C}^n:\; s(z)=0\}$ and $\omega(z)=(h(z)/s(z))dz$ with $h(z)$ and $s(z)$ holomorphic and $dz=dz_1\wedge\ldots \wedge dz_n$, then the residue form is an $(n-1)$-form\footnote{More precisely, we have a $(n-1,0)$-form.} given by
\be
{\rm res}_j[\omega] = (-1)^{j-1}\left.\left(\frac{h(z)}{s'_{z_j}}\right)\right|_{S}dz_{[j]}
\ee
at the points where $s'_{z_j}\neq 0$. Let us explain the different objects entering this formula. The formula is defined for any $j$ such that the condition is satisfied. $dz_{[j]}$ is a $n-1$ form obtained by wedging all $dz_i$'s except for $dz_{j}$. The function $h(z)/s'_{z_j}$ is evaluated on $S(z)=0$ and therefore it depends on only $n-1$ variables. Finally, $s'_{z_j}$ is the partial derivative with respect to $z_j$.

Now one can define the $(n-m)$-form
\be
{\rm res}^m[\omega] = {\rm res}_m \circ \;\cdots\; \circ {\rm res}_1[\omega]
\ee
called the {\it composite residue form}.

In simple cases, like when one has a map $f:\mathbb{C}^n\to \mathbb{C}^n$ with isolated zeroes, the composite residue form ${\rm res}^n[\omega]$ coincides with ${\rm res}(\omega)$ in (\ref{mapp}) and different orders of computing the composite residue form give rise to the same answer up a sign, but this is not true in general. Using this notation, we can return to our simple example of the function $\frac{1}{z_1 (z_1 + z_2 z_3)}$, putting  $s(z) = z_1
(z_1+z_2z_3)$ and $h(z)=1$. Then, ${\rm res}_{z_1}(\omega) = 1/z_2z_3$. The following composite residue form gives
\be
{\rm res}^3[\omega] = {\rm res}_{z_3} \circ {\rm res}_{z_2} \circ {\rm res}_{z_1} [\omega] = 1.
\ee
Mathematicians have also identified residue theorems involving composite residue forms. Our real interest in making a connection with the mathematical literature here is to exploit these results, since as we have seen residue theorems have some deep physical content in our picture. We will take up this subject at greater length in future work.

We find it intriguing that the notion of composite residue has made
{\it two} appearances in our story so far, first in defining certain
kinds of leading singularities as residues in the loop integrals,
and now, in identifying certain leading singularities as residues in
the integral over the $\tau_\gamma$ variables. Needless to say it
would be fascinating if this were more than a co-incidental fact!

\section{Amplitudes and the Schubert Calculus}
\label{sec:SchubCalc}

We have repeatedly mentioned that the pieces entering in our formula all have natural geometric interpretations in terms of Grassmannians. However, we have not made use of this fact in any of our computations. It is clear that a deeper connection with the algebraic geometry of Grassmannians could shed new light into our formula. In this section we give a simple example of the use of the homology and its ring structure in order to compute the number of solutions to a given set of equations coming from distinct minor factors.

The first step is to define the basis of homology for a Grassmannian $G(k,n)$. The ring structure is obtained by using as multiplication the intersection of cycles. The cycles in the basis are called Schubert cycles and the ring structure gives rise to the Schubert calculus.

In this discussion we closely follow \cite{GH}. Let us introduce a set of subspaces of $\mathbb{C}^n$ given by $V_i=\{e_1,\ldots , e_i\} \subset \mathbb{C}^n$, where $e_i$ is the unit vector in the $i^{\rm th}$ direction. A basis for the homology of the Grassmannian is determined by looking at planes $C\in G(k,n)$ whose intersection with our ``reference" subspaces $V_i$ is of a given specified dimension. The set of subspaces $V_i$ is called a ``flag" $V=(V_1\subset V_2 \subset \cdots \subset V_{n-1}\subset V_n = \mathbb{C}^n)$. More precisely, we are interested in cycles of the form
\be
\sigma_{a_1,\ldots, a_k} = \{ C \in G(k,n): \; {\rm dim}(C \cap V_{n-k+i-a_i}) \geq i\}
\ee
where $\{(a_1,\ldots ,a_k)\}$ ranges over all nonincreasing sequences of integers between $0$ and $n-k$. These are called Schubert cycles.

The intersection of Schubert cycles is well understood and has a fascinating connection to Young tableaux as one could have imagined from the restriction on the set of $a_i$'s. From the restriction it is clear that Schubert cycles are in one-to-one correspondence with Young tableaux. The whole intersection theory is encoded in the tensor product formula of irreducible representations of unitary groups. In other words, consider cycles $\sigma_\mu$ and $\sigma_\lambda$ where $\mu$ and $\lambda$ are two Young tableaux. We know that
\be
\mu \otimes \lambda = \oplus_\rho c_{\mu,\lambda}^\rho \; \rho
\ee
where the sum is over all Young tableaux and $c_{\mu,\lambda}^\rho$ are the Littlewood-Richardson coefficients.

Back to Schubert calculus, the intersection of $\sigma_\mu$ and $\sigma_\lambda$ is given by the formula
\be
\sigma_\mu \cdot \sigma_\lambda = \sum_\rho c_{\mu,\lambda}^\rho \; \sigma_\rho.
\ee
%

%\subsection{Counting and $N^3MHV$}

Now we are ready to make the connection with the computation of residues we have to perform to get physical information. Note that when we set a given minor to zero we get a hypersurface in $G(k,n)$. This hypersurface turns out to coincide with $\sigma_1$ for some choice of flag\footnote{We use the usual mathematical notation where $\sigma_{a,0,\ldots ,0}$ is denoted by $\sigma_a$.}. The cohomology ring of the Grassmannian does not depend on the choice of the flag. Therefore when we talk about a given cycle we are implicitly taking about the corresponding class.

The dimension of $\sigma_1$ is one less than the dimension of the Grassmannian, i.e, of codimension one. If we set another determinant to zero we get another representative of the class of $\sigma_1$. Imposing both equations at the same time gives a cycles of codimension 2 which is homologous to the self-intersection of $\sigma_1$, i.e, $\sigma_1^2$.
At this point we have to remember that we are not working directly on the full $G(k,n)$. The constraints coming from momentum conservation restricts our computation to a smaller Grassmannian, i.e., $G(k-2,n-4)$. This reduction was discussed in detail in section \ref{subsec:GeoPic}.

In order to compute residues we need to take enough self intersections of $\sigma_1$ so that we find cycles of dimension $zero$. Starting with $\sigma_1$ of dimension $D-1$ where $D=(k-2)(n-k-2)$ is the dimension of $G(k-2,n-4)$, one has to compute $\sigma_1^D$.
Recall that $\sigma_1$ is associated with the tableaux consisting of a single box, i.e., the fundamental representation. Now we are interested in how many times the tableaux with $k-2$ rows and $n-k-2$ columns appears\footnote{A simple way of computing the codimension of a cycle given its tableaux is to count the number of boxes.}. In the case at hand the tableaux of interest corresponds to a cycle of codimension $(k-2)(n-k-2)$, i.e, a point!
The computation of the corresponding Littlewood-Richardson coefficient is a classic computation in enumerative geometry and gives
\be
\label{magic}
\# = \frac{1!2!\ldots (k-3)!D!}{(n-k-2)!(n-k-1)!\ldots (n-5)!}.
\ee

Let us test the formula against the cases we know. For $k=3$ and all
$n$ we find $\#=1$. This is indeed the case as any choice of $n-5$
determinants set to zero has only $1$ solution since  all equations
are linear. For $k=4$ and $n=8$ we find $\# = 2$. This is precisely
the curious fact we noticed in previous sections. Also for $k=4$
we have checked that when $n=9$ we get $\# = 5$ solutions. The most
impressive agreement is for the number of solutions in the N$^3$MHV,
i.e., $k=5$, $n=10$ amplitude. In this case, we have to find all
solutions of a system of 9 cubic polynomial equations in 9
variables. With the help of Mathematica we have checked that there
is precisely $\# = 42$ solutions!

\section{Discussion}

We have identified an object that autonomously computes the leading singularities of quantum scattering amplitudes in an entirely different way than the usual local formulation of quantum field theory. There is a great deal left to be understood; we end with a discussion of open problems, speculations and directions for future work.

\subsection{Identifying Contours}

The most immediate open question we confront is how to identify contours associated with leading singularities for completely general $k$ and $n$. As we have seen, the amplitudes are associated with equivalence classes of contours which are related by sums of residues that vanish due to the global residue theorem. We have seen a beautiful pattern to these equivalence classes for NMHV amplitudes. Beginning with the 8 pt N$^2$MHV amplitude the story gets even more interesting, involving ``composite" residues in addition to the usual ones; in fact as we pointed out for large $n = 2k$, all the residues must be composite, so understanding their physical significance will surely be important. But obviously, apart from simply identifying the contours associated with known leading singularities, we need to find an independent principle which makes the correct associations.

At 1-loop, all that is needed to completely determine the amplitude is a prescription for associating a contour $\Gamma$ with every scalar box; our conjecture then says that the full amplitude is
\be
\includegraphics[scale=0.7]{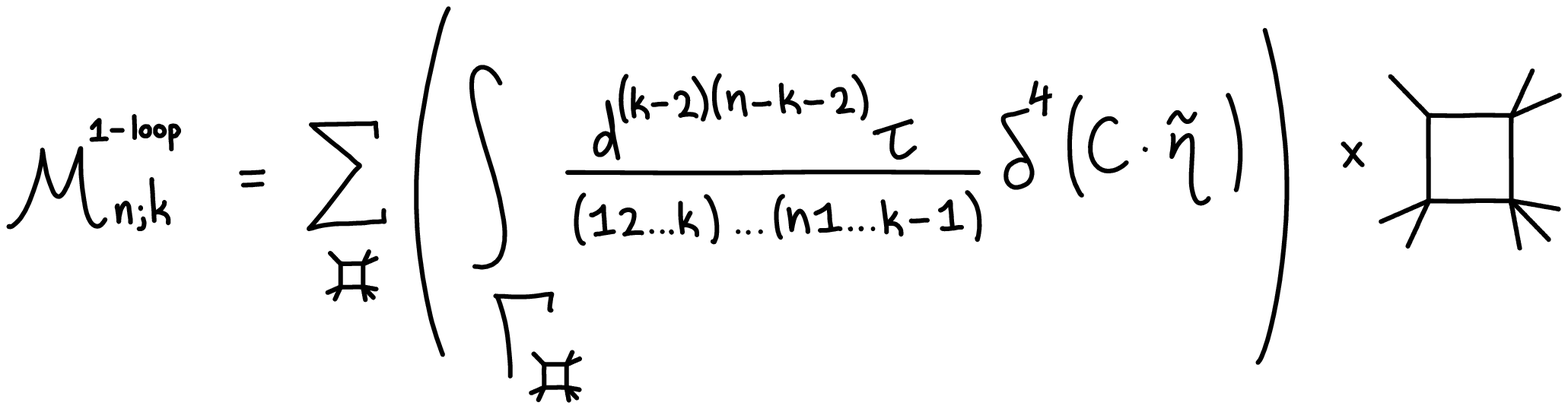} \nonumber
\ee
At higher loops, we need to learn both how to systematically associate residues with leading singularities as well as how to use the leading singularities to re-assemble the full amplitude. It is quite likely that these two steps are closely related to each other.
A possible clue is that the procedure for identifying residues is highly reminiscent of the way in which the leading singularities themselves are defined. At 1-loop one begins with a loop integral, thought of as a contour integral over the whole real axis for the 4 loop momenta, and then defines the leading singularity as being associated with exactly the same integral but evaluated on the two $T^4$'s associated with the two solutions to the quadruple cut equations. At 2 loops, one encounters ``composite" leading singularities entirely analogous to our ``composite" residues. This perhaps suggests that our expression for the leading singularities is in fact the reduction of some even larger integral representation of the full loop-level amplitude, and that the restriction to  $T^{4 \ell}$ contours which picks out leading singularities also specifies the correct contours in our conjecture. There are many other indications of a beautiful structure in the full loop amplitudes.  For instance at 1-loop, it is only the full amplitude (and not each individual scalar box) that exhibits the remarkable duality between Wilson loops and scattering amplitudes \cite{Drummond:2007aua, Drummond:2007au,
Drummond:2007cf, Drummond:2008aq, Alday:2007hr, Brandhuber:2007yx,
Berkovits:2008ic, Beisert:2008iq, McGreevy:2008zy}.

\subsection{Dual Superconformal Invariance and Momentum Twistors}

As we have stressed our basic expression for the amplitude makes the
action of the cyclic and parity symmetries manifest. As we have
learned over the past few years, SYM amplitudes also have a
remarkable dual superconformal invariance, which together with the
usual superconformal symmetry close into a Yangian algebra
\cite{DrummondYangian}, demonstrating the same integrable structure
in scattering amplitudes as has already been observed in the
spectrum of anomalous dimensions \cite{Beisert:2003tq}. While this
symmetry has been understood as a consequence of a fermionic
T-duality \cite{Berkovits:2008ic}, it would be nice to understand it even more directly.
Given that our proposal appears to unify all functions of the
kinematical invariants that appear in scattering amplitudes in one
object, it is very natural to explore whether it might make dual
superconformal invariance ``obvious". Indeed, perhaps the real
interpretation of our object is as a generating function for all
possible Yangian invariants which also knows about all the
remarkable relations among them.

Along these lines, it is perhaps worth pointing out some completely obvious re-writings of our initial twistor-space formula. We begin by writing the $\delta^{4|4}$'s as an integral over $k$ auxiliary
twistor
variables ${\cal Z}_\alpha$, and also writing each of the inverse determinant factors as a bosonic gaussian integral.
This gives
us a bosonic integral with a cubic action of the form \be
{\cal L}_{n;k}({\cal W}_a) =
\int d X^{(j)}_{\alpha} dY^{(j)}_\beta d {\cal Z}_\alpha dC_{\alpha a} e^{iS} \, , \, S = X^{(j)}_\alpha Y^{(j)}_\beta
C_{\beta, j + \alpha - 1} + {\cal W}_a C_{\alpha a} {\cal Z}_\alpha
\ee
This has the form of an action for a spin chain, with interactions involving sets of $k$ nearest-neighbors.

Properly interpreting dual conformal transformations in our formulation should also help us understand how our picture is related to the recent work of Hodges \cite{Hodges:2009hk}. His picture for NMHV amplitudes is of a volume integral in the dual twistor space, keeping the cyclic symmetries manifest, with the BCFW expression expressing the decomposition of this volume into tetrahedra, and unphysical poles associated with the internal faces of the polytope, which cancel in the sum representing the whole volume.
This seems very closely related to our picture in ordinary momentum or twistor space, where the amplitude is associated with a cyclically invariant homology class, even though any given contour integral representation does not have the symmetries manifest. On the other hand there are clear differences--our proposal also keeps parity manifest, and even for NMHV amplitudes, combines tree and loop information in the same object. Perhaps Hodges picture is more naturally associated with a ``T-dual" of the twistor string theory.

\subsection{Pure Yang-Mills?}
One of the striking features of maximally supersymmetric theories, even at tree-level, is that all the different helicities are unified in a single supermultiplet and the amplitudes can be treated as a single object with external states depending on Grassmann parameters. In the usual formulation of scattering amplitudes, the relations amongst the different helicity amplitudes that follow from SUSY seem utterly mysterious without knowledge of their supersymmetric origin. However our conjecture, written in its non-supersymmetric form, retains the same essential feature provided by SUSY: the integrand is a universal object, and the different helicity configurations arise from performing the integration over different charts on the Grassmannian. This perhaps suggests that SUSY is not playing a particularly crucial role in the story, and that there may be an extension of our formula for computing the subleading singularities as well, corresponding at one-loop to the triangle, bubble and rational pieces. Finding a generating function for all rational terms in scattering amplitudes would be particularly exciting both from a purely theoretical point of view, since these objects clearly have a fascinating and deep structure, as well as for practical purposes, since computing the rational parts of 1-loop amplitudes is crucial for state-of-the-art NLO calculations in QCD \cite{Berger:2008ag}.

\subsection{Emergent Space-Time and Unitarity}

We are still missing a real understanding of the physics behind our conjecture. A clue is perhaps provided by the nature of the space in which it is formulated. Instead of space-time, for $n$-particle scattering the dual is naturally formulated in an $n$-dimensional space. Thinking along ``holographic" lines, we may have expected a theory living on the boundary of spacetime or a close cousin like twistor space. However we are finding that our dual picture isn't associated with a space in which particles live at all; instead most of the action takes place in the Grassmannians $G(k,n)$. All the kinematical information associated with spacetime goes into specifying special directions in this $n$-particle space, which pick out out a smaller $G(k-2,n-4)$ Grassmannian naturally embedded in the larger one. Thus, while our duality is trivially ``holographic" given that we are working directly with on-shell variables, space-time emerges in a less directly ``holographic" way than we are accustomed to in AdS/CFT, beginning
instead from a picture where there is no space of any sort and the arena in which the dynamics takes place is determined by the number of particles.

One might be tempted to loosely think of this $n$-particle space as analogous to a classical phase space, but the similarity seems superficial. Dynamics in classical phase space is local. By contrast, leading singularities are contour integrals in $G(k-2,n-4)$, and they are associated with topological information. Indeed we can focus on the
$G(k-2,n-4)$ space, ``punctured" by removing the zero locus defined by the vanishing of the minor factors; the contours defining the amplitudes are then associated with non-trivial homology classes in this space. This gives several equivalent representations to a given amplitude, which helps us see the emergence of space-time by enforcing the cancelation of unphysical poles and infrared equations at loop level.

We emphasize the emergence of space-time since this is something one would expect to see in a dual theory of the S Matrix, but in fact it is possible that unitarity should also be thought of as emergent in this picture. After all, every discussion of scattering amplitudes begins by imagining all momenta as incoming, which immediately removes the unitary ``matrix" interpretation of scattering! An even stronger indication that unitarity is not a fundamental ingredient in the dual we have been studying is that the object which is being directly computed is the
leading singularity and not the unitarity cut. Consider the 1-loop amplitude; given the leading singularities we can construct an object that has the correct quadruple cut, which (with maximal supersymmetry) also correctly reproduces all the subleading cuts. The first subleading cut is the triple cut, and the ``double-cut" (which is the cut associated with unitarity) does not seem particularly privileged. The ``double-cut" only acquires special significance working in $(3,1)$ signature with real momenta, since both the quadruple cut and triple cuts vanish in this case.

A perhaps too-radical thought inspired by these observations is that the dual theory cares neither for space-time nor unitarity and computes ``amplitudes" as determined by the leading singularities. The very special analytic structure of the ``amplitudes" allows a local space-time description; restricting to spinor helicities satisfying a reality property $\tilde \lambda = \lambda^*$ gives $(3,1)$ signature and the possibility of a causal interpretation. The unitarity cut equations are satisfied as a rather incidental fact, and therefore the ``amplitudes" have a probabilistic interpretation. In this way both space-time and the usual quantum mechanical rules may arise from a more primitive starting point.

\subsection{${\cal N} = 8$ SUGRA}

Finally, we make some obvious comments about the extension of our work to ${\cal N}=8$ SUGRA \cite{Cremmer:1978ds, Cremmer:1978km, Cremmer:1979up}, which is bound to be much more interesting \cite{Bern:1998ug, Bern:1998sv, Bern:2005bb, BjerrumBohr:2005xx,
BjerrumBohr:2006yw, Bern:2006kd, Bern:2007xj, BjerrumBohr:2008vc,
BjerrumBohr:2008ji, BjerrumBohr:2008dp, Bern:2009kd, Bern:2008pv, Berkovits:2006vc, Green:2006yu, Howe:2002ui, Bern:2007hh,
 Kallosh:2007ym, Kallosh:2008mq, Kallosh:2008ru,
Bern:2009kf}.
 As stressed in \cite{simplest}, gravity amplitudes have a far richer structure than in Yang-Mills theory. They are governed by much larger ``obvious" symmetries: instead of cyclic invariance, supergravity amplitudes have permutation symmetry, and instead of having the massless scattering amplitudes only defined at the origin of moduli space as for ${\cal N} = 4$ SYM, the ${\cal N}= 8$ SUGRA S Matrix is defined everywhere on moduli space and is non-trivially acted on by the $E_{7(7)}$ symmetry \cite{simplest, Kallosh:2008rr, Kallosh:2008ic}. There are also indications of many more non-trivial relations between gravity amplitudes than exist for Yang-Mills. To begin with, gravity amplitudes are softer at complex infinity than gauge amplitudes, and this gives rise to extra relations for tree gravity amplitudes beyond those given by the BCFW recursions relations. A large number of such relations are needed in order to guarantee a remarkable property of loop-level gravity amplitudes: the soft limits are not renormalized beyond tree level. This is obvious from the usual local description but somewhat miraculous using on-shell methods, and its validity requires yet further non-trivial relations between tree gravity amplitudes.

All of these observations suggest a far richer structure controlling gravity amplitudes already at tree level, and yet there has always been a note of discouragement in pursuing this thought: even the tree MHV gravity amplitudes look more complicated than the beautifully simple Parke-Taylor amplitude. We believe this is because we have yet to see supergravity amplitudes expressed in a form that makes the full permutation symmetry manifest, and that the analog of our ${\cal N} = 4$ conjecture for ${\cal N} = 8$ SUGRA will make it clear that supergravity is indeed the ``simplest" theory in this dual formulation \cite{simplest}. There are already very encouraging indications that the link representation for gravity gives rise to a remarkable new form for tree MHV gravity amplitudes \cite{Nguyen:2009jk}.

There is an opportunity to study questions related to ${\cal N}=8$ SUGRA by looking at subleading trace structures in ${\cal N}=4$ SYM. For instance, the soft $\frac{1}{z^2}$ behavior of gravity amplitudes under BCFW deformation is also present for ${\cal N}=4$ SYM amplitudes, when non-adjacent legs are deformed. All of the structure we have seen in this paper, including the remarkable identities between different BCFW representations of tree amplitudes, have instead held only for planar BCFW diagrams associated with deforming adjacent legs. This suggests that even Yang-Mills theory should exhibit a rich structure in its non-planar sector, and understanding this will be relevant to ${\cal N}=8$ SUGRA as well.

\section*{Acknowledgments}

We thank Mark Spradlin and Nastja Volovich for very inspiring
discussions regarding a link representation form of the ``connected
prescription" for twistor string theory, and in particular for
pointing out that, up to annoying sign factors, the connected
prescription suggests a different way of combining the BCFW terms of
the 6 particle amplitude into a single object. We also thank James
Drummond, Andrew Hodges, Lionel Mason, Roger Penrose and David
Skinner for several days of enjoyable conversations. We finally
thank Fernando Alday, Jake Bourjaily, Davide Gaiotto, Juan
Maldacena, Ravi Vakil and Edward Witten for discussions and
correspondence. N.A.-H. is supported by the DOE under grant
DE-FG02-91ER40654, F.C. was supported in part by the NSERC of Canada
and MEDT of Ontario, and J.K. is supported by a Hertz foundation
fellowship and an NSF fellowship.

\appendix

\section{Relation to the Link Representation}

In this appendix we show how that our conjecture is related to what we called the ``link representation" of scattering amplitudes in \cite{ArkaniHamed:2009si}. Amongst other things, this will allow us to interpret (and finally be rid of) one of the annoying features of representing scattering amplitudes in twistor space--the presence of ``infinity twistor signs" \cite{ArkaniHamed:2009si, lioneldavid, Bargheer:2009qu} that obscure  manifest conformal invariance. It has always seemed vaguely like these signs are only present to ensure correct contour orientations, and we will see how this arises very precisely.

Let us illustrate how this works for the simplest non-trivial case of the 6pt NMHV amplitude. As explained in detail in \cite{ArkaniHamed:2009si}, assigning $i = 1,3,5$ to ${\cal W}'s$ and $I=2,4,6$ to ${\cal Z}'s$, we can write the full superamplitude as $M_6 = (1 + g^2 + g^4) U_6$ where
the twistor diagram for $U_6$ is
\be
\includegraphics[width=6cm]{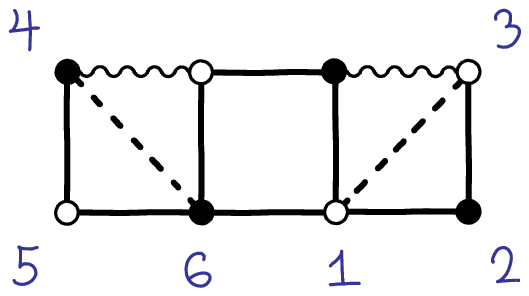}
\ee
This also allows us to write an expression for the superamplitude in what we called ``the link representation" in twistor space:
\be
U_6 = {\rm sgn}(\langle 46 \rangle [13]) \int d^9c_{Ii} U_6(c_{Ii}) e^{i c_{Ii} {\cal W}_i {\cal Z}_I}
\ee
with
\be
U(c_{Ii}) = \delta (c_{25}) \frac{1}{c_{45} c_{65}} \times \frac{1}{c_{61} c_{43} (c_{61} c_{43} - c_{41} c_{63})} \times \frac{1}{c_{21} c_{23}}
\ee
Since all the $\tilde \mu_i, \mu_I$ dependence of amplitudes in this representation is in the $e^{i c_{Ii} {\cal W}_i {\cal Z}_I}$ factor, it is trivial to go back to momentum space. Using the (say) $\tilde \eta$ representation for all particles we have
\begin{equation}
U_6 =  {\rm sgn}(\langle 46 \rangle [13]) \int d^9 c_{Ii} U_6(c_{Ii}) \delta^2(\lambda_i - c_{Ii} \lambda_I) \delta^2(\tilde \lambda_I + c_{Ii} \tilde \lambda_i) \delta^4(\tilde \eta_I + c_{Ii} \tilde \eta_i)
\end{equation}
As explained in the text, the bosonic delta functions have linearized momentum conservation, since
\begin{equation}
\lambda_i - c_{Ii} \lambda_I = 0, \, \tilde \lambda_I + c_{Ii} \tilde \lambda_I = 0 \implies \lambda_i \tilde \lambda_i + \lambda_I \tilde \lambda_I = 0
\end{equation}
Thus the 12 bosonic delta functions turn into the 4 delta functions for momentum conservation plus 8 more delta functions restricting the $c_{Ii}$.
Note that $U_6$ also contains a $\delta c_{25}$ factor, so the remaining 8 $c_{Ii}$ are completely determined by the 8 delta functions.
For instance we can use the $\lambda_5$ and $\tilde \lambda_2$ equations to solve for $c_{65},c_{45},c_{21},c_{23}$,
\begin{equation}
c_{65} = \frac{\langle 5 4 \rangle}{\langle 6 4 \rangle}, \, c_{45} = \frac{\langle 5 6 \rangle}{\langle 4 6 \rangle}, \, c_{21} = \frac{[23]}{[13]}, c_{23} = \frac{[21]}{[31]}
\end{equation}
and then use, say, the $\tilde \lambda_4, \tilde \lambda_6$ equations to solve for the rest of the $c$'s,
\begin{equation}
c_{41} = \frac{\langle 6 | (p_5 + p_4) |3]}{\langle 4 6 \rangle [1 3]}, c_{43} =  \frac{\langle 6 | (p_5 + p_4) |1]}{\langle 4 6 \rangle [3 1]}; c_{61} = \frac{\langle 4 | (p_5 + p_6) |3]}{\langle 6 4 \rangle [3 1]}, c_{63} = \frac{\langle 4 | (p_5 + p_6) |1]}{\langle 6 4 \rangle [1 3]}
\end{equation}
From here it is straightforward to recover momentum space amplitudes. To obtain the $(+-+-+-)$ amplitude, we integrate $\int d^4 \tilde \eta_I$, which simply gives 1, and set the remaining $\tilde \eta_i \to 0$. Taking account the Jacobian in pulling out the overall momentum conserving $\delta$ function and inserting the solutions for the $c_{Ii}$ into $U_6$ gives us the momentum space amplitude. We can also easily obtain the $(-+-+-+)$ amplitude, by integrating $\int d^4 \tilde \eta_i$, and setting $\tilde \eta_I \to 0$. The integrals yield a factor of det$^4(c_{Ii})$, so we find
\begin{eqnarray}
U_6^{-+-+-+} = {\rm det}^4(c_{Ii}) U_6^{+-+-+-}  &=& \left(\frac{[2|(1 + 3)|5 \rangle}{\langle 4 6 \rangle [1 3]}\right)^4 U_6^{+-+-+-} \nonumber \\ =
\frac{[2|(1+3)|5 \rangle^4}{[12][23]\langle 45 \rangle \langle 56 \rangle (p_4 + p_5 + p_6)^2}
&\times& \frac{1}{\langle 6 |5 + 4|3] \langle 4|5
 + 6 |1]}
\end{eqnarray}
and of course
\be
M_6^{-+-+-+} = (1 + g^2 + g^4) U_6^{-+-+-+-}
\ee
As an aside we can quickly derive the identities that follow from the cyclic symmetries of the superamplitude. For instance cyclicity implies that $M_6^{+-+-+-}(\lambda_i,\tilde \lambda_i) = M_6^{-+-+-+}(\lambda_{i+1}, \tilde \lambda_{i+1})$. Given the expressions we have just given for these two amplitudes we find a 6 term identity
\begin{eqnarray}
(1 + g^2 + g^4)  \frac{\langle 4 6 \rangle^4 [1 3]^4}{[12][23]\langle 45 \rangle \langle 56 \rangle (p_4 + p_5 + p_6)^2} \frac{1}{\langle 6 |5 + 4|3]
\langle 4|5 + 6 |1]} \nonumber \\ = (1 + g^2 + g^4) \frac{[3|(2+4)|6 \rangle^4}{[23][34]\langle 56 \rangle \langle 61 \rangle (p_5 + p_6 + p_1)^2}
\frac{1}{\langle 1 |6 + 5|4] \langle 5|6+ 1 |2]}
\end{eqnarray}
Of course this is just one component of an identity as a function of
the Grassmann parameters $\tilde \eta_i$. This is precisely the same
as the identity $M_{{\rm BCFW}} = M_{{\rm P(BCFW)}}$ we mentioned in
the introduction.

At any rate, the whole tree amplitude is the sum of three terms; we would like to understand them as all arising from a single object. The $\delta(c_{25})$ is a clue; perhaps we could somehow replace this with a factor $1/(c_{25})$ and think of $U_6$ as computing a residue on the pole where $c_{25} = 0$. The sgn$\langle 46 \rangle [13]$ factor appears to be an annoying obstruction to this identification, but in fact as we will see it has an important role to play to ensure everything works out perfectly.

We are actually very close to the desired result, but to see this, we need a small piece of inspiration: we have to
recognize that the factor $c_{45} c_{21}$ is really the same as
$c_{45} c_{21} - c_{25} c_{41} \equiv \tilde c_{63}$ when  evaluated on the support of $\delta(c_{25})$,
and similarly for $c_{65} c_{23} = c_{65} c_{23} - c_{25} c_{63} \equiv \tilde c_{41}$, these are of the same form as the already appearing factor $c_{14} c_{36} - c_{16} c_{34} \equiv \tilde c_{25}$.
In fact we were directly motivated to make this identification by studying the localization properties of this amplitude in twistor space, but it would take us too far afield to explain that here. For now we simply note that with the identification we can write $U_6$ in a more symmetrical looking form as
\be
U_6 = \int d^9 c_{Ii} e^{i c_{Ii} W_i Z_I} \, {\rm sgn}(\langle 4 6 \rangle [13]) \delta(c_{25}) \frac{1}{c_{41} c_{63}} \frac{1}{\tilde c_{25} \tilde c_{41} \tilde c_{63}}
\ee
where
\be
\tilde c_{Ii} = \epsilon_{ijk} \epsilon_{IJK} c_{jJ} c_{kK}
\ee
This form already motivates the loose association
\be
U_6 \sim {\rm Res} \left[ \frac{1}{c_{25} \tilde c_{61} c_{41} \tilde c_{25} c_{63} \tilde c_{41}}\right]_{c_{25} = 0}
\ee
which we will make completely precise in a moment. Note however before proceeding that the factors in the denominator are in perfect agreement with the consecutive minors of of the matrix
\be \label{alter} C =
\left(\begin{array}{cccccc} c_{21} & 1 & c_{23} & 0 & c_{25} & 0 \\
c_{41} & 0 & c_{43} & 1 & c_{45} & 0 \\ c_{61} & 0 & c_{63} & 0 &
c_{65} & 1
\end{array}\right).
\ee
Namely
\be
\label{cmap}
c_{41} = (612), c_{63} = (234), c_{25} = (456), \tilde c_{25} = (123),  \tilde c_{41} = (345), \tilde c_{63} = (561)
\ee
In order to do complete the transformation of our integral involving $\delta(c_{25})$ into the calculation of a residue, let us go back to momentum space, and repeat the steps we described in section 2 in the analysis of our formula.
Looking at the space of link variables on the support of the bosonic delta functions,
\begin{equation}
\lambda_i - c_{Ii} \lambda_I = 0, \, \tilde \lambda_I + c_{Ii} \tilde \lambda_i = 0
\end{equation}
we have 12 equations, but 4 are redundant given 4-momentum conservation, so there are 8 equations for 9 unknowns. The set of $c_{Ii}$ solving these
equations lie on a line in the 9 dimensional $c_{Ii}$ space. We can parametrize these solutions e.g. by solving for all the $c$'s as a function of one of them, or we can parametrize them more elegantly as
\begin{equation}
c_{Ii}(\tau) = c^*_{iI} + \epsilon_{ijk} \epsilon_{IJK} [jk] \langle J K \rangle \tau
\end{equation}
where $c^*_{iI}$ is any specific solution.
Here we use the fact that for three two-dimensional vectors e.g. $\lambda_{I,J,K}$,
\be
\epsilon_{IJK} \lambda_I \langle J K \rangle = 0
\ee
We can then write the product of the $\delta^2$ functions as the momentum conserving delta function multipled by an integral that localizes the $c_{Ii}$ to this
line:
\begin{equation}
\prod_i \delta^2(\lambda_i - c_{Ii}\lambda_I) \prod_I \delta^2(\tilde \lambda_I + c_{Ii} \tilde \lambda_i) = \delta^4(\sum_a \lambda_a \tilde \lambda_a) \int d \tau \delta^9 \left(c_{Ii} - c_{Ii}(\tau) \right)
\end{equation}
(it is easy to check that the Jacobian is 1).

Using this we can re-write $U^{+-+-+-}_6$, explicitly factoring out the momentum conserving delta function, as
\be
U^{+-+-+-}_6 = \delta^4(\sum_a p_a) \int d \tau  \frac{{\rm sgn}(\langle 4 6 \rangle [13]) \delta(c_{25}(\tau))}{c_{41}(\tau) c_{63}(\tau) \tilde
c_{25}(\tau) \tilde c_{41}(\tau) \tilde c_{63}(\tau)}
\ee
Now, something very nice happens, which allows us to finally interpret--and be rid of!--the pesky sign factors! We can always shift the origin of $\tau$ so that
\begin{equation}
c_{25}(\tau) = \langle 4 6 \rangle [13] \tau
\end{equation}
Then doing the $\tau$ integral simply gives us
\be
U^{+-+-+-}_6  = \delta^4(\sum_a p_a) \frac{1}{\langle 4 6 \rangle [13]} \times \frac{1}{(c_{41} c_{63} \tilde c_{25} \tilde c_{41} \tilde c_{63})(\tau = 0)}
\ee
But we can immediately recognize
\be
\frac{1}{\langle 4 6 \rangle [13]} = \oint_{c_{25}(\tau) = 0} \frac{1}{c_{25}(\tau)}
\ee
and therefore we have finally identified $U_6$ as a residue, which
\be
U_6^{+-+-+-} = \delta^4(\sum_a p_a) \oint_{c_{25}(\tau) = 0} \frac{1}{(c_{25} c_{63} c_{41} \tilde c_{25} \tilde c_{63} \tilde c_{41})(\tau)}
\ee
which we can re-write in precisely the form of our conjecture using equation (\ref{cmap}) as
\be
U_6^{+-+-+-} = \delta^4(\sum_a p_a) \oint_{c_{25}(\tau) = 0} \frac{1}{[(123)(234)(345)(456)(561)(612)](\tau)}
\ee
Note that in the end the role of the annoying sgn factors was just to tell us to consistently define all residues with the same orientation!

As discussed in the text, the BCFW form of the full tree amplitude is given by a contour integral enclosing the poles where $(612),(234),(456)$ vanish, while the $P(BCFW)$ form of the amplitude used the other three poles, and the remarkable 6 term identity equating these two forms just follows from Cauchy's theorem.

\section{An Explicit 7 particle Computation}

In this appendix we give an explicit derivation of a class residues for the 7 particle NMHV amplitude, associated with the three-mass boxes. This only involves solving linear equations and is completely straightforward. We include it here so the reader can transparently see what is involved; as mentioned in the text for the cases with higher $k$ and $n$ we leave the computation of residues to Mathematica.
%computing the residue corresponding to the coefficient of the three-mass box integral at $n=7$ particles. In appendix , we present very general formulas that can be used to compute any residue for $k=3$ and any $n$.
Up to cyclic permutations, there is a unique three-mass box for the $7$ particle amplitude.  We will compute the box with particles  $(2),(3,4),(5,6),(7,1)$ at each of the four corners.  We find it convenient to use the gauge fixing
\be
C = \left(\begin{array}{ccccccc}
1 & 0 & 0 & c_{14} & c_{15} & c_{16} & c_{17}  \\
0 & 1 & 0 & c_{24} & c_{25} & c_{26} & c_{27}  \\
0 & 0 & 1 & c_{34} & c_{35} & c_{36} & c_{37}   \end{array} \right)
\ee

We now have to examine the linear equations $\lambda_i - c_{Ii} \lambda_I = 0, \, \tilde \lambda_I + c_{Ii} \tilde \lambda_i$ = 0, as well as the linear equations involved with setting minors to zero. Here
$I = 1,2,3$ and $i=4,5,6,7$. We can solve these equations in any order we wish, and for convenience we will first impose the $\lambda$ equations, then the equation setting the minors to zero, and finally the $\tilde \lambda$ equation.

Let us begin with the $\lambda_i$ equations:

\be
\lambda_i = c_{i1}\lambda_1 + c_{i2}\lambda_2 + c_{i3}\lambda_3
\ee
These are two equations for three unknowns. We parameterize the solution as
\be
\label{pari}
c_{i1}(\tau_i) = \frac{\langle i~3\rangle -\langle 2~3\rangle \tau_i}{\langle 1~3\rangle}, \quad c_{i2}(\tau_i)=\tau_i, \quad c_{i3}(\tau_i) = \frac{\langle i~1\rangle -\langle 2~1\rangle \tau_i}{\langle 3~1\rangle}.
\ee
The three-mass box coefficient will correspond to the residue that occurs when the determinants $(234)$ and $(712)$ vanish.  This allows us to solve for
\be
\tau_4 = \frac{\langle 4~3 \rangle}{\langle 2~3\rangle}, \quad \tau_7 = \frac{\langle 7~1 \rangle}{\langle 2~1 \rangle}
\ee
Next let's impose the $\tilde \lambda_I + c_{Ii} \lambda_i = 0$ equation. This constraint translates into two equations which fix the remaining two $\tau$'s and introduce a dependence on the anti-holomorphic brackets--one might have expected six equations, but four turn into momentum conservation on the support of the other two.  For example we can choose to impose
\be
\label{lasty}
\tilde\lambda_{2} = \sum_{i\neq\{1,2,3\}}\tau_{i}\;\tilde\lambda_i,
\ee
and since $\tilde\lambda$ are two-component objects these constitute two equations.  The solutions are
\be
%\tau_5 = \frac{[2~6]}{[5~6]} + \frac{[6~4] \langle 4~3 \rangle}{[5~6] \langle 2~3 \rangle} + %\frac{[6~7]\langle 7~1 \rangle}{[5~6] \langle 2~1 \rangle} \\
%\tau_6 = \frac{[5~2]}{[5~6]} + \frac{[4~5]\langle 4~3 \rangle}{[5~6]\langle 2~3 \rangle} + \frac{[5~7] %\langle 1~7 \rangle}{[5~6] \langle 2~1 \rangle}
\tau_5 &=& \frac{\text{[76]} \langle 71\rangle }{\text{[65]} \langle
   21\rangle }-\frac{\text{[64]} \langle 43\rangle
   }{\text{[65]} \langle 23\rangle
   }-\frac{\text{[62]}}{\text{[65]}} \\
\tau_6 &=& \frac{\text{[75]} \langle 71\rangle }{\text{[56]} \langle
   21\rangle }-\frac{\text{[54]} \langle 43\rangle
   }{\text{[56]} \langle 23\rangle
   }-\frac{\text{[52]}}{\text{[56]}}
\ee
To obtain our final expression for the box coefficient, we need to evaluate the determinants $(2,3,4)$, $(3,4,5)$, $(4,5,6)$, and $(5,6,7)$ on the pole, and compute the associated Jacobians from the delta function constraints, the poles, and the transition to the standard momentum conservation delta function.  Plugging in for $\tau_i$ in the determinants, we obtain
\be
(3,4,5)&=& \frac{\langle 43\rangle [6|1+7|2\rangle
   }{[65]\langle 32\rangle \langle 21\rangle
   }\\
(4,5,6)&=&\frac{\langle 21\rangle  ([2|5+6|4\rangle
   \langle 32\rangle +
   t_4^{[3]} \langle
   43\rangle )+[7|5+6|4\rangle  \langle
   32\rangle  \langle 71\rangle }{[65] \langle
   21\rangle  \langle 31\rangle  \langle 32\rangle } \\
(5,6,7)&=& \frac{\langle 21\rangle  (-[2|5+6|7\rangle
   \langle 32\rangle -[4|5+6|7\rangle  \langle
   43\rangle )-
   t_5^{[3]}  \langle
   32\rangle  \langle 71\rangle }{[65]\langle
   21\rangle  \langle 31\rangle  \langle 32\rangle }\\
(6,7,1)&=& \frac{\langle 17\rangle [5|3+4|2\rangle
   }{[65]\langle 32\rangle \langle 21\rangle
   }
\ee
while $(1,2,3)=1$ and $(2,3,4)=(7,1,2)=0$. A short computation shows that the Jacobian takes the very simple form
\be
\mathcal{J}&=& \frac{1}{\text{[56]} \langle 12\rangle  \langle 23\rangle }
\ee
Thus, our final expression for the three-mass box coefficient at $n=7$ is
\be
L_{3,7}= \delta^4\left(\sum_{i=1}^7\lambda_i\tilde\lambda_i\right) \frac{{\cal J}\; \prod_{m=1}^{3}\delta^4(\tilde\eta_m+c_{i,m}(\tau_{5},\tau_{6})\tilde\eta_i)}{(3,4,5)(4,5,6)(5,6,7)
(6,7,1)}
\ee
In order to extract the box coefficient corresponding to various helicity assignments for the external particles, we simply perform the appropriate grassman integrations.  For instance, if want the three-mass box coefficient for the split helicity configuration $M(1^- 2^- 3^- 4^+ 5^+ 6^+ 7^+)$, then because of our particular gauge fixing choice, we can simply set all the $\tilde\eta$'s to zero, giving
\be
L_{\{1,2,3\},7}= \delta^4\left(\sum_{i=1}^7\lambda_i\tilde\lambda_i\right) \frac{{\cal J}}{(3,4,5)(4,5,6)(5,6,7)
(6,7,1)}
\ee
If instead we want the three-mass box coefficient for the mostly alternating configuration $M(1^- 2^+ 3^- 4^+ 5^- 6^+ 7^+)$, then we need to flip the helicities of particles 2 and 5 relative to the split helicity case.  For this reason we grassman integrate $L_{3,7}$ over $\int d^4 \tilde\eta_2 d^4 \tilde\eta_5 e^{\eta_2 \tilde\eta_2 + \eta_5 \tilde\eta_5}$, and then set all the $\eta$'s and $\tilde\eta$'s to zero.  Because of the simple dependence on the grassman variables, this integral simply brings down a factor of $c_{25}^4$, so
\be
L_{\{1,3,5\},7}&=& \delta^4\left(\sum_{i=1}^7\lambda_i\tilde\lambda_i\right) \frac{{\cal J} c_{25}^4}{(3,4,5)(4,5,6)(5,6,7)
(6,7,1)} \\
c_{25} &=&\frac{[6|7+2|1\rangle \langle 23\rangle + [6|4|3\rangle \langle 21\rangle}{[65]\langle 12\rangle\langle 23\rangle}
\ee
where we have solved for $c_{25}$ in terms of the $\tau_i$.  Box coefficients for all other helicity combinations of the NMHV seven-particle can be computed in this fashion.

\section{IR equations at 1-loop}

Here we give a short description of IR equations at 1-loop, for more
details see \cite{IReq, catani}. One-loop partial amplitudes in Yang
Mills are IR divergent. Due to color ordering, the divergences can
only depend on the momentum of two consecutive external particles
and it must be proportional to the tree-level amplitude. More
explicitly and in dimensional regularization,
\be
\left. M^{\rm 1 \hbox{-} loop}_n\right|_{\rm IR}  = -\frac{1}{\epsilon^2}\sum_{i=1}^n (-s_{i,i+1})^{\epsilon} M^{\rm tree}_n.
\ee
As discussed in previous sections, one-loop amplitudes can be written as linear combinations of scalar box integrals. Each integral can have IR divergences that depend on the kinematical invariants of the particular box under consideration. Only the four-mass box integral is completely finite.

Let us denote each box integral by $I(K_1,K_2,K_3,K_4)$ as in section \ref{subsec:4m} and the main kinematical invariants by $s=(K_1+K_2)^2$ and $t=(K_2+K_3)^2$. Each $K_i$ is the sum of consecutive external momenta. If $K_i$ is equal to the momentum of a single particle then it is null, if it is the sum of two or more then we say that it is massive. We denote null momenta by lower case letters $(p,q,r)$ while massive momenta by capital letters $(P,Q,R)$. Using our notation one-mass, two-mass-easy, two-mass-hard and three-mass boxes are respectively given by: $I(p,q,r,P)$, $I(p,P,q,Q)$, $I(p,q,P,Q)$, and $I(p,P,R,Q)$.

We now list the IR divergent structure of each of the four classes of IR singular boxes:
\begin{eqnarray}
\label{rhs}
I(p,q,r,P) & = & -\frac{1}{\epsilon^2}\left( (-s)^{-\epsilon} + (-t)^{-\epsilon} - (-P^2)^{-\epsilon} \right), \cr
I(p,P,q,Q) & = & -\frac{1}{\epsilon^2}\left( (-s)^{-\epsilon} + (-t)^{-\epsilon} - (-P^2)^{-\epsilon} - (-Q^2)^{-\epsilon}\right), \cr
I(p,q,P,Q) & = & -\frac{1}{\epsilon^2}\left( \frac{1}{2}(-s)^{-\epsilon}+(-t)^{-\epsilon}-\frac{1}{2}(-P^2)^{-\epsilon}-\frac{1}{2}(-Q^2)^{-\epsilon}\right),\cr
I(p,P,R,Q) & = &  -\frac{1}{\epsilon^2}\left( \frac{1}{2}(-s)^{-\epsilon} + \frac{1}{2}(-t)^{-\epsilon} - \frac{1}{2}(-P^2)^{-\epsilon} - \frac{1}{2}(-Q^2)^{-\epsilon}\right).
\end{eqnarray}

We would like to propose a distinction between two classes of IR equations. There is the class that relates the one-loop coefficients to the tree-level amplitude and the class that relates one-loop coefficients purely among themselves. In the following we will argue that this separation is meaningful. Moreover, we will find that the latter class is a subset of more purely one-loop identities which turn our to be derivable from a generalization of the residue theorem.

The first class is obtained by collecting all terms in the expansion in scalar box integrals that are proportional to a given $-\frac{1}{\epsilon^2}(-s_{i,i+1})^{\epsilon}$ divergence. The sum of the corresponding coefficients weighted by the numerical factors in (\ref{rhs}) must then be equal to $M_n^{\rm tree}$.
The second class is obtained by collecting all terms in the expansion in scalar box integrals that are proportional to $-\frac{1}{\epsilon^2}(-(p_i+p_{i+1}+\ldots + p_{i+m})^2)^{\epsilon}$ with $m>1$. The sum of the corresponding coefficients weighted by the numerical factors in (\ref{rhs}) must now vanish.

The first class is used to obtain the $\frac{1}{2}[{\rm BCFW} + {\rm P(BCFW)}]$ form of recursion relations for tree amplitudes. An example of the second class which we use in the text is provided by the seven particle amplitude. Consider for instance the equation coming from $t_{123}$,
\begin{eqnarray}
0 = & -B_{(1)(2)(3)(4567)} + B_{(7)(1)(23)(456)} + B_{(3)(4)(567)(12)} -\frac{1}{2}B_{(4)(5)(67)(123)} -\frac{1}{2}B_{(6)(7)(123)(45)} \nonumber \\ & -B_{(4)(56)(7)(123)} + B_{(1)(23)(4)(567)} + B_{(3)(456)(7)(12)} +\frac{1}{2} B_{(3)(45)(67)(12)} + \frac{1}{2} B_{(1)(23)(45)(67)}. \nonumber
\end{eqnarray}
As we discuss in the text, this highly non-trivial identity is exactly equivalent to a residue theorem.

\end{document}